\documentclass[11pt]{article}
\usepackage[a4paper,margin=2.3cm]{geometry}
\usepackage[T1]{fontenc}
\usepackage{mathptmx}
\usepackage{amsmath,amssymb}
\usepackage{graphicx}
\usepackage{booktabs}
\usepackage{multirow}
\usepackage{xcolor}
\usepackage[super,sort&compress]{natbib}
\usepackage{xurl}
\usepackage[colorlinks=true,linkcolor=blue!50!black,citecolor=blue!50!black,urlcolor=blue!50!black]{hyperref}
\usepackage{caption}
\usepackage{longtable}
\usepackage{placeins}
\usepackage{pdflscape}   %
\usepackage{adjustbox}
\usepackage{authblk}
\hypersetup{pdftitle={CrystalJev: calibrated foundation-model probabilities decide what to compute next in materials discovery},
  pdfkeywords={CrystalJev, System-1 models, atomistic foundation models, universal interatomic potentials, calibration, conformal prediction, value of information, materials discovery, Matbench Discovery}}
\DeclareCaptionLabelSeparator{nbar}{\,$|$\,}
\renewcommand{\figurename}{Fig.}

\newcommand{\ehat}{\hat{E}_{\mathrm{hull}}}
\newcounter{snote}
\newcommand{\snote}[2]{\refstepcounter{snote}\label{#2}\subsection*{Supplementary Note \thesnote: #1}}

\title{CrystalJev: calibrated foundation-model probabilities\\ decide what to compute next in materials discovery}
\author[1,2,3,*]{Peng Kang}
\author[1,2]{Yaxuan Sun}
\author[1,2,3]{Zhen Li}
\author[1,2,3]{Yu Liu}
\author[1,2,3,*]{Lei Zheng}
\author[1,2,3]{Huibin Xu}
\affil[1]{School of Materials Science and Engineering, Beihang University, Beijing 100191, China}
\affil[2]{National Key Laboratory of Artificial Intelligence for Material Science, Beihang University, Beijing 100191, China}
\affil[3]{Tianmushan Laboratory, Beihang University, Hangzhou 311115, China}
\affil[*]{Corresponding authors: Peng Kang (\href{mailto:pengkang@buaa.edu.cn}{pengkang@buaa.edu.cn}) and Lei Zheng (\href{mailto:zhenglei@buaa.edu.cn}{zhenglei@buaa.edu.cn})}
\date{}

\begin{document}
\maketitle

\begin{abstract}
\noindent Atomistic foundation models screen hypothetical crystals by thresholding predicted energies, an output that states neither how likely a candidate is to be stable nor whether more computation would change the decision. CrystalJev reads one forward pass of a frozen potential on the unrelaxed structure and returns calibrated probabilities, conformal guarantees and a rule for escalating candidates to relaxation or density-functional theory (DFT). Across 65 Matbench Discovery models, threshold calls hide cost assumptions, and ordering slower computation by its value for the decision, rather than by probability, recovers most of its benefit at a tenth of its cost. In a registered prospective test with 2,100 DFT calculations on ionic substitutions, forecasts calibrated on existing data over-stated the stable fraction of new candidates by 0.9--2.2 percentage points, because the population had shifted, not the labelling protocol; a few hundred labels from the new pool removed this bias more reliably than label-free corrections.
\end{abstract}

Computational materials discovery is a funnel of decisions. Generative models and substitution rules propose candidate crystals by the million\cite{merchant2023gnome,zeni2025mattergen,wang2021wbm}; universal interatomic potentials --- foundation models trained on millions of first-principles calculations --- relax them and place them on the convex hull of known phases\cite{batatia2025macemp,deng2023chgnet,rhodes2025orbv3,wood2025uma}; and the candidates predicted to be stable go to density-functional theory (DFT), a few of them on to the laboratory\cite{szymanski2023alab}. Matbench Discovery\cite{riebesell2025matbench} ranks these models by how often a thresholded energy, a predicted hull distance $\ehat\le0$, recovers DFT stability. This practice leaves two decisions unexamined. One is what an output means: a team deciding how many calculations to fund or how many hits a shortlist will deliver needs a probability that means what it says, and threshold metrics cannot supply one\cite{bartel2020critical,deng2025softening}. The other is how much effort a candidate deserves: a relaxation costs about thirty forward passes of the potential and a DFT calculation orders of magnitude more, yet every candidate is treated alike.

Human experts manage the same trade-off with two modes of thought\cite{kahneman2011thinking}: a fast System~1 that settles most questions at a glance and a slow System~2 engaged when the stakes justify the effort. Artificial intelligence has long asked when computation is worth its cost\cite{russell1991right,hay2012selecting,booch2021thinking}, and `System-one models' such as Jev now return typed decisions with calibrated probabilities in one call\cite{typesafe2026jev}. In materials discovery, one forward pass of a potential is a System~1, a relaxation a System~2 and a DFT calculation a System~3. The same division runs through data-driven materials design, where a machine-learned surrogate decides which energetic materials, high-entropy ceramics, polymers, thermoelectrics or electromagnetic absorbers are worth the slow work of synthesis and measurement\cite{kang2020energetic,wang2025silicate,zhang2025polyimide,wan2025mgthermo,bai2026thermoelectrics,du2026graphene,fan2026absorber}. Unlike human intuition, a machine System~1 can be calibrated, and a calibrated System~1 can say for itself when slower computation is worth its cost.

The pieces exist separately. Ensembles, evidential learning and probabilistic hulls give foundation potentials error bars\cite{bilbrey2025uq,liu2026heterogeneous,xu2026evidential,novick2024probabilistic}, and a classifier on frozen representations can flag predictions not to trust\cite{mehdi2026probe}, but neither an uncertain energy nor a trust flag is a calibrated answer to a campaign's question, and benchmarks of uncertainty rarely test the decisions it should support\cite{tran2020uq,varivoda2023uq,dale2025activelearning}. Learning to defer and model cascades pass hard inputs to an expert or a larger model\cite{madras2018predict,mozannar2020consistent,jitkrittum2023when}, and multi-fidelity screening chooses which fidelity to query in a search for the best material\cite{fare2022multi}. Calibration degrades under distribution shift\cite{ovadia2019can}; label-free remedies range from importance weighting and weighted conformal prediction\cite{shimodaira2000improving,tibshirani2019covariate,jin2023weighted} to label-shift estimation\cite{saerens2002adjusting,lipton2018detecting}, and prediction-powered inference turns a few labels and many predictions into valid estimates\cite{angelopoulos2023prediction}. Classical decision analysis says when to test before acting\cite{howard1966information,pauker1980threshold}. Frozen foundation-model representations are increasingly reused\cite{radova2025frozen,pyzerknapp2025foundation}, but without a contract on what their outputs mean, a rule for when to compute more, or a test of either on new candidates.

Here we introduce CrystalJev, a System-1 decision layer for atomistic foundation models (Fig.~\ref{fig:overview}). A candidate passes once, unrelaxed, through a frozen potential; small heads on the state of that pass answer typed questions with probabilities calibrated on a small random slice of the candidate pool; conformal prediction turns these into finite-sample guarantees\cite{vovk2005alrw,angelopoulos2023gentle,jin2023selection}; and the value of information decides which candidates deserve a slower system. We show that the stability calls of 65 Matbench Discovery models hide cost assumptions, that a calibrated probability makes explicit what to compute next, so that slower computation spent where decisions are open recovers most of its value at a fraction of its cost, and, in a registered prospective test with 2,100 new DFT calculations, how far calibration transfers to new candidates and what fixing it costs.

\section*{Results}

\subsection*{A calibrated System~1 turns one pass into decisions}

\begin{figure}[tbp]
\centering
\includegraphics[width=\textwidth]{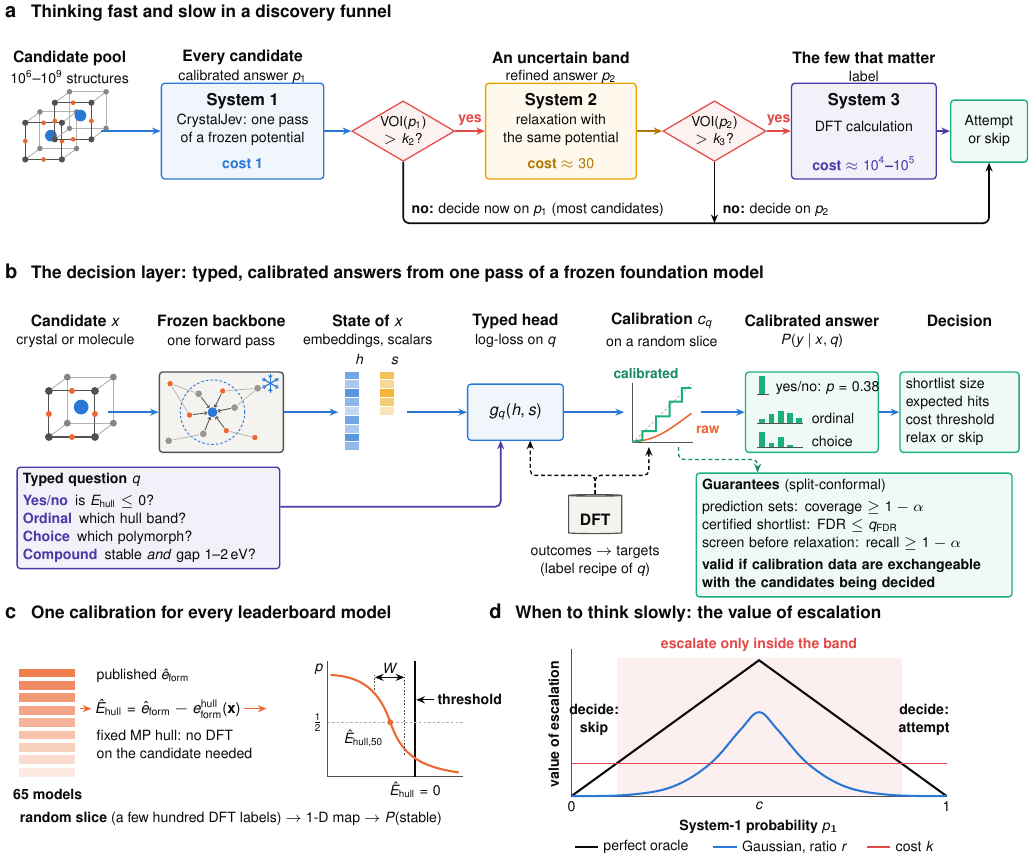}
\caption{\textbf{CrystalJev: thinking fast and slow with a frozen foundation potential.} \textbf{a}, A discovery funnel with deliberate escalation. Every candidate receives one forward pass of a frozen interatomic potential (System~1); the calibrated answer decides directly when it is clear, and sends the candidate to a relaxation (System~2) or a DFT calculation (System~3) only when the value of information exceeds the extra cost (costs in forward passes). \textbf{b}, The decision layer. Pooled node embeddings $h$ and physics scalars $s$ of the single pass form the state on which a head $g_q$ answers each typed question $q$ (yes/no, ordinal, choice or compound), trained with log-loss on targets from the question's label recipe; a map $c_q$ fitted on a slice exchangeable with the candidates calibrates the answer, which then carries split-conformal guarantees. \textbf{c}, The same calibration applied to the published predictions of any Matbench Discovery model gives its probability of stability, its 50\% point $\hat E_{\mathrm{hull},50}$ and the width $W$ of its indecision band. \textbf{d}, The value of escalating a candidate is bounded by that of a perfect oracle, a tent peaking at the attempt cost $c$, so escalation at cost $k$ can pay only inside a band of System-1 probabilities; for a Gaussian refinement it is governed by one information ratio $r$.}
\label{fig:overview}
\end{figure}

CrystalJev separates what current practice fuses: the state of a candidate, the question asked about it and the effort spent on it (Fig.~\ref{fig:overview}). The state is read from one forward pass of a frozen potential on the unrelaxed structure: pooled node embeddings and ten physics scalars such as the single-point energy, forces and stresses (Methods). Each typed question --- yes/no, ordinal, choice or compound --- gets a small head that trains in minutes on a CPU and is calibrated on a random slice of labelled candidates exchangeable with those being decided. Calibration turns answers into decisions. If a hit is worth 1 and an attempt costs $c$, a candidate with calibrated probability $p$ is worth attempting exactly when $p\ge c$, and a shortlist is expected to deliver $\sum_i p_i$ hits; conformal procedures turn the same probabilities into guarantees for prediction sets, certified shortlists and recall-controlled screens; and the probability states how much a relaxation (System~2) or DFT (System~3) could change the decision (Fig.~\ref{fig:overview}d).

\subsection*{Threshold calls hide cost assumptions}

\begin{figure}[tbp]
\centering
\includegraphics[width=\textwidth]{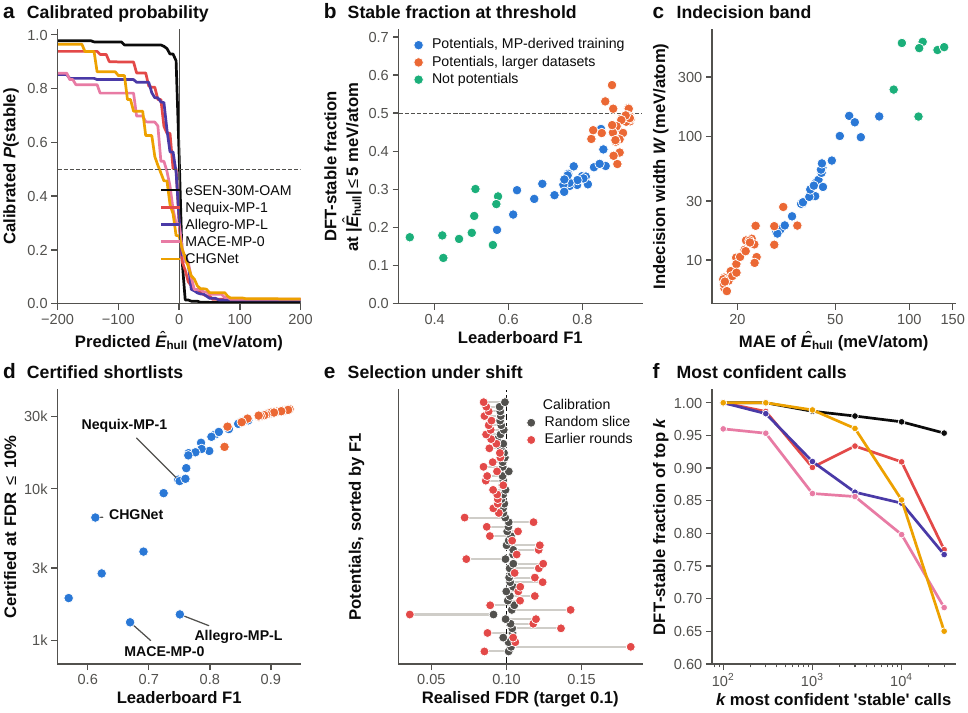}
\caption{\textbf{What the stability calls of 65 Matbench Discovery models are worth.} WBM unique prototypes; calibration maps fitted on a random 5\% slice (five seeds). \textbf{a}, Calibrated probability of stability against predicted hull distance (isotonic map) for five representative models. \textbf{b}, DFT-stable fraction among candidates predicted within $\pm5$~meV/atom of the hull, without any calibration map. \textbf{c}, Width of the indecision band $0.2\le p\le0.8$ against mean absolute error. \textbf{d}, Candidates certified at a false-discovery rate of at most 10\% by conformal selection (interatomic potentials; mean over 20 calibration draws). \textbf{e}, Realised false-discovery rate of conformal selection calibrated on a random slice (grey) or on WBM rounds 1--2 and applied to rounds 4--5 (red); potentials sorted by F1. \textbf{f}, DFT-stable fraction among each model's $k$ most confident `stable' calls. Colours in \textbf{a} and \textbf{f}: models; in \textbf{b}--\textbf{d}: training data.}
\label{fig:audit}
\end{figure}

Because the benchmark's Materials Project (MP) hull\cite{jain2013mp} is fixed and contains no candidate, any Matbench Discovery prediction fixes a predicted hull distance --- the energy above the convex hull of known phases --- that can be mapped to a calibrated probability (Methods). Maps fitted on a random 5\% slice of the WBM candidates\cite{wang2021wbm} for all 65 models (25 interatomic potentials trained on MP-derived data, hereafter MP-trained, 30 trained on larger datasets and 10 other models) reproduce every published F1 score to within 0.0007 (Fig.~\ref{fig:audit}). The threshold does not sit where the probability is one half: among candidates predicted within 5~meV/atom of the hull, only 0.19--0.46 are stable for MP-trained potentials and 0.12--0.30 for models that are not potentials (Fig.~\ref{fig:audit}a,b). Much of this is a property of the candidates: 1.6 times as many lie in the 25~meV/atom above the hull as in the 25~meV/atom below it, so errors carry more unstable structures across the threshold than stable ones. If hull distances have a local density proportional to $e^{\kappa E}$ near the hull and a model's errors there have bias $b$ and spread $s$, then $P(\text{stable}\mid\ehat=t)=\Phi\bigl((b-\kappa s^2-t)/s\bigr)$ (Methods), which predicts each model's calibrated probability at the threshold with a Pearson correlation of 0.96 (Extended Data Fig.~\ref{fig:value}a).

Read as a decision rule, the threshold is a hidden cost assumption: $\ehat\le0$ maximises value only if a failed attempt costs 0.18--0.48 of a hit for MP-trained potentials. At other costs the calibrated rule is worth more --- a median of 12.8 hits per 1,000 candidates for attempts as expensive as a synthesis ($c=0.7$) --- and it forecasts its own yield to within 0.3\%, whereas the hard rule implicitly promises that every shortlisted candidate is stable and delivers 24\%--93\% of that promise (Extended Data Fig.~\ref{fig:value}b); 200 random DFT labels suffice for a nearly optimal map (Extended Data Fig.~\ref{fig:value}c). And MP-trained potentials whose F1 scores differ by less than 0.01 can differ about 9-fold in the size of the shortlist that conformal selection\cite{jin2023selection} certifies at a 10\% false-discovery rate (Fig.~\ref{fig:audit}d,f; Supplementary Note~\ref{sn:audit}).

\subsection*{One pass is a cheap first opinion}

\begin{figure}[tbp]
\centering
\includegraphics[width=\textwidth]{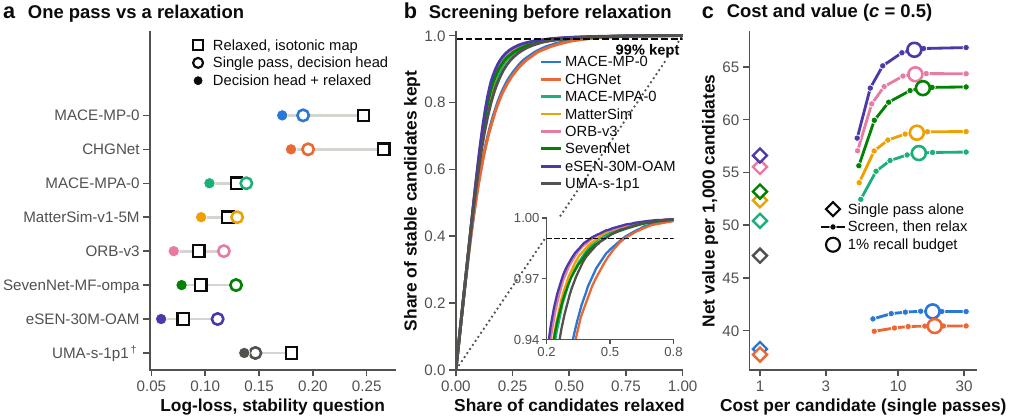}
\caption{\textbf{One forward pass across eight foundation potentials.} WBM test candidates, three seeds. \textbf{a}, Log-loss of the stability question for each backbone's published relaxed prediction with an isotonic map, for the single-pass decision head and for the head given the relaxed prediction as one more input ($^\dagger$UMA-s-1p1, which has no published WBM predictions: our relaxations of 5,000 random test candidates of the first split, all rules on these candidates, with the single pass and the relaxation combined by a logistic regression on the slice). \textbf{b}, Screening curves of the single-pass head: share of stable candidates kept against share of candidates sent to relaxation in order of decreasing single-pass probability (dashed line: 99\% of stable candidates kept; inset: detail near 99\%). \textbf{c}, Cost per candidate in forward passes (one pass, plus a relaxation of thirty passes for each candidate sent on) against realised net value at $c=0.5$, for the conformal screen at recall budgets $\alpha$ from 0.2 to 0 (right end: relax every candidate); large open circles: $\alpha=0.01$; diamonds: the single pass alone. Colours in \textbf{b},\textbf{c}: backbones as in \textbf{a}.}
\label{fig:backbones}
\end{figure}

CrystalJev itself decides from one unrelaxed forward pass. We trained heads on the WBM candidates of 45\% of the chemical systems, calibrated them on a random 5\% slice of the others and tested them on the rest, against each backbone's own relaxation recalibrated on the same slice (the calibrated relaxation; Methods). For the two MP-trained potentials behind the original leaderboard, the single pass decides better than the relaxation, at a thirtieth of its cost: log-loss 0.191 against 0.247 for MACE-MP-0\cite{batatia2025macemp} and 0.196 against 0.266 for CHGNet\cite{deng2023chgnet} (Fig.~\ref{fig:backbones}a; Supplementary Table~\ref{tab:backbones}). For the five newer potentials, whose relaxations are far more accurate, it is worse, by 7--42\% in log-loss (all outside 95\% intervals), and a head given the relaxed prediction as one more input improves on the relaxation for all seven backbones. The single pass is paid for in labels: it matches an isotonically recalibrated relaxation only after 10,000 (CHGNet) to 30,000 (MACE-MP-0) training labels, whereas recalibrating a relaxed prediction needs a few hundred (Extended Data Fig.~\ref{fig:decision}d).

Stronger probabilistic models of the energy, trained on the same state, labels and slice, change this picture only at the margin. After recalibration on the slice, natural-gradient boosting\cite{duan2020ngboost}, Gaussian-process regression\cite{rasmussen2006gaussian} and a conformal predictive system on boosted regressions\cite{vovk2019nonparametric} decided worse than the head on every backbone (log-loss higher by 0.009--0.243); a deep ensemble of heteroscedastic networks\cite{lakshminarayanan2017simple} decided better for CHGNet and MACE-MP-0 (by up to 0.006) and worse for the six newer potentials (by up to 0.014; Extended Data Fig.~\ref{fig:baselines}e; Supplementary Note~\ref{sn:onepass}). Averaging the logits of the two readers, recalibrated on the slice, beat both on every backbone (post hoc; log-loss 0.003--0.015 below the head's). The same state answers other questions --- whether DFT will find a metal (AUROC 0.977--0.989), screening questions on Matbench property datasets and molecular questions on QM9 --- with heads that beat regress-then-threshold baselines for 20 of 24 Matbench question--backbone pairs (Extended Data Fig.~\ref{fig:typed}; Supplementary Note~\ref{sn:typed}).

\subsection*{Calibrated probabilities say what to compute next}

\begin{figure}[tbp]
\centering
\includegraphics[width=\textwidth]{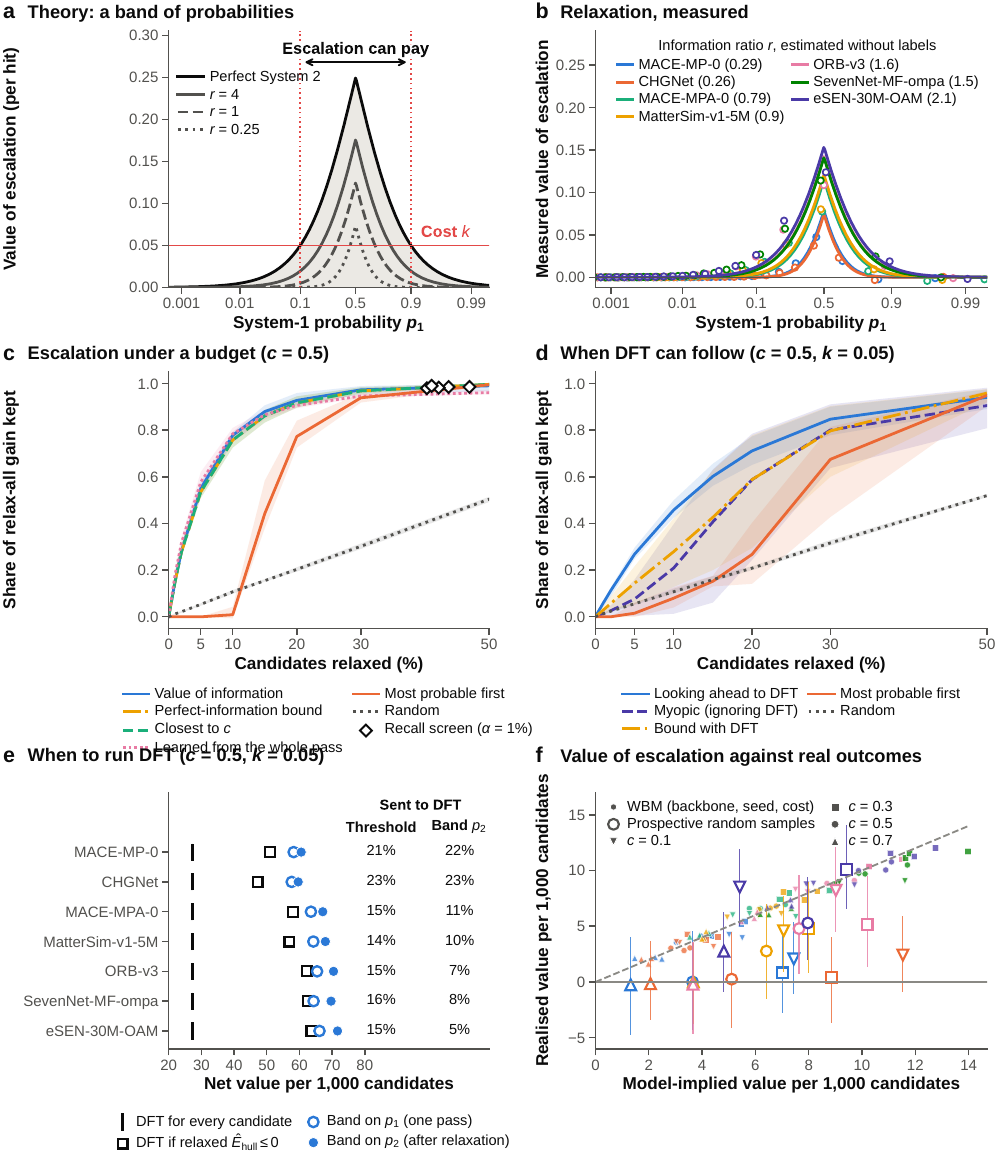}
\caption{\textbf{Calibrated probabilities say what to compute next.} WBM test candidates, seven potentials with published relaxed predictions, three seeds; System~1 is the single-pass head and System~2 the head given the relaxed prediction. \textbf{a}, Value of escalating a candidate against its System-1 probability $p_1$ ($c=0.5$): the perfect-information bound (Eq.~\ref{eq:oracle}) and the Gaussian closed form for several information ratios $r$; escalation costing $k$ can pay only inside the marked band. \textbf{b}, Measured value of escalation (points) and the closed form with $r$ estimated without labels. \textbf{c}, Share of the gain of relaxing every candidate retained when a fraction of them is relaxed, by escalation order (mean and range over potentials; diamonds: the conformal screen at a 1\% recall budget). \textbf{d}, The same when DFT ($k=0.05$) can follow. \textbf{e}, Net value per 1,000 candidates when DFT can precede an attempt: computing every candidate, computing those with a relaxed hull distance of at most zero, and the three-way band on $p_1$ and on $p_2$ (percentages: share sent to DFT). \textbf{f}, Realised against model-implied value of relaxing every candidate (small markers: WBM, by potential, seed and cost; large open markers with 95\% intervals: the prospective random samples, backbones with registered relaxations).}
\label{fig:voi}
\end{figure}

For the newer potentials a relaxation still adds information, so the question becomes which candidates to relax. Deciding at once on a candidate with calibrated probability $p_1$ is worth $\max(p_1-c,0)$; consulting a slower system, which updates it to $p_2$, is worth the expectation of $\max(p_2-c,0)$, less its cost $k$. Because a calibrated update does not shift the probability on average, the value of information\cite{howard1966information}
\begin{equation}
\mathrm{VOI}(p_1)=\mathbb{E}\bigl[\max(p_2-c,0)\mid p_1\bigr]-\max(p_1-c,0)
\label{eq:voi}
\end{equation}
is non-negative and never exceeds the classical value of perfect information,
\begin{equation}
0\le\mathrm{VOI}(p_1)\le\min\{p_1(1-c),\,(1-p_1)\,c\},
\label{eq:oracle}
\end{equation}
whose break-even points are Pauker and Kassirer's thresholds for testing before acting\cite{pauker1980threshold} (Methods). Escalation can therefore pay only inside the band $k/(1-c)<p_1<1-k/c$, whatever the slower system (Fig.~\ref{fig:voi}a).

What matters in practice is the order in which candidates are escalated. Relaxing the most probable candidates first, as a screening funnel does, recovers at most 4\% of the gain of relaxing all of them at a 10\% budget ($c=0.5$), because those candidates would have been attempted anyway. Any order that targets the decision boundary --- the parameter-free bound of Eq.~\ref{eq:oracle}, the distance $|p_1-c|$ or the estimated value of information --- recovers 76\%--78\% on average (seven potentials, three seeds; Fig.~\ref{fig:voi}c); only when costs are asymmetric does the estimated value of information do clearly better (84\% against 65\% for $|p_1-c|$ at $c=0.7$). When a DFT calculation can follow, the order must look ahead: a relaxation is then worth resolving a candidate into an attempt, a calculation or a discard, and ordering by this value recovers 43\%--52\% of the gain at a 10\% budget (DFT at a twentieth of a hit, $c=0.3$--0.7), against 19\%--28\% for orders that ignore the DFT option (Fig.~\ref{fig:voi}d). DFT itself attains the bound: discard below $k/(1-c)$, attempt above $1-k/c$ and compute in between; on the single-pass probability this band beats computing every candidate whose relaxed hull distance is at most zero for all seven potentials (58--66 against 47--64 net hits per 1,000 candidates; Fig.~\ref{fig:voi}e).

How much a relaxation is worth depends on the potential. For Gaussian uncertainties the value of information has a closed form in $p_1$, $c$ and one information ratio $r$, the precision of System~2 relative to the uncertainty of System~1 (Gaussian preposterior analysis\cite{raiffa1961applied}; Methods). Estimated without labels, $r$ rises from 0.26 (CHGNet) and 0.29 (MACE-MP-0) to 1.6 (ORB-v3\cite{rhodes2025orbv3}) and 2.1 (eSEN-30M-OAM), so relaxation is more informative for potentials trained on more data, and with this $r$ the closed form predicts the realised value of relaxing every candidate to within a factor of 1.4 (Fig.~\ref{fig:voi}b). The realised value of escalation --- the change in net value that the relaxation produced on each candidate's actual label --- agrees with the value implied by the calibrated probabilities on average (median ratio 0.98 across potentials, seeds and costs; Fig.~\ref{fig:voi}f and Extended Data Fig.~\ref{fig:voireal}); on new candidates it fell short (below).

\subsection*{Guarantees hold where the calibration data come from}

Every guarantee above assumes calibration data exchangeable with the candidates, and discovery breaks this assumption in three ways. Labels are plentiful where decisions are not: a MACE-MP-0 head trained and calibrated on MP and applied to the later WBM rounds covers only 0.87 of the true labels with nominal 90\% conformal sets, whereas calibrating on a random slice of the candidates restores 0.90 (Extended Data Fig.~\ref{fig:decision}c). Calibrating on early WBM rounds and deciding on later ones breaks false-discovery control for 20 of the 25 MP-trained potentials (Fig.~\ref{fig:audit}e). Label-free corrections do not repair this: importance weights from a classifier that tells the rounds apart (AUROC 0.74) raise the number of potentials that significantly exceed the nominal rate from 8 to 15 of 25, and re-estimating the stable fraction of the later rounds does not improve calibration, whereas 100 to 500 random labels from the later rounds restore control for all 25 (Extended Data Fig.~\ref{fig:shift}). Guarantees also hold on average over chemistries rather than within each --- one map per anion family (Mondrian calibration\cite{vovk2005alrw}) reduces the error in the stable fraction of the worst family from 1.7 to 0.5 percentage points --- and closed loops that recalibrate on their own acquisitions underestimate the stable structures left in the pool by 76\%--80\% (Extended Data Figs.~\ref{fig:baselines}d and~\ref{fig:loop}; Supplementary Note~\ref{sn:guarantees}).

\subsection*{A registered test: new candidates need new labels}

\begin{figure}[tbp]
\centering
\includegraphics[width=\textwidth]{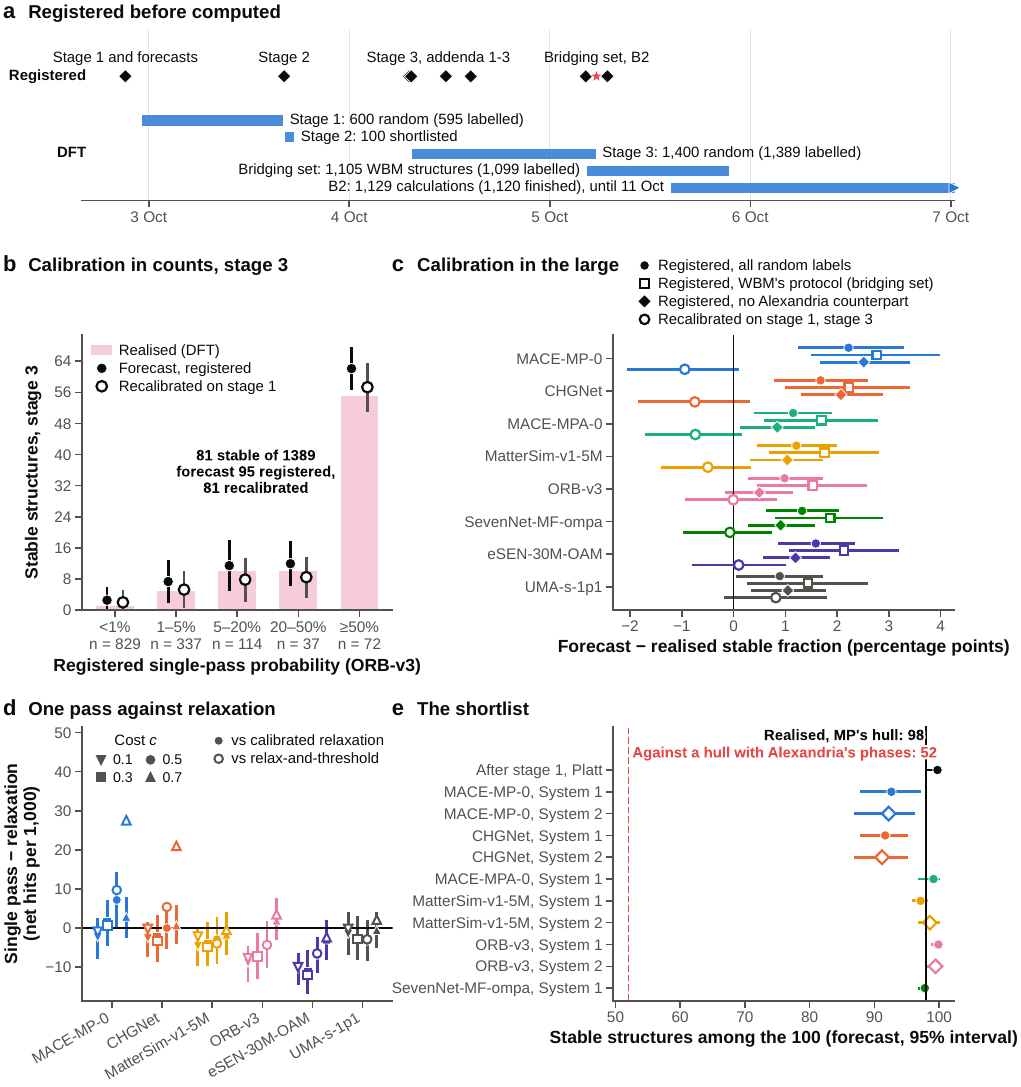}
\caption{\textbf{A registered prospective test with new DFT calculations.} \textbf{a}, What was registered (diamonds: stage~1 with the forecasts of six backbones; stage~2; stage~3 followed, before any of its labels, by three addenda --- shift remedies, eSEN-30M-OAM and UMA-s-1p1; the bridging set and B2; star: public posting of all digests) and computed (bars). \textbf{b}, Stable structures found by DFT (bars) and forecast (markers, $\pm1.96$ s.d.) among the 1,389 stage-3 candidates, by bin of the registered single-pass probability of ORB-v3: registered forecasts calibrated on WBM only (filled) and recalibrated on the 595 stage-1 labels and frozen before stage~3 (open). \textbf{c}, Forecast minus realised stable fraction for each backbone: registered single-pass forecasts on all 1,984 random candidates, the same with the realised fraction converted to WBM's labelling protocol by the bridging set (open squares; intervals also resample the bridging structures), on candidates without an Alexandria counterpart, and the stage-1 recalibration on stage~3 (95\% intervals that resample parent prototypes). \textbf{d}, Realised net value per 1,000 stage-3 candidates of the single-pass rule $p_1\ge c$ minus that of the registered System-2 rule, the relaxed prediction with its WBM-calibrated map (filled; paired 95\% intervals), and minus that of relax-and-threshold, the hard rule $\ehat\le0$ on the relaxed prediction (open), at the four registered attempt costs. \textbf{e}, The shortlist (stage~2): stable structures forecast at registration by every backbone and after stage~1 (ORB-v3 recalibrated on all stage-1 labels before the shortlist's DFT), and realised against MP's hull and against a hull completed with Alexandria's phases; lines: 95\% intervals of the yield if outcomes were independent. Colours in \textbf{c}--\textbf{e}: backbones.}
\label{fig:prospective}
\end{figure}

To test the layer where it is meant to be used, we generated 8,000 new candidates, absent from MP and WBM, by data-mined ionic substitution into MP prototypes\cite{hautier2011substitution} --- a generator related to WBM's but not the same --- over the 50 elements whose DFT energies we can put on MP's scale (Methods). DFT then labelled three samples, each registered before its calculations (Fig.~\ref{fig:prospective}a): a uniform random sample of 600 candidates; the 100 candidates a screening campaign would compute next, those ranked highest by the primary backbone, ORB-v3; and a second random sample of 1,400, for which every backbone's decisions, the shift remedies above and the forecasts of eSEN-30M-OAM and UMA-s-1p1 were frozen before any label was collected (Extended Data Table~\ref{tab:registered}). Our labels are put on MP's energy scale by per-element shifts fixed in advance (Extended Data Fig.~\ref{fig:align}). New candidates matter because WBM is no longer unseen: 99.4\% of its unique prototypes have a structural counterpart in the Alexandria database\cite{schmidt2024alexandria}, from which the training data of the newest potentials were drawn, whereas 83\% of our candidates have none (Extended Data Fig.~\ref{fig:overlap}a).

Of the 1,984 random candidates with labels, 113 (5.7\%) proved stable. Calibrated on WBM alone, the registered single-pass forecasts of all eight backbones over-stated this fraction, by 0.9--2.2 percentage points (16--39\% of the stable count), with 95\% intervals that exclude zero for all eight, while ranking the candidates with AUROC 0.931--0.971 (Fig.~\ref{fig:prospective}c); on the second random sample alone, registered in full before any of its labels, it was 1.0--2.0 points. It persists on the 1,635 candidates without an Alexandria counterpart (0.5--2.5 points), so it is not memorisation. Either the pool differs from WBM or our labels differ from WBM's. A registered bridging set tells the two apart: the WBM structures of the calibration slice whose chemistry matches the pool, recomputed with our protocol (1,099 of 1,105 labelled). On these structures the two protocols found nearly the same stable fraction (difference 0.4 percentage points, 95\% interval $-$0.6 to 1.4). Converted to WBM's protocol as registered, the random candidates were no more often stable (5.2\% against 5.7\%), and the over-statement remained, at 1.4--2.8 points with intervals that exclude zero for all eight backbones: by the registered rule, the population had shifted, not the labelling protocol (Fig.~\ref{fig:prospective}c; Supplementary Note~\ref{sn:prospective}). Single relaxations do, however, understate stability near the hull: when the 461 candidates near the hull or shortlisted were relaxed again from rattled and machine-learned starts (registered set B2), the stable fraction rose by 0.7 points, which can explain at most as much of the over-statement, leaving 0.2--1.6 points, significant for three of the eight backbones (Extended Data Table~\ref{tab:registered}).

New labels removed the bias most reliably. Recalibrated on the 595 stage-1 labels and frozen before stage~3 was computed, the forecasts of its 1,389 candidates erred by $-$0.9 to 0.8 points, with intervals that include zero for all eight (Fig.~\ref{fig:prospective}b,c). Of the label-free remedies, the label-shift re-estimate over-corrected ($-$2.5 to $-$0.7 points) and the importance-weighted map removed only part of the bias (0.3 to 1.4 points left, intervals including zero for five of the eight; Extended Data Fig.~\ref{fig:shift}d), because a classifier tells the pool from WBM almost perfectly (AUROC 0.975--0.976), leaving an effective calibration sample of only 541--550 of WBM's 10,774 slice structures. With stable candidates this rare, a fix needs a few hundred labels: refitting only the intercept on 200 random stage-1 labels brought the median absolute error to 0.6--1.2 points, and shrinking the refitted intercept towards the registered forecast reduced the harm of small samples --- with 25 labels, 20\%--42\% of random draws ended worse than no recalibration, against 58\%--83\% without shrinkage (post hoc; Extended Data Fig.~\ref{fig:shift}e).

Against relaxation calibrated on the same WBM slice, the registered System-2 rule, the single pass gave up value at low attempt costs for the most accurate relaxers ($-$11 net hits per 1,000 candidates for eSEN-30M-OAM at $c=0.3$, interval $-$17 to $-$6) and matched it at $c=0.7$ for all six backbones with registered relaxations (paired differences $-$3 to 3; Fig.~\ref{fig:prospective}d). Against relax-and-threshold, the hard rule on the relaxed prediction, calibration paid where relaxations call too many candidates stable (28 and 21 more net hits per 1,000 candidates for MACE-MP-0 and CHGNet at $c=0.7$). The realised value of escalation fell short of its model-implied value for four of the five backbones analysed, though within sampling error for 12 of the 20 backbone--cost combinations, and matched it for eSEN-30M-OAM (post hoc; Fig.~\ref{fig:voi}f): value-of-information guidance is only as good as the calibration of the probabilities it is given.

With the stage-1 labels as calibration set, the conformal screen at $\alpha=0.1$ kept 88\%--95\% of the stable stage-3 candidates while passing on 13\%--27\% of all candidates; calibrated on WBM's slice it kept only 74\%--85\%, short of the 90\% target for all eight backbones. These guarantees hold on average over calibration sets. One set with 32 stable candidates buys little more: a screen that holds for this set with probability 0.9 (a training-conditional guarantee\cite{vovk2012conditional}) passes on 21\%--64\% of all candidates, and conformal selection with calibration-conditional $p$-values\cite{bates2023testing} certifies no candidate at false-discovery rates up to 0.2 (post hoc; Supplementary Note~\ref{sn:prospective}).

Of the 100 shortlisted candidates, 98 proved stable against MP's hull, but against a hull completed with Alexandria's phases only 52 remain stable, and the stable fraction of the second random sample falls from 5.8\% to 1.7\% (Fig.~\ref{fig:prospective}e): the yield measures agreement with the benchmark's definition of stability rather than discovery. The primary backbone's forecast, 99.9~$\pm$~0.3 hits, was over-confident because outcomes were not independent (design effect 2.1; post hoc): a shortlist built from one model's ranking concentrates that model's errors.

\section*{Discussion}

These results argue for a different division of labour in computational discovery. A frozen foundation potential, read once, already knows enough to settle most candidates, and what it cannot settle it can delegate deliberately, because a calibrated probability states how much a slower computation could change the decision. Ordering slow computation by probability, as a screening funnel does, spends it where it changes nothing; ordering it by its value for the decision spends it where decisions are open, and when DFT can follow, the order must look ahead to it. The information ratios we measured, higher for potentials trained on more data, suggest that this allocation will matter more, not less, as potentials improve and as candidate pools grow toward the trillions of compositions that integrated database frameworks now hold\cite{huang2026trillion}.

The prospective test shows what such a layer needs in practice. Calibration fitted on existing data carried over to new candidates from a related generator, but not exactly --- the candidates had shifted, not the labelling protocol --- and label-free corrections repaired the difference only in part or over-corrected it, whereas a few hundred random labels from the new pool repaired it. When stable candidates are rare, the number of stable calibration examples, not the quality of the calibration, limits what conformal methods can guarantee with one calibration set. The practical recipe is to compute a random sample of the pool being decided about --- a few hundred candidates at the stable fractions of substitution campaigns, more for strong guarantees --- to recalibrate with shrinkage, to calibrate out of sample inside closed loops and to state the population over which a guarantee is claimed. Stability benchmarks could report, next to F1, a proper score after calibration on a disclosed slice and the size of a shortlist certified at a fixed false-discovery rate; the audit code, to be released with this work, computes both from any prediction file. Benchmarks also need fresh candidates now that WBM's are no longer unseen.

Our study has limitations. Stability labels are DFT energies relative to a fixed MP hull, so DFT and hull uncertainty near the hull\cite{hegde2023dftuncertainty} caps the attainable calibration, and against a hull completed with Alexandria's phases only 52 of the 98 shortlisted `stable' candidates remain stable. The single-pass heads were trained on about $10^5$ in-domain labels; with few labels, and for the most accurate potentials at low attempt costs, a recalibrated relaxation remains the better choice. Our candidates come from ionic substitution over 50 elements; candidates from generative models are another shift, and experimental outcomes\cite{leeman2024alab,jakob2026disorder} the strongest. We tested frozen backbones; training them on decision losses, as System-one models for software are trained\cite{typesafe2026jev}, is a natural next step.

A calibrated System~1 turns a foundation model from a simulator into a decision-maker that knows when to ask for help --- the kind of component that agent frameworks for materials research, which plan work across distributed data and compute\cite{kang2026openaaas} or carry the literature through to experiments\cite{ma2026mxllm}, can delegate to. Crystal stability was a convenient place to start, because its data are abundant and its slow systems well defined; the logic of calibrated answers and deliberate escalation applies wherever fast models and slow computations or measurements share a budget, as in machine-learned microstructure--property mapping of titanium and steels\cite{liang2026twinning,wang2026steels}.

\section*{Methods}
\label{sec:methods}

\paragraph{Data.} We use the Matbench Discovery release files\cite{riebesell2025matbench}: the MP reference set\cite{jain2013mp} (154,718 entries with energies and relaxed structures) and the WBM candidate set\cite{wang2021wbm} (256,963 structures generated by chemical substitution in five rounds; 215,488 unique prototypes, of which 15.5\% lie on or below the MP hull after the leaderboard's 1-meV rounding). All evaluations use the unique prototypes, as the leaderboard does. Every decision is made from the initial, unrelaxed WBM structure; DFT-relaxed structures, band gaps and symmetry labels define labels only. Leaderboard predictions are the published WBM formation energies of all 65 models with published predictions in October 2026.

\paragraph{Probabilities from leaderboard predictions.} The MP hull used by Matbench Discovery is fixed and contains no WBM entry. Hence, for any predicted formation energy $\hat e_\mathrm{form}$,
\begin{equation}
\ehat \;=\; E_\mathrm{hull}^\mathrm{DFT} + \bigl(\hat e_\mathrm{form} - e_\mathrm{form}^\mathrm{DFT}\bigr) \;=\; \hat e_\mathrm{form} - e_\mathrm{form}^\mathrm{hull}(\mathbf{x}),
\label{eq:identity}
\end{equation}
where $e_\mathrm{form}^\mathrm{hull}(\mathbf{x})$, the formation energy of the MP hull at the candidate's composition $\mathbf{x}$, is available before the candidate is computed. We turn $\ehat$ into a probability with one-dimensional maps fitted on a disclosed random 5\% slice of the pool (five seeds): Platt scaling\cite{platt1999probabilistic}, isotonic regression\cite{zadrozny2002isotonic} and the `error-bar' practice $p=\Phi(-\ehat/\sigma)$, with $\sigma$ the model's root-mean-square error. Each map exposes the calibrated probability at the threshold, $p_0=p(\ehat=0)$, the 50\% point $\hat E_{\mathrm{hull},50}$ and the width $W$ of the band $0.2\le p\le0.8$; we report isotonic maps, which fit best, and the map-free stable fraction among candidates predicted within $\pm5$~meV/atom of the hull.

\paragraph{The exponential-tilt closed form.} If the true hull distances of the pool have a local density proportional to $e^{\kappa E}$ near the hull and a model's errors there are Gaussian with bias $b$ and spread $s$, Bayes' rule makes the true hull distance given $\ehat=t$ Gaussian with mean $t-b+\kappa s^2$ and standard deviation $s$, so that
\begin{equation}
P(\text{stable}\mid\ehat=t)=\Phi\!\left(\frac{b-\kappa s^2-t}{s}\right),
\label{eq:tilt}
\end{equation}
with 50\% point $\hat E_{\mathrm{hull},50}=b-\kappa s^2$ and value-maximising threshold $t_c=b-\kappa s^2-s\,\Phi^{-1}(c)$. We estimate $\kappa$ from the WBM hull-distance histogram near the hull, and $b$ and $s$ as robust estimates of the location and scale of each model's errors on candidates near the hull.

\paragraph{Backbones and the single-pass state.} Each backbone is evaluated once per structure: MACE-MP-0\cite{batatia2025macemp}, CHGNet\cite{deng2023chgnet}, MACE-MPA-0, MatterSim-v1-5M, ORB-v3\cite{rhodes2025orbv3} and SevenNet-MF-ompa. The state consists of pooled per-atom representations that enter each model's energy readout and ten physics scalars of the single pass, among them the single-point formation energy and hull distance (Eq.~\ref{eq:identity}), force and stress statistics and simple size descriptors. Relaxations follow the leaderboard's protocol.

\paragraph{Heads, calibration and splits.} Heads are gradient-boosted trees or linear probes, trained with log-loss and Platt-scaled on the calibration slice. Ordinal questions use cumulative binary heads with monotone differencing; choice questions use a softmax over a predicted energy (the backbone's own energy plus a learned residual) with a temperature fitted on held-out compositions. The WBM unique prototypes are split by chemical system: 45\% of the systems form the training set, and the remaining structures are divided at random into a calibration slice (5\% of all structures) and the test set, so that calibration and test data are exchangeable while both are chemically disjoint from training (three seeds). The relax-and-threshold reference recalibrates each backbone's published relaxed prediction with an isotonic map on the same slice. \emph{Probabilistic baselines} regress the hull distance on the same state, labels and splits: NGBoost\cite{duan2020ngboost} with a normal distribution, deep ensembles\cite{lakshminarayanan2017simple} of heteroscedastic multilayer perceptrons, Gaussian-process regression\cite{rasmussen2006gaussian} on a compressed state, and a split conformal predictive system\cite{vovk2019nonparametric} on boosted regressions of the hull distance and of the logarithm of its cross-fitted absolute residual. Their probability of stability is $\Phi(-\mu/\sigma)$ (averaged over members for the ensemble) or the conformal predictive distribution at zero, used raw and Platt-recalibrated on the slice; an ensemble of five multilayer-perceptron classifiers is a second decision head. Post hoc, a combined reader averages the logits of the head and of the recalibrated deep regression ensemble and is Platt-recalibrated on the slice; its decisions are compared with the head's by net value per 1,000 candidates and by conformal selection with the slice as calibration set.

\paragraph{Conformal procedures.} Split-conformal prediction sets at $\alpha=0.1$ use the non-conformity score $1-p_y$\cite{vovk2005alrw,angelopoulos2023gentle}. Conformal selection\cite{jin2023selection} computes conformal $p$-values $p_j=(1+\#\{i\in\mathcal{C}: y_i=0,\ s_i\ge s_j\})/(|\mathcal{C}|+1)$ from the calibration slice $\mathcal{C}$ and applies the Benjamini--Hochberg procedure at level $q$, which bounds the false-discovery rate of the certified shortlist when calibration data and candidates are exchangeable. The screen before relaxation discards candidates whose single-pass probability falls below $\tau_\alpha$, the $k$-th smallest probability among the $n_+$ stable structures of the slice with $k=\lfloor\alpha(n_++1)\rfloor$; a new stable candidate exchangeable with the slice is then discarded with probability at most $\alpha$. Mondrian variants fit one map and one procedure per anion family (halide if F, Cl, Br or I is present, else oxide, chalcogenide, pnictide or intermetallic, in this order of precedence).

\paragraph{Value of information.} Let $Y\in\{0,1\}$ be the label, $p_1=P(Y=1\mid S_1)$ the calibrated System-1 probability and $p_2=P(Y=1\mid S_1,S_2)$ the calibrated probability after System~2 is consulted; an attempt costs $c\in(0,1)$ in units of a hit, and escalation costs $k$. Propositions~1 and~2 specialise classical results on the value of sample and perfect information and on Gaussian preposterior analysis\cite{howard1966information,raiffa1961applied} to a calibrated yes/no decision; calibration is what makes them usable, because $p_1$ and $p_2$ can then be estimated, and the bounds checked, on held-out data.

\emph{Proposition 1.} $\mathrm{VOI}(p_1)=\mathbb{E}[u(p_2)\mid S_1]-u(p_1)$ with $u(p)=\max(p-c,0)$ satisfies $0\le\mathrm{VOI}(p_1)\le\min\{p_1(1-c),(1-p_1)c\}$, with equality on the right if $S_2$ reveals $Y$. \emph{Proof.} Non-negativity is Jensen's inequality for the convex $u$ with $\mathbb{E}[p_2\mid S_1]=p_1$. For the upper bound, $u(q)\le q(1-c)$ on $[0,1]$, so for $p_1<c$, $\mathrm{VOI}=\mathbb{E}[u(p_2)]\le p_1(1-c)$; for $p_1\ge c$, $\mathrm{VOI}=\mathbb{E}[u(p_2)]-(p_1-c)=\mathbb{E}[\max(c-p_2,0)]\le c\,\mathbb{E}[1-p_2]=c(1-p_1)$. If $p_2=Y$, $\mathbb{E}[u(Y)]=p_1(1-c)$ and both bounds are attained. Hence escalation at cost $k$ can pay only if $k/(1-c)<p_1<1-k/c$. $\square$

\emph{Proposition 2 (Gaussian refinement).} Suppose that given $S_1$ the hull distance is $E\sim\mathcal{N}(\mu_1,\sigma_1^2)$, so $p_1=\Phi(-\mu_1/\sigma_1)$, and $S_2$ observes $Z=E+\varepsilon$ with independent $\varepsilon\sim\mathcal{N}(0,\tau^2)$. With the information ratio $r=\sigma_1^2/\tau^2$, $a=\Phi^{-1}(p_1)$, $b=(\sqrt{1+r}\,a-\Phi^{-1}(c))/\sqrt{r}$ and $\rho=\sqrt{r/(1+r)}$,
\begin{equation}
\mathrm{VOI}(p_1;c,r)=\Phi_2(a,b;\rho)-c\,\Phi(b)-\max(p_1-c,0),
\label{eq:voigauss}
\end{equation}
where $\Phi_2$ is the standard bivariate normal distribution function. \emph{Proof.} The posterior after $Z$ is Gaussian with variance $\sigma_2^2=\sigma_1^2\tau^2/(\sigma_1^2+\tau^2)$ and mean $\mu_2$, which given $S_1$ is distributed as $\mathcal{N}(\mu_1,v^2)$ with $v^2=\sigma_1^4/(\sigma_1^2+\tau^2)$; $E-\mu_2$ is independent of $\mu_2$, so $(E,\mu_2)$ are jointly Gaussian with correlation $v/\sigma_1=\rho$. The attempt is made if and only if $p_2=\Phi(-\mu_2/\sigma_2)\ge c$, that is $\mu_2\le m=-\sigma_2\Phi^{-1}(c)$, and $\mathbb{E}[u(p_2)]=P(E\le0,\mu_2\le m)-c\,P(\mu_2\le m)$; standardising gives Eq.~\ref{eq:voigauss}, which depends on $(p_1,c,r)$ only. $r\to\infty$ recovers the oracle bound and $r\to0$ gives zero. $\square$ Eq.~\ref{eq:voigauss} agrees with Monte Carlo simulation to four decimals.

\emph{Measurement.} System~1 is the single-pass head; System~2 is the head given the backbone's published relaxed prediction as one more input. The martingale property requires $p_2$ to be calibrated conditionally on $p_1$, so $p_2$ is recalibrated by a logistic regression on $(\mathrm{logit}\,p_1,\mathrm{logit}\,p_2)$ fitted on the calibration slice. The estimated $\widehat{\mathrm{VOI}}(p_1)$ is a kernel regression of $u(p_2)$ on $\mathrm{logit}\,p_1$ over the slice minus $u(p_1)$; it needs System-2 outputs on the slice but no further labels. Measured curves average $u(p_2)-u(p_1)$ over equal-mass bins of $p_1$ on the test set; because its expectation given $p_1$ is $\mathrm{VOI}(p_1)$, $r$ is fitted by least squares to the values of the individual test candidates (mean over seeds), which avoids the bias of evaluating a steep curve at the medians of wide bins. Because this model-implied value relies on the calibration of $p_2$, we also measure the realised value of escalation, $v=(y-c)\,(\mathbf{1}[p_2\ge c]-\mathbf{1}[p_1\ge c])$, the change in net value that the label confers on the decision actually taken; if $p_2$ is calibrated, $\mathbb{E}[v\mid S_1]=\mathrm{VOI}(p_1)$, so the two curves must agree, and $r$ is fitted to each. \emph{Budgeted escalation.} With a budget of $B$ escalations and independent refinements, expected value is separable across candidates, so escalating the $B$ candidates with the largest value of information is optimal (an exchange argument); we compare this order with the most probable first, the closest to the decision boundary ($|p_1-c|$ or $|\mathrm{logit}\,p_1-\mathrm{logit}\,c|$ smallest), the perfect-information bound of Proposition~1 and random order, and with a learned estimate that uses more of the single pass than $p_1$: gradient-boosted regressions of $u(p_2)$ on $\mathrm{logit}\,p_1$, the ten physics scalars (forces, stresses and energies) and 16 principal components of the embedding, cross-fitted over two halves of the test set grouped by chemical system, which use System-2 outputs but no labels. \emph{A DFT option.} If DFT at cost $k$ can follow, deciding on a probability $p$ is worth $V(p)=\max(0,\,p-c,\,p(1-c)-k)$ (discard, attempt, or compute and attempt if stable), the perfect-information bound becomes $\min\{p(1-c),(1-p)c,k\}$, and a relaxation is worth $\mathbb{E}[V(p_2)\mid S_1]-V(p_1)$, which we estimate as above; realised values count $y-c$ for an attempt and $y(1-c)-k$ for a calculation. \emph{A label-free information ratio.} Under Proposition~2, $\Phi^{-1}(p_2)$ given $p_1$ is normal with mean $\sqrt{1+r}\,\Phi^{-1}(p_1)$ and variance $r$, so $r$ can be estimated without labels as the variance of $\Phi^{-1}(p_2)$ among test candidates with $|\Phi^{-1}(p_1)|<0.25$; we compare the closed form with this $r$ against the realised value of relaxing every candidate. \emph{DFT allocation.} With attempt cost $c$ and DFT cost $k$, discarding, attempting directly and computing first are worth $0$, $p-c$ and $p(1-c)-k$; the optimal rule is the band of Proposition~1 for a perfect oracle.

\paragraph{Typed questions.} Electronic questions on WBM use the PBE band gap of the relaxed structure\cite{wang2021wbm}: metal (gap $=0$), photovoltaic window ($1\le$ gap $\le2$~eV) and a five-level gap question ($0$, $(0,1]$, $(1,2]$, $(2,3]$, $>3$~eV), with the splits of the stability question. Compound questions (stable and in the photovoltaic window; stable and non-metallic) are answered by a dedicated head, by the product of the two calibrated marginal heads, and by that product recalibrated on the slice. If the marginal answers are calibrated conditionally on the state $s$, then $\mathbb{E}[p_A p_B]-P(A\wedge B)=-\mathbb{E}\bigl[\mathrm{Cov}(\mathbf{1}_A,\mathbf{1}_B\mid s)\bigr]$: the product under-forecasts the compound event exactly when the two events are positively associated given what the layer knows. Matbench property questions\cite{dunn2020matbench} threshold the datasets' targets (band gap, bulk and shear moduli, refractive index, highest phonon density-of-states peak, exfoliation energy and perovskite formation energy; thresholds in Supplementary Note~\ref{sn:typed}) and use one MatterSim-v1-5M or ORB-v3 pass on the dataset structures, with five-fold cross-validation and 10\% of each training fold held out for calibration. The baselines, on the same features, are a gradient-boosted regression of the target whose prediction minus the threshold is Platt-calibrated, the same regression with its empirical residual distribution propagated through the threshold, and a head on element fractions. A regression can decide worse because, with residual $r=y-\hat f(x)$, the answer to `is $y\ge t$?' is $P(r\ge t-\hat f(x)\mid x)$: a monotone map of $\hat f(x)$ can equal it only if the residual distribution is the same for all inputs with the same prediction, whereas a head trained with log-loss on the question has $P(y\ge t\mid x)$ itself as its population minimiser. QM9 molecules\cite{ramakrishnan2014qm9} (130,831 after removing those that failed the original consistency check), in their B3LYP geometries, are passed once through MACE-OFF23\cite{kovacs2025maceoff} or ORB-v3; yes/no questions threshold the HOMO--LUMO gap ($\ge8$~eV, $\le5$~eV) and the dipole moment ($\ge4$~D), with the same protocol and baselines. The choice question draws sets of five molecules of one formula within the test fold and asks which has the lowest $U_0$; the choice head scores each molecule by the backbone's energy plus a correction learned on the training fold and fits its softmax temperature on sets drawn within the calibration fold.

\paragraph{Shift, Mondrian calibration and closed loops.} For the shift experiment, a head trained on MP (split 90/10 by chemical system into training and calibration) is calibrated on three sources of increasing proximity to the candidates --- MP calibration systems, first-round WBM substitutions of the training systems and the random WBM slice --- and evaluated on the WBM rounds 2--5 structures of the test set. Budgeted loops draw a pool of 30,000 WBM unique prototypes and a 500-structure random seed; in each of 20 rounds a logistic head on single-pass features (a compressed pooled embedding plus the physics scalars) is refitted, calibrated and used to acquire 250 structures, whose labels are then revealed. Calibration maps are fitted on the most recent 30\% of acquired labels, on all acquired labels with five-fold cross-fitted predictions, on the random seed only, or on a 10\% uniform random sample of every batch; five seeds. \emph{Remedies.} Three remedies were tested on the shift between WBM substitution rounds for every MP-trained leaderboard potential, with calibration sets of 5\% of WBM drawn from rounds 1--2 and test samples of 2,000 structures from rounds 4--5 (20 draws). Covariate-shift weights $w(x)=p_\mathrm{test}(x)/p_\mathrm{cal}(x)$ come from a cross-fitted logistic classifier of the round on element fractions, the numbers of elements and sites and the model's prediction\cite{shimodaira2000improving}, clipped at their 99th percentile; they enter a weighted Platt map, weighted conformal screens\cite{tibshirani2019covariate} and weighted conformal selection\cite{jin2023weighted}. Label shift is corrected by re-estimating the stable fraction of the test sample by expectation--maximisation and adjusting the probabilities to it\cite{saerens2002adjusting}. The third remedy spends $n$ random DFT labels of the later rounds, as the calibration set of the conformal procedures or to refit the intercept of the Platt map. All three were also registered for the third prospective stage, fitted on WBM's calibration slice.

\paragraph{Training data and reference phases.} We compared WBM and our candidates with the Alexandria PBE database\cite{schmidt2024alexandria} (releases 2023.12.29, 2024.12.15 and 2025.07.02; 5.78~million distinct entries), from which ORB-v3 drew training structures directly\cite{rhodes2025orbv3} and from which sAlex and OMat24, used by MACE-MPA-0, SevenNet-MF-ompa, eSEN-30M-OAM and UMA, were derived after removing structures with WBM's prototypes\cite{barroso2024omat24}. A structure has a counterpart if an Alexandria entry of the same reduced formula is matched to it by pymatgen's StructureMatcher with default tolerances\cite{ong2013pymatgen}; WBM's DFT-relaxed structures were compared with Alexandria's relaxed entries, and each candidate's initial and, where computed, DFT-relaxed structure with all entries. For hull distances against MP and Alexandria together, Alexandria's phases within 0.1~eV/atom of its hull in the candidates' subsystems were put on MP's energy scale by per-element shifts fitted to 255,499 WBM--Alexandria pairs of the same structure (median absolute residual 0.4~meV/atom) and given the MP2020 corrections; each candidate's own counterparts are left out, so that it competes with the phases into which it could decompose rather than with itself.

\paragraph{Prospective test.} \emph{Candidates.} Candidates are data-mined ionic substitutions\cite{hautier2011substitution} into near-hull MP prototypes, a generator related to WBM's\cite{wang2021wbm} but not the same; they have two to four elements and small cells, drawn from the 50 elements for which DFT labels can be put on MP's energy scale (those whose PAW potentials available to us carry MP's symbols and titles, which leaves out, among others, S, Cl, P, Li, Na, Mg, Fe, Co, Mn, Cr, Mo, Y, Zr, Ga, Ge and As); only compositions absent from MP and WBM are kept. The pool is the 8,000 most probable candidates (5,780 compositions in 4,221 chemical systems; median 10 sites).
\emph{Registration.} For each backbone, heads were trained on all WBM unique prototypes except a random 5\% calibration slice and applied unchanged to the pool; System-2 forecasts use the backbone's own relaxation of each candidate (MACE-MP-0, CHGNet, MatterSim-v1-5M and ORB-v3). All forecasts, the sampling seed, the stage-1 DFT sample --- a uniform random sample of the pool, split at random into a calibration slice of 200 and a test sample of 400 candidates --- the stage-2 shortlist --- the 100 remaining candidates with the highest single-pass probability of the primary backbone, ORB-v3, fixed in advance --- and the analysis plan were written to a registration file whose SHA-256 digest (\texttt{8f359667a6acda7c}\ldots) was recorded outside the computing cluster at 2026-10-02 21:12:07 UTC, before any DFT calculation of a pool member. After stage~1, and before the shortlist's DFT, the primary backbone's single-pass probability was recalibrated on all stage-1 labels and the updated forecast of the shortlist's yield registered in the same way (\texttt{07d33c3f6708c63a}\ldots, 2026-10-03 16:12:04 UTC).
\emph{Stage 3.} After the stage-2 labels, a uniform random sample of 1,400 of the remaining pool members was registered with every backbone's decisions for it frozen on the stage-1 labels: the Platt and isotonic maps, the conformal screens ($\alpha=0.05$, 0.1, 0.2) with thresholds from the stage-1 stable candidates, and conformal selection ($q=0.1$, 0.2) with the stage-1 labels as calibration set (\texttt{117beb36d621d968}\ldots, 2026-10-04 07:10:06 UTC). Its calculations ran on three clusters with byte-identical PAW files and inputs written once; candidates were assigned to clusters at random and unstarted calculations were later moved between clusters to balance the load. 24 stage-1 structures recomputed from fresh inputs on every cluster validated the pooling of labels (registered criterion: median absolute difference in final energy below 1~meV/atom and no changed label). Before any stage-3 label was collected we registered the shift remedies above, fitted on WBM (\texttt{748d1f689b3c740a}\ldots), the single-pass forecasts of eSEN-30M-OAM and UMA-s-1p1 with the same frozen decisions (\texttt{4913d95a7c30cc86}\ldots) and their System-2 forecasts from relaxations of the labelled candidates (\texttt{0788e409ad786f12}\ldots). The stage-3 results reported here use the labels collected on 5 October 2026, before the registered stopping time (2026-10-08 12:00 UTC): 11 calculations had failed and zero was still running (Supplementary Note~\ref{sn:prospective}).
\emph{Bridging set.} To tell a difference in labelling protocol from a shift of the population, we registered on 2026-10-05 04:18:50 UTC (\texttt{e865729ea521a66d}\ldots), after the stage-3 labels and before any of its calculations, a census of the 1,105 WBM unique prototypes of the calibration slice shared by every registered forecast whose chemistry matches the pool (elements among the 50, two to four elements, at most 12 sites, an anion family of the pool; 207 of them stable by WBM's labels), computed from WBM's initial structures with the stage-3 protocol. Its primary analyses are the agreement of WBM's and our labels on the same structures (sensitivity $Se$, false-positive rate FPR and the difference $D$ of the stable fractions) and the protocol-adjusted calibration of the registered forecasts on the random candidates of stages 1 and 3, whose stable fraction under WBM's protocol is estimated from ours, within three groups of the primary forecast, as $(\pi_\mathrm{ours}-\mathrm{FPR})/(Se-\mathrm{FPR})$. The over-statement is attributed to the labelling protocol if the interval of $D$ excludes zero and the adjusted intervals include zero for at least six of the eight backbones, to a shift of the population in the converse case, and to both otherwise. Failed calculations, and any unfinished at the registered stopping time (2026-10-09 12:00 UTC), are reported and excluded, with bounds. The digests of all registration files were posted publicly at 2026-10-05 05:35 UTC (\url{https://gist.github.com/kpleo/69d0855be72bc7830025aec1dfbbc116}), after the labels of stages 1--3 and before any label of the bridging set was collected; the earlier digests were recorded by the authors at the times given.
\emph{Protocol-robust labels (B2).} To test whether our labels near the hull depend on the starting structure, on symmetry or on magnetic order, we registered on 2026-10-05 06:53:42 UTC (\texttt{307ffbf22414bea0}\ldots), before any of its calculations, a census of the 461 candidates within 50~meV/atom of the hull in the random samples and the shortlist, each relaxed again with the stage-3 protocol from our relaxed structure with every coordinate displaced by $\mathcal{N}(0,(0.05~\text{\AA})^2)$ and symmetry switched off, and from eSEN-30M-OAM's relaxation of the unrelaxed candidate; two antiferromagnetic starts for each of the 75 spin-polarised candidates that are stable or shortlisted; and SCAN meta-GGA local hulls (VASP 6.1.0 has no r2SCAN), with the competing phases of each chemical system recomputed, for the 18 candidates whose labels hinge on the energy-correction scheme (mixed-anion $+U$, halogen-only and 5d binary halides). Its primary analyses are the change of each candidate's hull distance when the lowest of its energies replaces the original, and the stable fraction and calibration of the registered forecasts with these robust labels; stopping time 2026-10-12 12:00 UTC. Its digests, with those of all earlier registration files, were posted publicly at 2026-10-05 08:57 UTC (\url{https://gist.github.com/kpleo/fa7e8aec40b9006de808b3ffca38aae0}), before any B2 calculation started.
\emph{DFT labels.} Calculations use VASP\cite{kresse1996efficient} with MP's GGA(+U) protocol and start, as in WBM, from the unrelaxed candidate structure. The projector-augmented-wave potentials\cite{blochl1994paw,perdew1996pbe} available to us carry MP's symbols and titles but belong to a later release, so total energies differ from MP's by amounts that are, to first order, constants of the elements. We therefore fitted one shift per element (and one per $+U$ element in oxides and fluorides) by robust least squares to recomputed MP elemental phases, binary $+U$ oxides and a first control set of MP and WBM structures, and fixed the shifts (SHA-256 \texttt{b88706bf90887082}\ldots, 2026-10-02 23:08:28 UTC) before any prospective calculation was collected. A second control set validated them, and a third, chosen to match the candidate pool and computed from WBM's initial structures like the candidates, measured the residual label noise: aligned hull distances reproduce WBM's to a median of 3.4~meV/atom, with WBM's label for 92 of 98 structures and a tail set by relaxations that reach different minima from the same start (Supplementary Note~\ref{sn:dft}). Hull distances include the MP2020 corrections\cite{wang2021mp2020} and are taken against the patched phase diagram of the MP 2023-02-07 entries, which reproduces the WBM labels of Matbench Discovery exactly; a candidate is stable if its hull distance, rounded to 1~meV/atom, is at most zero. Calculations that did not finish (five of the 600 stage-1 calculations) are reported and excluded; all 100 shortlisted calculations finished.
\emph{Analysis.} As registered: on the stage-1 test sample, calibration in the large (forecast minus realised stable fraction), expected calibration error, log-loss and Brier score of every backbone's registered System-1 forecast and, for the four backbones that relax, System-2 forecast, against the same forecasts recalibrated on the slice (Platt and isotonic); realised net value at $c=0.1$, 0.3, 0.5 and 0.7 of the rule $p_1\ge c$, the rule $p_2\ge c$ and relax-and-threshold; recall of the conformal screen calibrated on the slice ($\alpha=0.05$, 0.1); and the false-discovery proportion of conformal selection ($q=0.1$, 0.2) among the test sample with the slice as calibration set. Stage~2: the realised number of stable structures in the shortlist against every registered forecast. Secondary: the stage-1 analyses without the $+U$ chemistries. Stage~3, as registered: the same calibration, value and conformal analyses for the registered forecasts on stage~3 and on all random labels, for the forecasts recalibrated on stage~1 and for the shift remedies, with intervals clustered by parent prototype; secondary: candidates without an Alexandria counterpart, without $+U$ chemistries, and labels against a hull completed with Alexandria's phases. Not part of the registered plan, and reported as such: bootstrap intervals, AUROC and the five-bin implementation of the expected calibration error (all implemented in the evaluation script before the registration); sensitivity analyses of stage~1 to the DFT labels, which resample the residuals of the third control set (added after the registration but before any stage-1 label was collected); and post hoc analyses: paired bootstrap intervals of the differences in value between rules for stages 1 and 2 (registered for stage~3); for the shortlist, DFT relaxations of its two failures without symmetry from MACE-MP-0's relaxed structure and from randomly displaced starts; and, for stage~3, the registered forecasts recalibrated on random subsets of the stage-1 labels (25 to 400 candidates, 200 draws each), by refitting the intercept of the logit or the full Platt map, by refitting the intercept with a normal prior of standard deviation 0.5 on the logit scale (posterior mode; 0.25 and 1 as sensitivity), or by a prediction-powered estimate of the pool's stable fraction\cite{angelopoulos2023prediction}; a training-conditional screen that misses at most 10\% of stable candidates with probability 0.9 over calibration sets\cite{vovk2012conditional} and conformal selection with calibration-conditional $p$-values\cite{bates2023testing} (beta quantiles of the calibration order statistics with a Bonferroni correction over the first 100 ranks, and the Dvoretzky--Kiefer--Wolfowitz bound); the label-noise sensitivity analysis of stage~1 repeated on stage~3, bounds for its unlabelled calculations and two one-sided tests of the remedies at a margin of 1 percentage point; random re-splits of the conformal calibration sets; and the realised value of escalation on the random samples. Deviations from the registered plan are listed in Supplementary Note~\ref{sn:prospective}, and every registered analysis with its outcome in Extended Data Table~\ref{tab:registered}.

\paragraph{Metrics and statistics.} Proper scoring rules: log-loss (natural logarithm) and Brier score\cite{brier1950,murphy1973partition}; expected calibration error with 10 equal-mass bins of the probabilities sorted with ties broken by candidate identifier, which makes it deterministic\cite{naeini2015ece} (the stage-1 evaluation registered five bins without a tie rule, whose value for tied forecasts depends on their order by up to 0.01); calibration intercept and slope of a logistic regression of the label on the forecast's logit; discrimination: AUROC; ordinal questions: ranked probability score; choice questions: top-1 accuracy. Net value per 1,000 candidates is $1000\,\overline{(y-c)\,\mathbf{1}[\text{attempt}]}$, $(1-c)$ times the net benefit of decision-curve analysis at threshold probability $c$\cite{vickers2006decision}. Results on WBM are means over three seeds (five for the audit maps and the closed loops). Candidates substituted into the same parent prototype are not independent, so intervals in the prospective test are bootstrap 95\% intervals that resample parent structures with all their candidates (1,000 resamples)\cite{davison1997bootstrap}; intervals that resample candidates are reported alongside, and the ratio of the two variances is the design effect of the clustering (10,000 resamples in the sensitivity analysis and for paired differences in value between rules).

\paragraph{Software and hardware.} All computations, including the DFT calculations, ran on CPU clusters.

\section*{Data availability}
This manuscript is currently under review. The data supporting its findings will be made publicly available upon publication.

\section*{Code availability}
This manuscript is currently under review. The code will be made publicly available upon publication.

\section*{Acknowledgements}
This work was supported by the Tianmushan Laboratory Research Project (Projects No.~TK2024D006 and No.~TK2023C021) and by the ``Pioneer'' and ``Leading Goose'' R\&D Program of Zhejiang (Project No.~2024SSYS0084). We thank the National Key Laboratory of Artificial Intelligence for Material Science and Tianmushan Laboratory for providing the computational resources used in this work. The authors are fully responsible for the ideas, reasoning and rigor of this work; a large language model assisted with derivations, code, data analysis and writing, under the authors' direction and subject to their review.

\section*{Author contributions}
P.K. and L.Z. conceived and designed the study, developed the methods, performed the computations and analyses, and wrote the manuscript. Y.S., Z.L., Y.L. and H.X. assisted with the computations, the analyses and the revision of the manuscript. All authors discussed the results and approved the final manuscript.

\section*{Competing interests}
The authors declare no competing interests.

\clearpage
\setcounter{figure}{0}\setcounter{table}{0}
\renewcommand{\figurename}{Extended Data Fig.}
\renewcommand{\tablename}{Extended Data Table}
\renewcommand{\theHfigure}{ED.\arabic{figure}}\renewcommand{\theHtable}{ED.\arabic{table}}
\section*{Extended Data}
\FloatBarrier

\begin{figure}[htbp]
\centering
\includegraphics[width=\textwidth]{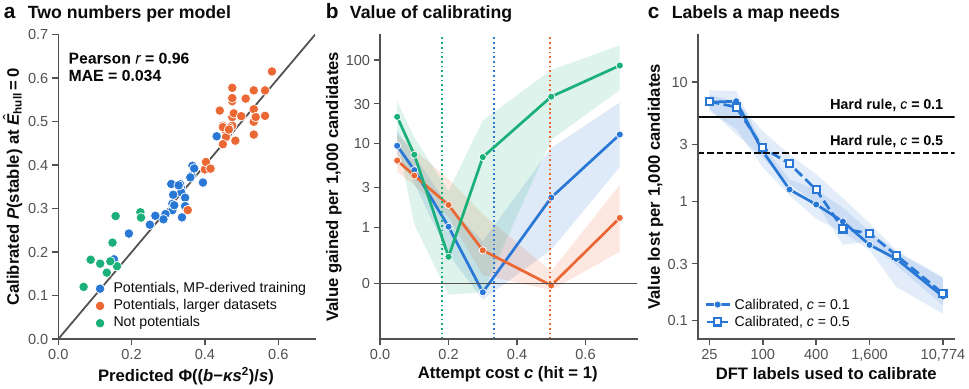}
\caption{\textbf{Where the leaderboard threshold sits, what it costs and what fixing it takes.} \textbf{a}, Calibrated probability of stability at the threshold $\ehat=0$ against the exponential-tilt closed form (Eq.~\ref{eq:tilt}), with $\kappa$ fitted once to the WBM population and each model's error bias $b$ and spread $s$ near the hull; 65 models. \textbf{b}, Net value gained per 1,000 candidates by the calibrated rule $p\ge c$ over the threshold, against the attempt cost $c$ in units of a hit (median and 10--90\% range per group of models; colours as in \textbf{a}); dotted lines: each group's median implied cost $c^\ast$, at which the threshold is optimal. \textbf{c}, Value lost by the calibrated rule relative to the best monotone recalibration in hindsight, against the number of random DFT labels used to fit the isotonic map (25 MP-trained potentials; median and interquartile range over ten random halves of the pool); horizontal lines: the threshold rule at the same cost.}
\label{fig:value}
\end{figure}

\begin{figure}[htbp]
\centering
\includegraphics[width=\textwidth]{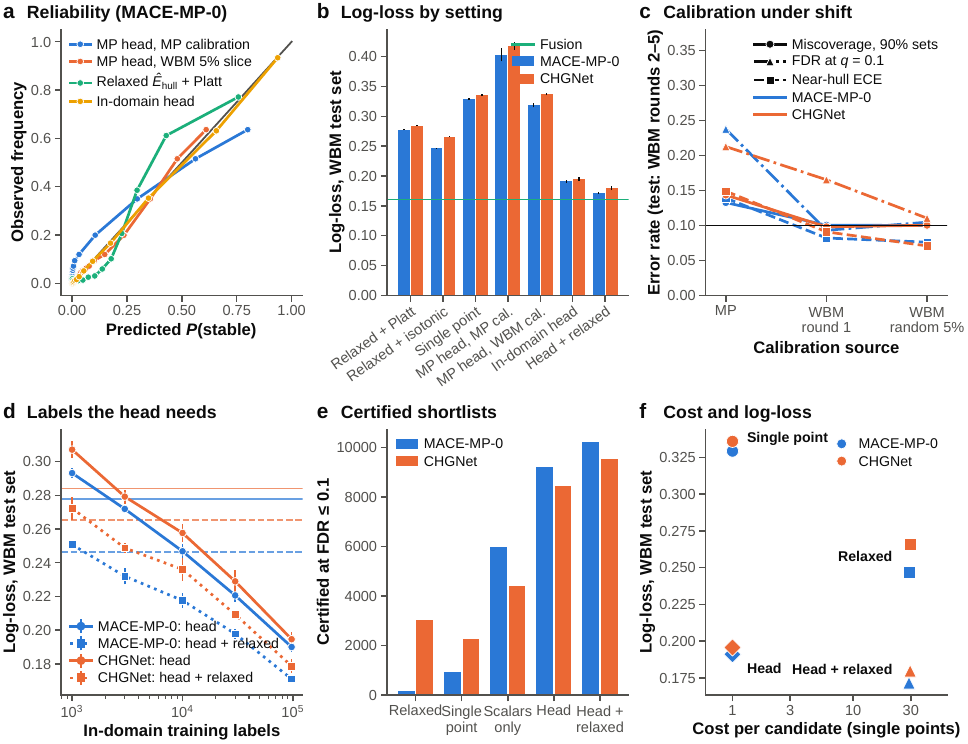}
\caption{\textbf{Transfer, shift and label efficiency of the decision layer on MACE-MP-0 and CHGNet.} WBM test candidates, 3 seeds. \textbf{a}, Reliability of single-pass probabilities of stability (MACE-MP-0, first seed) for a head trained on MP and calibrated on MP or on the WBM slice, for the relaxed hull distance with Platt scaling and for the in-domain head. \textbf{b}, Log-loss by regime and backbone (cal.: calibration); horizontal line: fusion of both backbones. \textbf{c}, Shift ladder for heads trained on MP and evaluated on WBM rounds 2--5 as the calibration source moves towards the candidates: miscoverage of 90\% conformal sets, near-hull expected calibration error and realised false-discovery rate of conformal selection at $q=0.1$; horizontal line: the 0.1 target. \textbf{d}, Log-loss against the number of in-domain training labels (two seeds); horizontal lines: relaxed prediction with Platt (solid) or isotonic (dashed) calibration. \textbf{e}, Candidates certified at a false-discovery rate of at most 0.1. \textbf{f}, Backbone cost per candidate in forward passes against log-loss.}
\label{fig:decision}
\end{figure}

\begin{figure}[htbp]
\centering
\includegraphics[width=\textwidth]{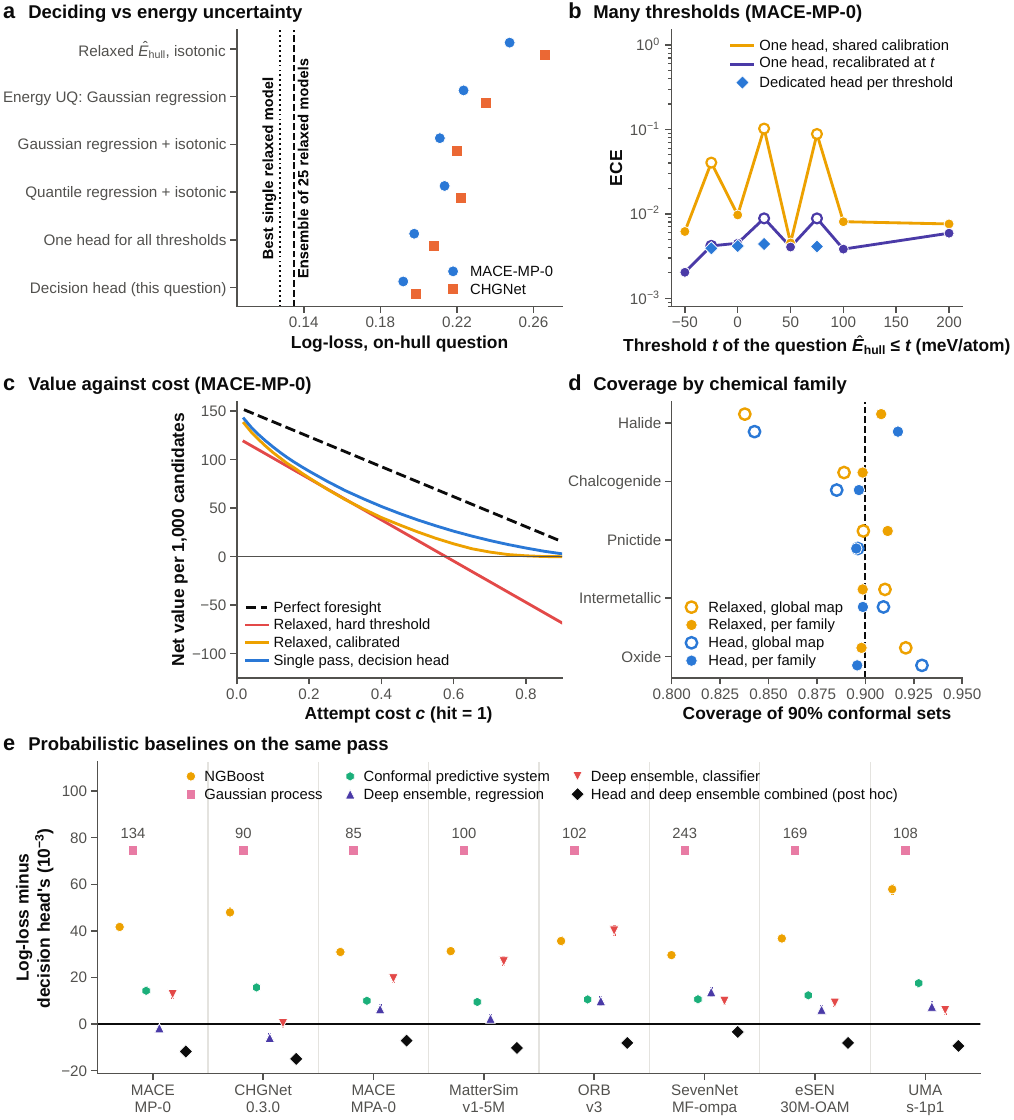}
\caption{\textbf{Deciding directly versus propagating energy uncertainty.} WBM test candidates, three seeds. \textbf{a}, Log-loss of the stability question for decision rules built on the same single-pass features, labels and calibration slice, and for the backbone's relaxed prediction; vertical lines: an ensemble of the 25 MP-trained leaderboard potentials (mean and spread of their relaxed predictions with a boosted map) and its best member (isotonic map). \textbf{b}, Expected calibration error of a single threshold-conditioned head against the queried threshold, with one shared calibration map or recalibrated at the queried threshold, and of heads trained for each threshold; open symbols: thresholds not seen in training. \textbf{c}, Realised net value per 1,000 candidates against the attempt cost $c$; dashed line: perfect foresight. \textbf{d}, Coverage of 90\% conformal sets within each anion family, averaged over MACE-MP-0 and CHGNet, with one calibration map for all candidates (open) or one per family (filled). \textbf{e}, Probabilistic baselines on the same single-pass state, labels, splits and slice: log-loss minus that of the decision head for regressions of the hull distance read off as the probability below zero (NGBoost, Gaussian-process regression and a deep ensemble of heteroscedastic networks, each Platt-recalibrated on the slice), a conformal predictive system, a deep ensemble of classifiers and, post hoc, the head and the deep regression ensemble combined (average logit, Platt-recalibrated on the slice) (95\% intervals: bootstrap over chemical systems, pooled over seeds; values beyond the axis are printed at its top; Supplementary Table~\ref{tab:mlbaselines}).}
\label{fig:baselines}
\end{figure}

\begin{figure}[htbp]
\centering
\includegraphics[width=\textwidth]{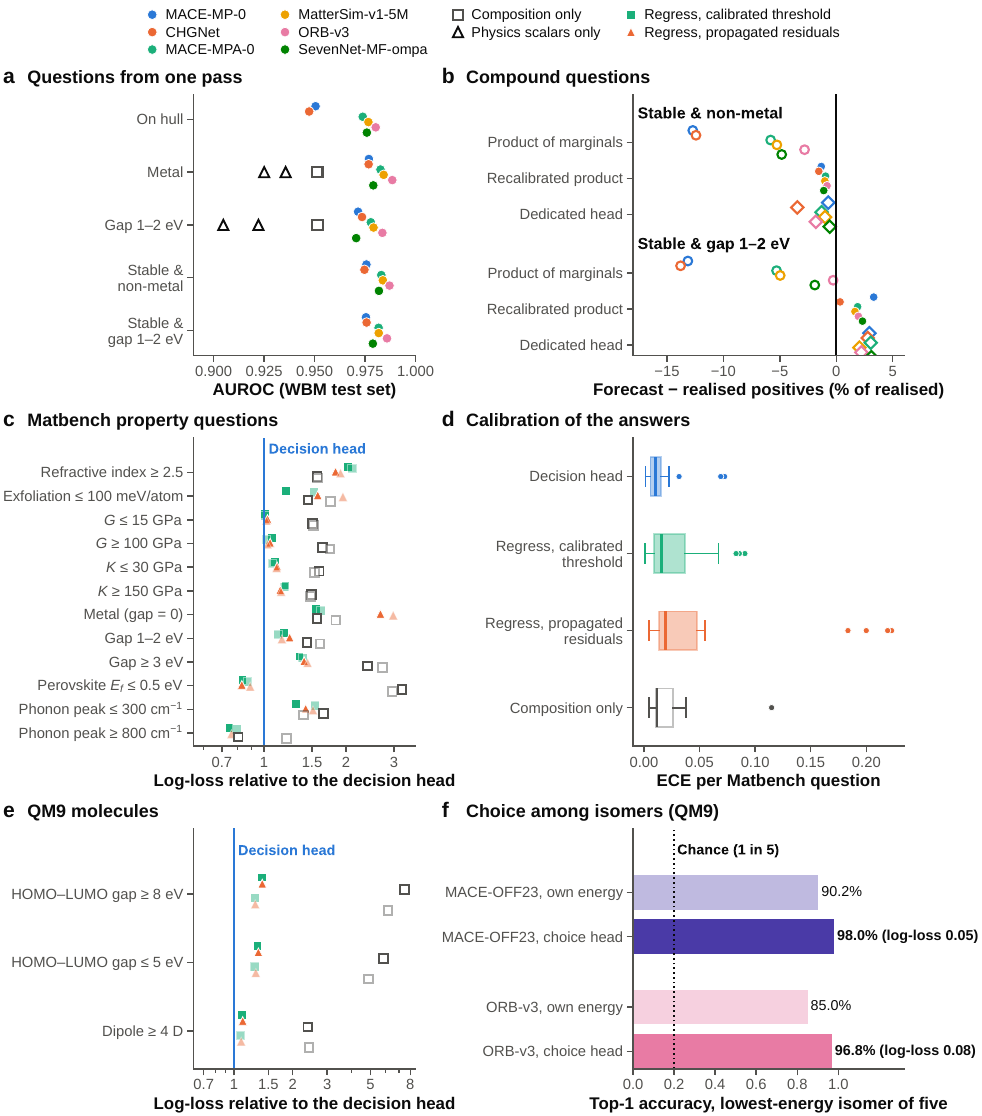}
\caption{\textbf{One pass, many questions.} Key at the top: backbones (\textbf{a},\textbf{b}) and baselines (\textbf{a},\textbf{c}--\textbf{e}). \textbf{a},\textbf{b}, WBM test candidates, three seeds. \textbf{a}, AUROC of stability, electronic-structure and compound questions answered from one forward pass of each backbone on the unrelaxed structure (filled), from element fractions only (open squares) and from the ten physics scalars only (open triangles; range over backbones). \textbf{b}, Signed error of the forecast number of candidates satisfying a compound question, relative to the realised number, for the product of two calibrated marginal answers, that product recalibrated on the slice and a head trained on the compound. \textbf{c}, Log-loss of screening questions on Matbench property datasets (five-fold cross-validation; one MatterSim-v1-5M pass and, faded, one ORB-v3 pass) for a regression with a calibrated threshold, a regression with propagated residuals and a composition-only head, relative to the head trained on the question (vertical line; values above 1 are worse). \textbf{d}, Expected calibration error of the same answers. \textbf{e}, As \textbf{c} for yes/no questions on 130,831 QM9 molecules (MACE-OFF23 and, faded, ORB-v3). \textbf{f}, Choice among five isomers of one formula: top-1 accuracy of the backbone's own energy and of the choice head (log-loss in brackets).}
\label{fig:typed}
\end{figure}

\IfFileExists{fig_voi_realised.pdf}{\begin{figure}[htbp]
\centering
\includegraphics[width=\textwidth]{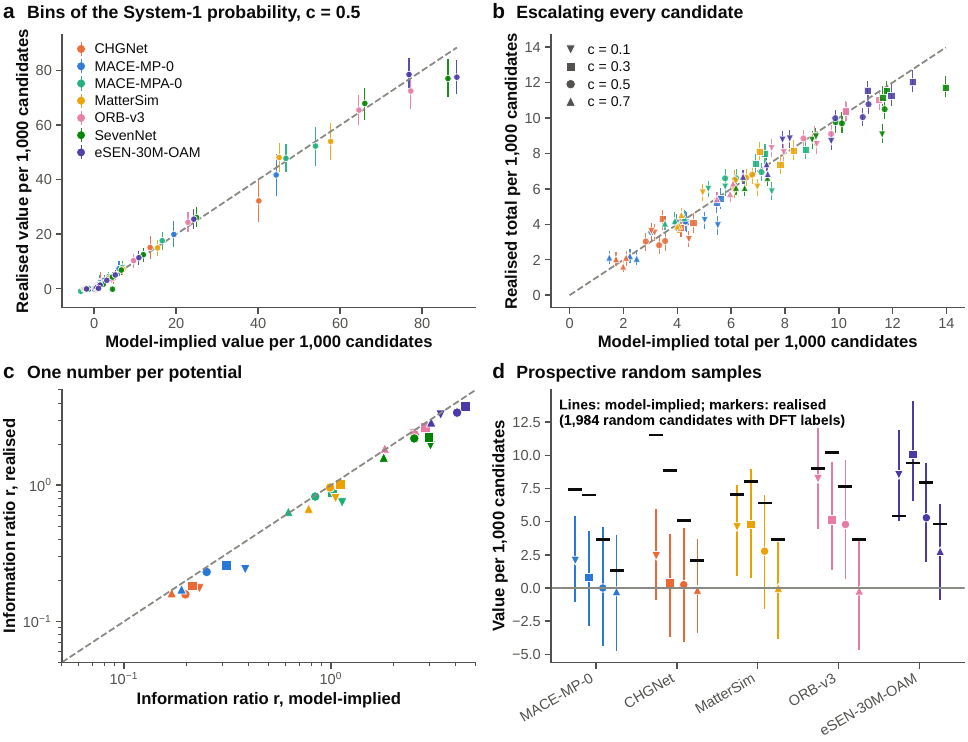}
\caption{\textbf{The value of escalation against real outcomes.} System~1 is the single-pass head and System~2 the head given the backbone's relaxed prediction, calibrated jointly on the slice (\textbf{a}--\textbf{c}, WBM test splits, three seeds; \textbf{d}, the prospective random samples, post hoc). \textbf{a}, Per equal-mass bin of the System-1 probability ($c=0.5$): realised value of consulting the relaxation, $(y-c)(\mathbf{1}[p_2\ge c]-\mathbf{1}[p_1\ge c])$, against the model-implied value $\max(p_2-c,0)-\max(p_1-c,0)$ (bootstrap 95\% intervals; dashed line: equality). \textbf{b}, Total value of escalating every candidate, per 1,000 candidates, realised against model-implied, for every potential, seed and attempt cost. \textbf{c}, Information ratio $r$ of the Gaussian closed form (Eq.~\ref{eq:voigauss}) fitted by least squares to the realised gains of the individual candidates against that fitted to their model-implied gains, per potential and attempt cost (mean over seeds). \textbf{d}, Realised (markers, 95\% intervals) and model-implied (lines) totals on the prospective random samples.}
\label{fig:voireal}
\end{figure}}{}

\IfFileExists{fig_shift.pdf}{\begin{figure}[htbp]
\centering
\includegraphics[width=\textwidth]{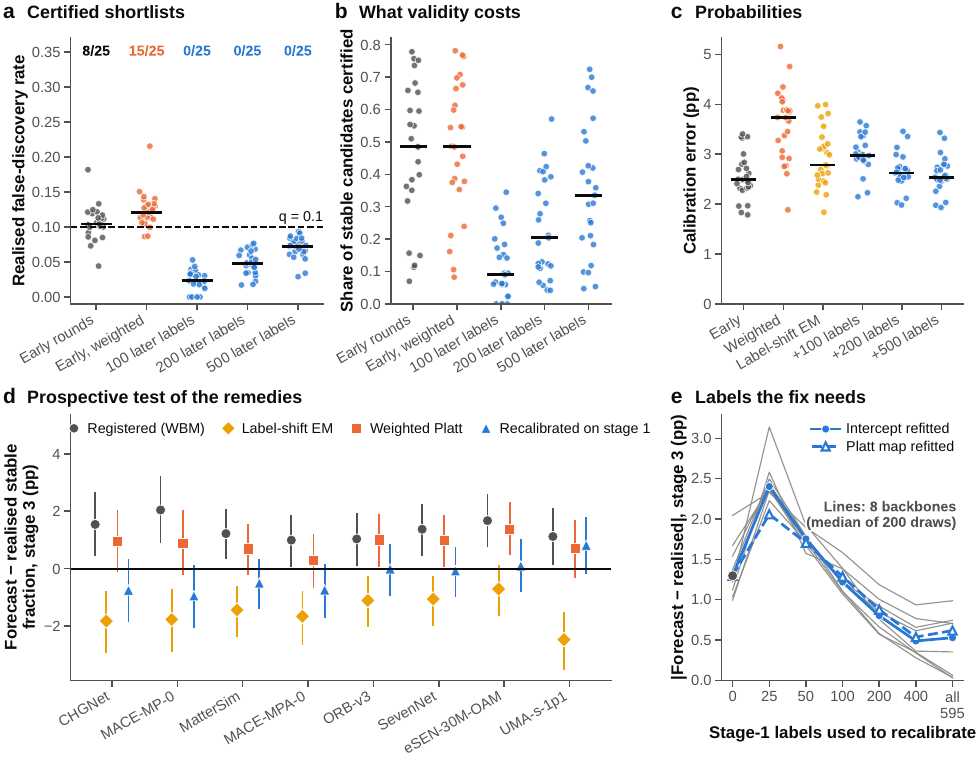}
\caption{\textbf{Remedies for the shift between WBM substitution rounds, and in the prospective test.} \textbf{a}--\textbf{c}, WBM: the 25 MP-trained leaderboard potentials of Fig.~\ref{fig:audit}e; calibration on 5\% of WBM drawn from rounds 1--2, decisions on 2,000 random structures of rounds 4--5 (20 draws; points: means per potential; bars: medians). \textbf{d},\textbf{e}, the prospective test (stage~3). \textbf{a}, Realised false-discovery rate of conformal selection at $q=0.1$, calibrated on the early rounds, on the early rounds with importance weights, or on 100, 200 or 500 random labelled structures of the later rounds; numbers: potentials whose mean exceeds $q$ (one-sided $z$-test over the draws, $z>2$). \textbf{b}, Share of the stable candidates certified by the same procedures. \textbf{c}, Expected calibration error of the early-round Platt map, of the same map with importance weights or with label-shift re-estimation of the stable fraction, and with its intercept refitted on 100 to 500 later-round labels. \textbf{d}, The same remedies tested prospectively, as registered before the stage-3 labels: forecast minus realised stable fraction of the stage-3 candidates for each backbone's registered forecast (calibrated on WBM only), its label-shift re-estimate and its importance-weighted map (both fitted on WBM's slice and the unlabelled pool), and its recalibration on the stage-1 labels (95\% intervals clustered by parent prototype). \textbf{e}, Post hoc: absolute forecast minus realised stable fraction of the stage-3 candidates after recalibrating each backbone's registered forecast on $n$ random stage-1 labels, by refitting the intercept of its logit or the full Platt map (median over 200 draws; grey lines: backbones, intercept refitted; $n=0$: the registered forecast).}
\label{fig:shift}
\end{figure}}{}

\begin{figure}[htbp]
\centering
\includegraphics[width=\textwidth]{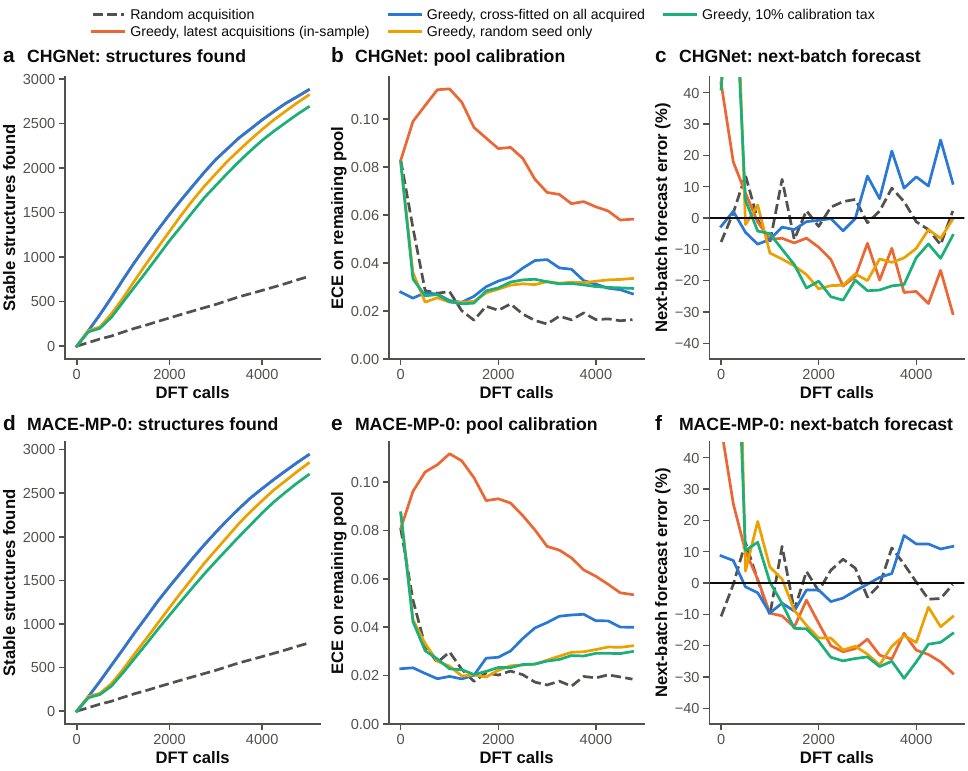}
\caption{\textbf{Calibration inside budgeted DFT loops.} Simulated campaigns on WBM with the hidden DFT labels as the oracle (top: CHGNet, bottom: MACE-MP-0 single-pass features; greedy acquisition unless stated, dashed lines: random acquisition; means over five seeds). \textbf{a},\textbf{d}, Stable structures found after a 500-structure random seed. \textbf{b},\textbf{e}, Expected calibration error on the remaining pool for calibration maps fitted on the most recent acquisitions, on all acquisitions with cross-fitted predictions, on the random seed or on a 10\% random sample of every batch. \textbf{c},\textbf{f}, Signed relative error of the forecast number of hits in the next batch.}
\label{fig:loop}
\end{figure}

\IfFileExists{fig_align.pdf}{\begin{figure}[htbp]
\centering
\includegraphics[width=\textwidth]{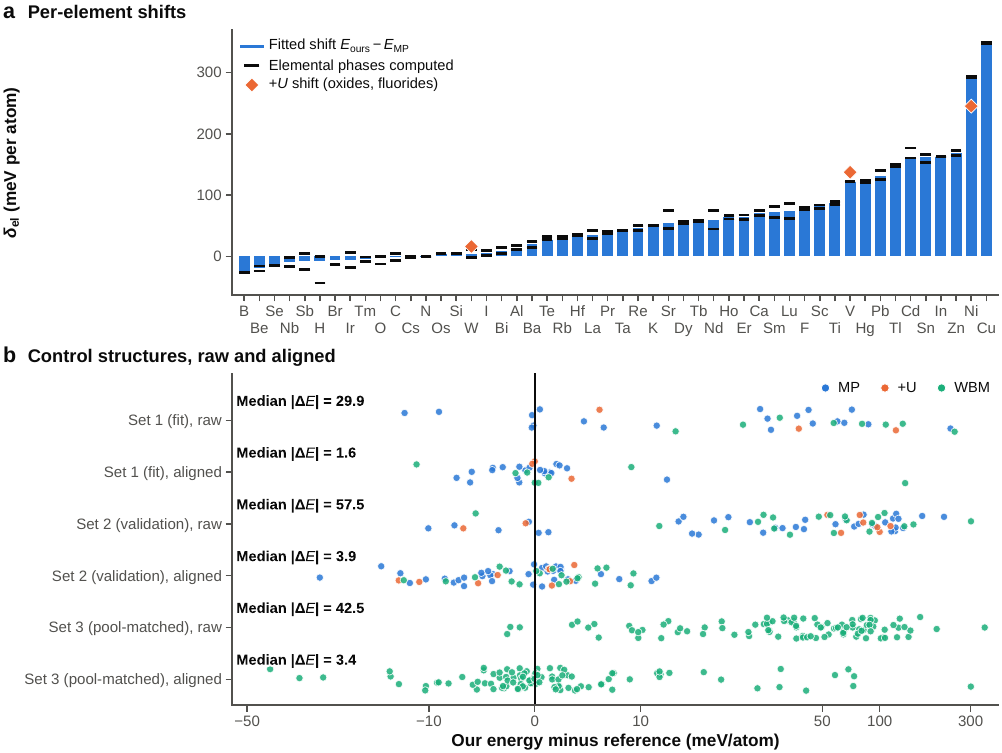}
\caption{\textbf{Putting new DFT energies on the Materials Project scale.} \textbf{a}, Per-element shifts fitted to recomputed MP elemental phases, binary $+U$ oxides and a first control set (bars), the differences of the individual elemental phases (dashes) and the $+U$ shifts of Ni, V and W in oxides and fluorides (diamonds). \textbf{b}, Our energy minus the reference energy, before and after alignment, for the first (fit), second (validation) and third (matched to the prospective pool) control sets: MP ground states, $+U$ compounds and WBM structures computed from their initial geometry, as the candidates were. Shifts were fixed before any prospective calculation was collected.}
\label{fig:align}
\end{figure}}{}

\IfFileExists{fig_overlap.pdf}{\begin{figure}[htbp]
\centering
\includegraphics[width=\textwidth]{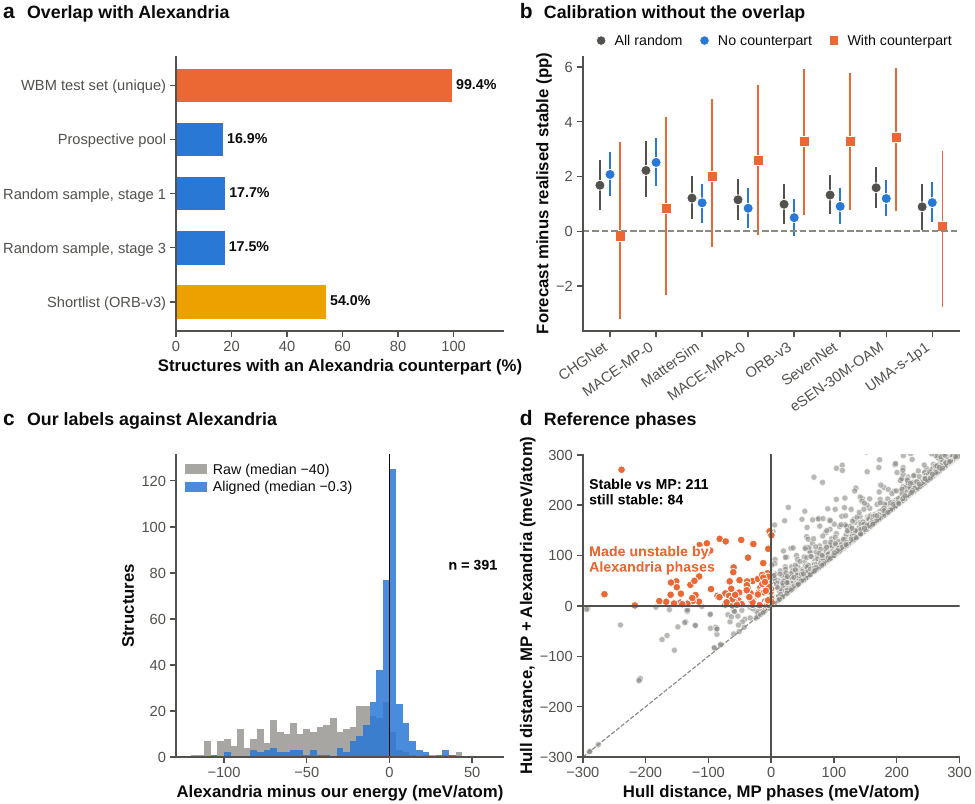}
\caption{\textbf{Training data and reference phases.} \textbf{a}, Share of structures with a counterpart in the Alexandria PBE database (pymatgen's StructureMatcher, any of three releases): WBM's unique prototypes (DFT-relaxed), our pool (initial or DFT-relaxed structures), the two random samples and the shortlist. \textbf{b}, Forecast minus realised stable fraction of the registered single-pass forecasts on all random labels (stages 1 and 3): all candidates, those without a counterpart and those with one (95\% intervals clustered by parent prototype). \textbf{c}, Alexandria's minus our energy for candidates whose DFT-relaxed structure matches an Alexandria entry, before and after our per-element alignment to MP's scale. \textbf{d}, Hull distance of our labelled candidates against MP's phases (the registered label) and against MP's plus Alexandria's phases near Alexandria's hull, put on MP's scale, with each candidate's own counterparts left out; highlighted: candidates stable against MP's phases only.}
\label{fig:overlap}
\end{figure}}{}

\FloatBarrier
\begin{table}[htbp]
\centering\scriptsize\setlength{\tabcolsep}{3pt}\renewcommand{\arraystretch}{1.0}
\caption{\textbf{Registered analyses of the prospective test and their outcomes.} Every analysis named in a registration file, with its result (ranges over backbones; percentage points, pp, for calibration in the large, CITL: forecast minus realised stable fraction; intervals: 95\%, resampling parent prototypes) and where it is reported (Fig.~\ref{fig:prospective}, Extended Data, ED, or Supplementary Note~\ref{sn:prospective}, SI). Analyses that no registration named are listed at the end as post hoc; registration digests and times are in Methods.}
\label{tab:registered}
\begin{tabular}{p{5.4cm}p{9.0cm}p{1.2cm}}
\toprule
Registered analysis & Outcome & Where\\
\midrule
\multicolumn{3}{p{15.6cm}}{\emph{Stage 1: random sample of 600 (calibration slice 200, test 400), registered 2026-10-02 21:12:07 UTC}}\\
Calibration of WBM-calibrated forecasts (CITL, ECE, NLL, Brier), against Platt and isotonic maps fitted on the slice & CITL 0.8 to 2.1 pp, intervals include zero for four of six; recalibration moved CITL towards zero for five & SI\\
Value at $c=0.1$--0.7 of System~1, System~2 and relax-and-threshold & System~1 8--19 net hits per 1,000 at $c=0.5$ & SI\\
Conformal screen ($\alpha=0.05$, 0.1) and selection ($q=0.1$, 0.2) with the slice & recall at $\alpha=0.1$ 87\%--100\%; selection non-empty for five backbones at $q=0.2$ & SI\\
Secondary: without $+U$ chemistries & CITL 0.2 to 2.2 pp & SI\\
\addlinespace
\multicolumn{3}{p{15.6cm}}{\emph{Stage 2: shortlist of 100, registered 2026-10-03 16:12:04 UTC (update after stage 1)}}\\
Realised stable structures against every registered forecast & 98 of 100, against forecasts of 92--100; 99.8 after stage~1 & Fig.~\ref{fig:prospective}e\\
\addlinespace
\multicolumn{3}{p{15.6cm}}{\emph{Stage 3: random sample of 1,400, registered 2026-10-04 07:10:06 UTC; 1,389 labelled}}\\
(1) Calibration of the registered forecasts on stage~3 and on all 1,984 random labels (CITL, ECE, NLL, Brier, AUROC) & stage~3: CITL 1.0--2.0 pp, all intervals above zero; NLL 0.088--0.138; calibration slope 0.82--1.01; all random labels: CITL 0.9--2.2 pp, AUROC 0.931--0.971 & Fig.~\ref{fig:prospective}c\\
(2) Stage-1 Platt and isotonic maps on stage~3 & CITL $-$0.9 to 0.8 (Platt) and $-$0.9 to 1.1 pp (isotonic); intervals include zero for all eight and seven of the eight & Fig.~\ref{fig:prospective}b,c\\
(3) Conformal screen with stage-1 thresholds & recall 90\%--98\%, 88\%--95\% and 70\%--91\% at $\alpha=0.05$, 0.1, 0.2 & SI\\
(4) Conformal selection with stage-1 calibration & non-empty for four ($q=0.1$) and seven ($q=0.2$) backbones; FDP 0.24--0.35 at $q=0.2$ & SI\\
(5) Value at four costs: System~1 (registered, recalibrated), System~2, relax-and-threshold, paired differences & System~1 minus System~2: $-$10 to $-$2 at $c=0.1$, $-$3 to 3 at $c=0.7$; minus relax-and-threshold at $c=0.7$: 28 and 21 (MACE-MP-0, CHGNet) & Fig.~\ref{fig:prospective}d\\
(6a) Without an Alexandria structural counterpart & all random labels: CITL 0.5--2.5 pp & ED Fig.~\ref{fig:overlap}b\\
(6b) Without an Alexandria protostructure counterpart & 16\% of the pool has one; random labels without one: CITL 0.7--2.3 pp & SI\\
(6) Without $+U$ chemistries & CITL 1.2--2.5 pp (1,238 candidates) & SI\\
(7) Labels against MP's plus Alexandria's phases & stable fraction 5.8\% $\to$ 1.7\%; CITL 5.2--6.2 pp & ED Fig.~\ref{fig:overlap}d\\
(8) eSEN-30M-OAM and UMA-s-1p1 by the registered protocol & included in (1)--(5); stage~3 alone CITL 1.7 and 1.1 pp & Fig.~\ref{fig:prospective}\\
Addendum 1: label-shift EM and weighted Platt forecasts & CITL $-$2.5 to $-$0.7 (EM) and 0.3 to 1.4 pp (weighted) & ED Fig.~\ref{fig:shift}d\\
Addendum 1: WBM-calibrated screens and selection, unweighted and weighted & screen recall at $\alpha=0.1$ 74\%--85\% (unweighted), 74\%--86\% (weighted); selection FDP at $q=0.2$ up to 0.32 and 0.29 & SI\\
Addenda 2--3: System~2 of eSEN-30M-OAM and UMA-s-1p1 & included in (1) and (5) & Fig.~\ref{fig:prospective}d\\
\addlinespace
\multicolumn{3}{p{15.6cm}}{\emph{Bridging set: 1,105 WBM structures with our protocol, registered 2026-10-05 04:18:50 UTC; digests public 2026-10-05 05:35 UTC; 1,099 labelled}}\\
(1)--(2) Agreement of WBM's and our labels; protocol-adjusted CITL on all random labels & WBM's minus our stable fraction 0.4 pp ($-$0.6 to 1.4); adjusted CITL 1.4--2.8 pp, intervals exclude zero for all eight: a shift of the population & Fig.~\ref{fig:prospective}c\\
(3)--(8) Secondary: by group, family and hull distance; residuals; label-noise models & CITL on the bridging structures $-$0.3 to 0.8 pp (our labels) & SI\\
\addlinespace
\multicolumn{3}{p{15.6cm}}{\emph{B2: 461 candidates near the hull or shortlisted, registered 2026-10-05 06:53:42 UTC; digests public 2026-10-05 08:57 UTC; 1,120 of 1,129 calculations finished}}\\
(A1) Hull distance from the lowest of the original, rattled and eSEN-started relaxations & lower by more than 1 and 10~meV/atom for 39.3\% and 8.7\% of 461; 13 unstable labels become stable & SI\\
(A2)--(A4) Stable fraction, CITL and shortlist yield with these labels & stable fraction +0.7~points (0.3 to 1.1); CITL 0.2--1.6 pp, intervals exclude zero for three of the eight; shortlist 98 hits & SI\\
(B1) Antiferromagnetic starts & lower by more than 1~meV/atom for 28 of 75; no stable call changes & SI\\
(C1) SCAN local hulls of 18 candidates & labels changed for 12 of 18 at 0~meV/atom and 10 at 50~meV/atom, all from stable to unstable (for 4, a competing phase failed: lower bounds) & SI\\
\addlinespace
\multicolumn{3}{p{15.6cm}}{\emph{Post hoc (not registered)}}\\
\multicolumn{3}{p{15.6cm}}{Bootstrap intervals and AUROC of stage~1; label-noise sensitivity of stages 1 and 3; bounds for the 11 unlabelled stage-3 calculations; equivalence tests of the remedies; labels needed for the fix; re-splits of the conformal calibration sets; realised value of escalation on the random samples; design effect of the shortlist and the diagnostic calculations of its failures.}\\
\bottomrule
\end{tabular}
\end{table}

\clearpage
\label{page:si}
\setcounter{figure}{0}\setcounter{table}{0}
\renewcommand{\figurename}{Supplementary Fig.}
\renewcommand{\tablename}{Supplementary Table}
\renewcommand{\theHfigure}{SI.\arabic{figure}}\renewcommand{\theHtable}{SI.\arabic{table}}
\section*{Supplementary Information}
\noindent This Supplementary Information accompanies the article. Supplementary Notes~\ref{sn:audit}--\ref{sn:prospective} give the full analyses summarised in the main text; Supplementary Tables report the quantities behind the figures. Figures and Extended Data figures are cited by their numbers in the article.

\snote{Auditing the Matbench Discovery leaderboard}{sn:audit}

Any model on the Matbench Discovery leaderboard can be read as a System-1 scorer because the MP hull used by the benchmark is fixed and contains no WBM entry: a predicted formation energy $\hat e_\mathrm{form}$ determines the predicted hull distance $\ehat=\hat e_\mathrm{form}-e^\mathrm{hull}_\mathrm{form}(\mathbf{x})$, where $e^\mathrm{hull}_\mathrm{form}(\mathbf{x})$ is the formation energy of the MP hull at the candidate's composition (Methods, Eq.~\ref{eq:identity}). We converted the published WBM predictions of all 65 models (25 interatomic potentials trained only on MP-derived data, hereafter \emph{MP-trained}; 30 potentials trained with larger external sets such as OMat24, sAlex or Alexandria; and 10 models that are not interatomic potentials) into probabilities with one-dimensional maps fitted on a random 5\% slice of the candidate pool, reproducing the published F1 and precision of every model to within 0.0007 (Methods).

\paragraph{The threshold is not where the probability is one half.} Without any calibration map, the fraction of DFT-stable structures among candidates predicted within $\pm5$~meV/atom of the hull is 0.19--0.46 for MP-trained potentials, 0.12--0.30 for non-potential models and 0.37--0.57 for potentials trained on larger sets (Fig.~\ref{fig:audit}a,b). For MP-trained potentials the hard rule $\ehat\le0$ draws its boundary through a population that is mostly unstable. Much of this is a property of the population: there are 1.6 times as many WBM structures in the first 25~meV/atom above the hull as in the 25~meV/atom below it, so errors carry more unstable structures across the threshold than stable ones, and even an \emph{unbiased} predictor with Gaussian errors of 30 or 100~meV/atom would see stable fractions of only 0.34 and 0.24 there. The isotonic map places the 50\% point between $-$54 and $-$0.3~meV/atom for MP-trained potentials and within 2.2~meV/atom of zero for potentials trained on larger sets.

\paragraph{A two-number account.} If the true hull distances of the pool have a local density proportional to $e^{\kappa E}$ near the hull and a model's errors there are Gaussian with bias $b$ and spread $s$, then $P(\mathrm{stable}\mid\ehat=t)=\Phi\bigl((b-\kappa s^2-t)/s\bigr)$ (Methods, Eq.~\ref{eq:tilt}). Fitting $\kappa=14.5$~eV$^{-1}$ once to the WBM population and taking $b$ and $s$ from each model's errors near the hull predicts the calibrated probability at the threshold of all 65 models with Pearson $r=0.96$ (Extended Data Fig.~\ref{fig:value}a), and, applied to each WBM substitution round separately ($\kappa$ from 11.5 to 16.7~eV$^{-1}$), with $r=0.94$ over all model--round pairs. Where a model's threshold should sit is a joint property of its errors and of the candidate population, not of either alone.

\paragraph{A threshold is a cost assumption.} Read as a decision rule, $\ehat\le0$ maximises value only if a failed attempt costs $c^\ast=0.18$--$0.48$ of a hit for MP-trained potentials (median 0.33; Extended Data Fig.~\ref{fig:value}b). At any other cost the calibrated rule $p\ge c$ is worth more: when attempts are expensive ($c=0.7$, as for a synthesis) it gains a median of 12.8 hits per 1,000 candidates for MP-trained potentials and 85.6 for non-potential models, whose hard calls then lose value outright. The calibrated rule also forecasts what it will deliver: the expected number of stable structures in the leaderboard shortlist is forecast to within 0.3\% (median over models), whereas the hard rule implicitly promises that every shortlisted candidate is stable and delivers 24\%--93\% of that promise. A few hundred labels suffice: an isotonic map fitted on 200 random labels from the pool comes within 0.004 in Brier score of the best monotone map in hindsight for the median MP-trained potential (Extended Data Fig.~\ref{fig:value}c), and 100 labels already beat the hard rule at $c=0.1$ for 25 of 25 MP-trained potentials.

\paragraph{F1 hides how far confident calls can be trusted.} Ranking models by the log-loss of their calibrated probabilities reproduces the F1 ranking closely (Kendall $\tau=0.98$), because both are dominated by the size of each model's errors. The number of candidates that can be certified at a 10\% false-discovery rate by conformal selection\cite{jin2023selection} is less well summarised (Fig.~\ref{fig:audit}d): between MP-trained potentials whose F1 differs by less than 0.01 it differs up to 7.8-fold in the mean over the audit's 20 calibration draws, and about 9-fold in the median over 50 further random slices of the same size (bootstrap interval 7.5--14; 3--34-fold in 90\% of single slices; Allegro-MP-L certifies fewer candidates than each of its three F1 neighbours in every slice), whereas at a 20\% false-discovery rate the same pairs differ at most 1.1-fold, because what matters is how deep each model's ranking stays clean (Fig.~\ref{fig:audit}f) --- only 86\% of MACE-MP-0's 1,000 most confident ``stable'' calls are stable, against 99\% for CHGNet.

\FloatBarrier
{\footnotesize\setlength{\tabcolsep}{3.2pt}
\begin{longtable}{lcrrrrrrrrr}
\caption{Calibration leaderboard of all audited Matbench Discovery models (WBM unique prototypes; 1-D maps fitted on a random 5\% slice, means over 5 seeds; conformal selection averaged over 20 draws). Sorted by isotonic NLL. Data: MP = potential trained only on MP-derived data (MP-trained); ext. = potential trained with larger external sets; non-pot. = not an interatomic potential. $p_{\pm5}$: DFT-stable fraction among candidates predicted within $\pm5$~meV/atom of the hull. ECE$_\sigma$: ECE of the error-bar map $\Phi(-\hat E_\mathrm{hull}/\sigma)$. $W$: width of the $0.2\le p\le0.8$ band of the isotonic map (-- where the map does not span the band). $N_{0.1}$: candidates certified at FDR $\le$ 0.1. FDR$_\mathrm{shift}$: realised FDR when calibrating on WBM rounds 1--2 and selecting from rounds 4--5.}\label{tab:audit}\\
\toprule
model & data & F1 & MAE & NLL$_\mathrm{iso}$ & ECE$_\mathrm{iso}$ & ECE$_\sigma$ & $p_{\pm5}$ & $W$ & $N_{0.1}$ & FDR$_\mathrm{shift}$\\
 & & & (meV) & & & & & (meV) & & \\
\midrule
\endfirsthead
\toprule
model & data & F1 & MAE & NLL$_\mathrm{iso}$ & ECE$_\mathrm{iso}$ & ECE$_\sigma$ & $p_{\pm5}$ & $W$ & $N_{0.1}$ & FDR$_\mathrm{shift}$\\
\midrule
\endhead
GRACE-3L-OAM-L & ext. & 0.925 & 18 & 0.074 & 0.002 & 0.101 & 0.49 & 7 & 33,143 & 0.090\\
EquiformerV3+DeNS-OAM & ext. & 0.931 & 18 & 0.075 & 0.002 & 0.104 & 0.48 & 6 & 33,488 & 0.085\\
TECE-OAM-RRA-1.0 & ext. & 0.929 & 18 & 0.076 & 0.002 & 0.103 & 0.49 & 7 & 33,174 & 0.087\\
EquFlashV2 & ext. & 0.929 & 18 & 0.076 & 0.002 & 0.101 & 0.48 & 6 & 33,159 & 0.088\\
TACE-OAM-RRA-Preview & ext. & 0.928 & 18 & 0.077 & 0.002 & 0.105 & 0.49 & 7 & 33,132 & 0.085\\
Prophet-OAME-MBD & ext. & 0.928 & 18 & 0.078 & 0.002 & 0.103 & 0.49 & 6 & 33,066 & 0.090\\
eSEN-30M-OAM & ext. & 0.925 & 18 & 0.078 & 0.002 & 0.105 & 0.51 & 7 & 33,031 & 0.088\\
PET-OAM-XL & ext. & 0.924 & 19 & 0.079 & 0.002 & 0.106 & 0.51 & 7 & 32,954 & 0.086\\
MatRIS-10M-OAM & ext. & 0.921 & 19 & 0.082 & 0.002 & 0.103 & 0.49 & 8 & 32,570 & 0.090\\
EquFlash & ext. & 0.918 & 19 & 0.083 & 0.002 & 0.101 & 0.48 & 7 & 32,720 & 0.093\\
eqV2 M & ext. & 0.917 & 20 & 0.085 & 0.002 & 0.104 & 0.49 & 8 & 32,468 & 0.089\\
TACE-OAM-L & ext. & 0.910 & 20 & 0.089 & 0.002 & 0.103 & 0.45 & 9 & 32,220 & 0.096\\
Nequip-OAM-XL & ext. & 0.906 & 20 & 0.092 & 0.003 & 0.102 & 0.45 & 10 & 31,956 & 0.096\\
ORB v3 & ext. & 0.905 & 24 & 0.092 & 0.002 & 0.119 & 0.48 & 10 & 31,819 & 0.085\\
SevenNet-Omni-i12 & ext. & 0.906 & 21 & 0.092 & 0.002 & 0.102 & 0.49 & 11 & 31,863 & 0.091\\
SevenNet-MF-ompa & ext. & 0.901 & 21 & 0.096 & 0.002 & 0.101 & 0.43 & 12 & 31,442 & 0.087\\
AlphaNet-v1-OAM & ext. & 0.901 & 24 & 0.096 & 0.002 & 0.118 & 0.40 & 11 & 31,844 & 0.094\\
Allegro-OAM-L & ext. & 0.895 & 22 & 0.097 & 0.002 & 0.103 & 0.37 & 12 & 31,289 & 0.086\\
Nequip-OAM-L & ext. & 0.893 & 22 & 0.100 & 0.002 & 0.101 & 0.47 & 14 & 30,833 & 0.098\\
DPA3-v2-OpenLAM & ext. & 0.890 & 22 & 0.101 & 0.002 & 0.099 & 0.43 & 14 & 30,748 & 0.091\\
TACE-v1-OAM-M & ext. & 0.889 & 22 & 0.102 & 0.002 & 0.101 & 0.43 & 14 & 30,581 & 0.094\\
DPA-3.1-3M-FT & ext. & 0.884 & 23 & 0.104 & 0.003 & 0.100 & 0.39 & 14 & 30,295 & 0.094\\
DPA3-v1-OpenLAM & ext. & 0.883 & 23 & 0.106 & 0.003 & 0.098 & 0.45 & 15 & 30,040 & 0.091\\
GRACE-2L-OAM-L & ext. & 0.883 & 22 & 0.107 & 0.002 & 0.099 & 0.51 & 14 & 30,270 & 0.094\\
GRACE-2L-OAM & ext. & 0.880 & 23 & 0.107 & 0.002 & 0.101 & 0.47 & 13 & 29,942 & 0.095\\
ORB v2 MPA & ext. & 0.880 & 28 & 0.114 & 0.003 & 0.101 & 0.57 & 13 & 30,371 & 0.072\\
MatterSim v1 5M & ext. & 0.862 & 24 & 0.119 & 0.003 & 0.094 & 0.53 & 19 & 29,022 & 0.087\\
EquiformerV3+DeNS-MP & MP & 0.863 & 29 & 0.127 & 0.004 & 0.114 & 0.36 & 17 & 28,252 & 0.118\\
MACE-MPA-0 & ext. & 0.852 & 28 & 0.128 & 0.003 & 0.102 & 0.45 & 19 & 27,483 & 0.089\\
DPA-4.0.1-Pro-MPtrj & MP & 0.857 & 29 & 0.136 & 0.003 & 0.111 & 0.40 & 16 & 28,070 & 0.108\\
DPA-4.0-Pro-MPtrj & MP & 0.850 & 30 & 0.138 & 0.003 & 0.114 & 0.46 & 18 & 27,477 & 0.104\\
MatRIS-10M-MP & MP & 0.847 & 31 & 0.140 & 0.004 & 0.117 & 0.37 & 19 & 26,812 & 0.122\\
GRACE-1L-OAM & ext. & 0.824 & 31 & 0.148 & 0.004 & 0.101 & 0.43 & 27 & 18,886 & 0.073\\
eSEN-30M-MP & MP & 0.831 & 33 & 0.150 & 0.003 & 0.115 & 0.36 & 23 & 24,758 & 0.121\\
NequIP-GNoME & ext. & 0.829 & 35 & 0.151 & 0.003 & 0.118 & 0.46 & 19 & 25,685 & 0.107\\
eqV2 S DeNS & MP & 0.815 & 36 & 0.159 & 0.003 & 0.125 & 0.31 & 29 & 23,699 & 0.124\\
MatRIS v0.5.0 MPtrj & MP & 0.809 & 37 & 0.162 & 0.004 & 0.122 & 0.33 & 29 & 22,819 & 0.122\\
AlphaNet-v1-MPtrj & MP & 0.799 & 41 & 0.164 & 0.004 & 0.138 & 0.33 & 33 & 17,756 & 0.119\\
DPA-3.1-MPtrj & MP & 0.803 & 37 & 0.168 & 0.003 & 0.120 & 0.33 & 29 & 22,013 & 0.105\\
DPA3-v2-MPtrj & MP & 0.786 & 39 & 0.177 & 0.005 & 0.124 & 0.31 & 32 & 20,097 & 0.109\\
Eqnorm MPtrj & MP & 0.787 & 40 & 0.181 & 0.004 & 0.122 & 0.32 & 37 & 18,301 & 0.124\\
HIENet & MP & 0.776 & 41 & 0.189 & 0.005 & 0.117 & 0.36 & 40 & 17,405 & 0.108\\
DPA3-v1-MPtrj & MP & 0.765 & 42 & 0.190 & 0.004 & 0.122 & 0.31 & 43 & 17,175 & 0.119\\
ORB v2 MPtrj & MP & 0.765 & 45 & 0.192 & 0.003 & 0.130 & 0.32 & 39 & 16,606 & 0.109\\
Nequip-MP-L & MP & 0.761 & 43 & 0.197 & 0.005 & 0.117 & 0.34 & 45 & 13,677 & 0.089\\
SevenNet-l3i5 & MP & 0.760 & 44 & 0.199 & 0.004 & 0.121 & 0.34 & 54 & 11,636 & 0.143\\
Nequix MP & MP & 0.751 & 44 & 0.200 & 0.005 & 0.127 & 0.29 & 56 & 11,240 & 0.120\\
Nequix MP PFT & MP & 0.748 & 44 & 0.202 & 0.005 & 0.121 & 0.31 & 51 & 11,473 & 0.118\\
Allegro-MP-L & MP & 0.751 & 44 & 0.204 & 0.005 & 0.120 & 0.33 & 60 & 1,484 & 0.036\\
SevenNet-0 & MP & 0.724 & 48 & 0.215 & 0.004 & 0.131 & 0.28 & 64 & 9,357 & 0.136\\
GRACE-2L-MPtrj & MP & 0.691 & 52 & 0.241 & 0.005 & 0.113 & 0.31 & 101 & 3,851 & 0.087\\
MACE-MP-0 & MP & 0.669 & 57 & 0.247 & 0.006 & 0.144 & 0.27 & 146 & 1,317 & 0.104\\
CHGNet & MP & 0.612 & 63 & 0.267 & 0.006 & 0.159 & 0.23 & 98 & 6,461 & 0.183\\
BAM-MP-core & MP & 0.623 & 60 & 0.276 & 0.006 & 0.125 & 0.30 & 130 & 2,765 & 0.106\\
M3GNet & MP & 0.569 & 75 & 0.289 & 0.006 & 0.199 & 0.19 & 145 & 1,902 & 0.085\\
EMA-GNN & non-pot. & 0.557 & 86 & 0.293 & 0.007 & 0.205 & 0.15 & 238 & 1,559 & 0.058\\
ESNet & non-pot. & 0.571 & 109 & 0.308 & 0.006 & 0.137 & 0.28 & 144 & 2,897 & 0.068\\
ALIGNN & non-pot. & 0.567 & 93 & 0.315 & 0.007 & 0.163 & 0.26 & 566 & 205 & 0.116\\
CGCNN+P & non-pot. & 0.500 & 113 & 0.338 & 0.008 & 0.214 & 0.19 & 578 & 232 & 0.120\\
MEGNet & non-pot. & 0.510 & 130 & 0.339 & 0.007 & 0.157 & 0.30 & 497 & 230 & 0.013\\
CGCNN & non-pot. & 0.507 & 138 & 0.344 & 0.008 & 0.165 & 0.23 & 524 & 193 & 0.112\\
Wrenformer & non-pot. & 0.466 & 110 & 0.354 & 0.007 & 0.250 & 0.17 & 515 & 230 & 0.193\\
BOWSR & non-pot. & 0.423 & 118 & 0.373 & 0.007 & 0.254 & 0.12 & -- & 41 & 0.000\\
AlchemBERT & non-pot. & 0.421 & 117 & 0.376 & 0.007 & 0.258 & 0.18 & -- & 140 & 0.137\\
Voronoi RF & non-pot. & 0.333 & 148 & 0.411 & 0.008 & 0.269 & 0.17 & -- & 49 & 0.009\\
\bottomrule
\end{longtable}
}

\snote{One forward pass versus a relaxation}{sn:onepass}

The audit calibrates what a model computes after a relaxation. CrystalJev instead decides from a single forward pass on the unrelaxed input. On the WBM unique prototypes we trained heads on 45\% of the chemical systems, calibrated them on a random 5\% slice of the remaining structures and evaluated them on the rest, so that calibration and test are exchangeable but chemically disjoint from training (Methods); the relax-and-threshold practice of the same backbone --- its published relaxed prediction with an isotonic map on the same slice --- is the reference.

\paragraph{Seven backbones that relax.} For the two MP-trained potentials that underlie the original leaderboard, the single pass decides better than the relaxation (Supplementary Table~\ref{tab:backbones}; Fig.~\ref{fig:backbones}a): log-loss 0.191 against 0.247 for MACE-MP-0 and 0.196 against 0.266 for CHGNet, with higher AUROC, at one forward pass instead of a relaxation of about thirty (median per-structure cost ratio 29 and 30). For the five newer potentials trained on larger data (MACE-MPA-0 (MPtrj and sAlex), MatterSim-v1-5M (MatterSim's own data), ORB-v3 (MPtrj, Alexandria and OMat24), SevenNet-MF-ompa (MPtrj, OMat24 and sAlex) and eSEN-30M-OAM (OMat24, sAlex and MPtrj)), whose relaxed predictions are much more accurate (isotonic log-loss 0.080--0.129), the single pass comes within 0.009--0.033 in log-loss of the far more accurate relaxed predictions of five newer backbones, and, given the relaxed prediction as one more input, improves on it for every one of the seven backbones studied. Every one of these differences, and the improvement of the head given the relaxed prediction, exceeds its 95\% interval when chemical systems rather than candidates are resampled (65,957 systems over the three seeds; design effects of the differences 1.2--1.9; the single pass's log-loss differs from the calibrated relaxation's by $-$23\% for MACE-MP-0 and $-$26\% for CHGNet, and by $+$7\% to $+$42\% for the newer potentials). The embeddings carry information beyond the energies: a boosted head on the ten physics scalars and the relaxed prediction alone reaches 0.214 (MACE-MP-0), against 0.172 with embeddings. Fusing two backbones helps further (log-loss 0.161 for MACE-MP-0 with CHGNet).

\paragraph{Deciding directly against propagating energy uncertainty.} The usual route to a probability of stability is to predict the energy with an uncertainty and propagate it through the threshold. On the same single-pass features, labels and slice (Extended Data Fig.~\ref{fig:baselines}a; Supplementary Table~\ref{tab:baselines}), a heteroscedastic Gaussian regression of $E_\mathrm{hull}$ overstates the stable fraction by 2.2 percentage points (MACE-MP-0) and, even after isotonic recalibration, reaches log-loss 0.211 against 0.192 for the head trained on the question; quantile regression behaves alike. The regression models describe the whole distribution of $E_\mathrm{hull}$ better (lower continuous ranked probability score) but spend that accuracy away from the threshold at which the decision is taken. This holds for regressions built with the same gradient-boosted trees as the head; a deep ensemble of networks regressing the hull distance can do better (below). An ensemble of the 25 MP-trained leaderboard potentials, a strong but expensive baseline, reaches 0.135 with a boosted map on the mean and spread of their relaxed predictions, about what its best single member reaches with an isotonic map (EquiformerV3+DeNS-MP, 0.128); the single pass on MACE-MP-0 matches the median member after relaxation. A single threshold-conditioned head can answer every question of the family $E_\mathrm{hull}\le t$, but only at thresholds seen in training unless it is recalibrated at the queried one (Extended Data Fig.~\ref{fig:baselines}b).

\paragraph{Probabilistic baselines.} To test the head against stronger probabilistic models, we fitted four regressions of the hull distance with a predictive distribution to the same single-pass state, labels, splits and slice for all eight backbones (Methods) --- NGBoost, Gaussian-process regression, a deep ensemble of five heteroscedastic networks and a split conformal predictive system on boosted regressions --- and an ensemble of five network classifiers trained on the label (Extended Data Fig.~\ref{fig:baselines}e; Supplementary Table~\ref{tab:mlbaselines}). After Platt recalibration on the slice (the conformal system is calibrated on it by construction), NGBoost, Gaussian-process regression and the conformal predictive system decided worse than the head on every backbone, by 0.030--0.058, 0.085--0.243 and 0.009--0.018 in log-loss (24 of 24 backbone--model pairs with 95\% intervals above zero, from a bootstrap over chemical systems pooled over three seeds; the head's log-loss is 0.112--0.196). The deep regression ensemble was the strongest baseline: better than the head for CHGNet and MACE-MP-0 (by 0.002--0.006) and worse for the other six backbones (by 0.002--0.014). Recalibration on the slice mattered as much as the learner: without it, the same ensemble lost 0.003--0.010. The ensemble of classifiers decided worse than the boosted head on seven of the eight backbones (its interval covers zero for the others) and worse than the regression ensemble on six: with networks, the hull distance is usually a richer training signal than its sign. Every model here decides from the same unrelaxed forward pass and the same slice; which learner reads the pass best depends on the backbone, and the deep ensemble gains most where the head's log-loss is highest, that is, where the pass says least about stability. The two readers are complementary. Averaging the logits of the head and of the recalibrated deep regression ensemble and recalibrating on the slice --- a combination designed after the results above, hence post hoc --- improved on the head on all eight backbones, by 0.003--0.015 in log-loss (3\%--8\%), and improved the decisions: 0.5--2.2 more net hits per 1,000 candidates at $c=0.5$ (the head realises 38--57), with intervals above zero for eight backbones, and 2\%--10\% more candidates certified by conformal selection at $q=0.1$, with realised false-discovery proportions of at most 0.11. A lower log-loss need not mean better decisions: on MACE-MP-0 and CHGNet, where the deep ensemble's log-loss is lower than the head's, its decisions were better only at $c=0.1$ and indistinguishable from the head's at $c=0.3$--0.7.

\paragraph{Energy-level baselines and the threshold-conditioned head.} All baselines of Supplementary Note~\ref{sn:onepass} use the same features, training subsample, calibration slice and gradient-boosting settings as the decision heads. The heteroscedastic Gaussian regression fits a mean model of $E_\mathrm{hull}$, a model of the logarithm of five-fold cross-fitted absolute residuals, and one variance scale $k$ that minimises the Gaussian negative log-likelihood on the calibration slice; $p=\Phi(-\mu/k\hat\sigma)$. Quantile regression fits seven quantile models (5--95\%), sorts them per candidate and reads $p=\hat F(0)$ by linear interpolation. The threshold-conditioned head is one gradient-boosted classifier on the features and the threshold $t$, trained on all (structure, threshold) pairs for $t\in\{-50,0,50,100,200\}$~meV/atom with a monotone constraint in $t$, followed by one Platt map shared across thresholds; its CDF is made monotone by a running maximum. The CRPS is integrated over $-0.25\le t\le0.5$~eV/atom. The leaderboard ensemble uses the published relaxed predictions of the 25 MP-trained potentials; the boosted map is a shallow gradient-boosted classifier on the mean and the logarithm of the standard deviation, fitted on the calibration slice.

\paragraph{The decision layer on MACE-MP-0 and CHGNet.} Extended Data Fig.~\ref{fig:decision} and Supplementary Table~\ref{tab:onhull} detail the regimes compared in the main text for the two MP-trained backbones: zero-shot transfer of a head trained on MP relaxed structures, in-domain post-training, and relax-and-threshold, with the label efficiency of in-domain training and the cost of a relaxation in single passes. Supplementary Table~\ref{tab:typed} reports the ordinal, choice and prototype questions for these backbones and Supplementary Table~\ref{tab:backbones_typed} for all six.

\begin{landscape}
\par\noindent\begin{minipage}{\linewidth}\captionsetup{type=table}\centering\small
\caption{The decision layer across backbones (stability question, WBM test set, means $\pm$ s.d. over three seeds; identical splits, labels and calibration slice). Relaxed: the backbone's published leaderboard prediction after relaxation, with an isotonic map on the slice. Single pass: gradient-boosted head on the pooled node features and physics scalars of one forward pass on the unrelaxed input, Platt-calibrated. $N_{0.1}$: candidates certified by conformal selection at FDR $\le0.1$. eSEN-30M-OAM and UMA-s-1p1 were evaluated with the same splits and heads (Methods); $^\dagger$UMA-s-1p1 has no published WBM relaxations: its relaxed columns use our relaxations of 5,000 random test candidates of the first split, with the isotonic map fitted on 5,000 relaxed candidates of its slice.}\label{tab:backbones}
\begin{adjustbox}{max width=\linewidth}\begin{tabular}{llccccccc}
\toprule
 & & \multicolumn{2}{c}{relaxed, isotonic} & \multicolumn{2}{c}{single pass} & single pass + relaxed & \multicolumn{2}{c}{$N_{0.1}$}\\
\cmidrule(lr){3-4}\cmidrule(lr){5-6}\cmidrule(lr){7-7}\cmidrule(lr){8-9}
backbone & training data & AUROC & NLL & AUROC & NLL & NLL & relaxed (Platt) & single pass\\
\midrule
MACE-MP-0 & MPtrj & 0.913\,$\pm$\,0.001 & 0.247\,$\pm$\,0.001 & 0.951\,$\pm$\,0.001 & 0.191\,$\pm$\,0.003 & 0.172\,$\pm$\,0.002 & 182 & 9,193\\
CHGNet & MPtrj & 0.895\,$\pm$\,0.002 & 0.266\,$\pm$\,0.001 & 0.947\,$\pm$\,0.002 & 0.196\,$\pm$\,0.003 & 0.180\,$\pm$\,0.003 & 3,021 & 8,432\\
MACE-MPA-0 & MPtrj, sAlex & 0.975\,$\pm$\,0.000 & 0.129\,$\pm$\,0.001 & 0.974\,$\pm$\,0.000 & 0.138\,$\pm$\,0.001 & 0.104\,$\pm$\,0.001 & 14,123 & 13,650\\
MatterSim-v1-5M & MatterSim & 0.977\,$\pm$\,0.001 & 0.121\,$\pm$\,0.002 & 0.977\,$\pm$\,0.000 & 0.130\,$\pm$\,0.001 & 0.096\,$\pm$\,0.000 & 14,934 & 14,085\\
ORB-v3 & MPtrj, Alexandria, OMat24 & 0.984\,$\pm$\,0.001 & 0.094\,$\pm$\,0.002 & 0.980\,$\pm$\,0.000 & 0.117\,$\pm$\,0.001 & 0.071\,$\pm$\,0.001 & 16,618 & 15,070\\
SevenNet-MF-ompa & MPtrj, OMat24, sAlex & 0.984\,$\pm$\,0.000 & 0.096\,$\pm$\,0.002 & 0.976\,$\pm$\,0.000 & 0.129\,$\pm$\,0.002 & 0.078\,$\pm$\,0.001 & 16,666 & 14,287\\
eSEN-30M-OAM & OMat24, sAlex, MPtrj & 0.986\,$\pm$\,0.000 & 0.080\,$\pm$\,0.002 & 0.982\,$\pm$\,0.000 & 0.112\,$\pm$\,0.001 & 0.059\,$\pm$\,0.001 & 17,183 & 15,437\\
UMA-s-1p1$^\dagger$ & OMat24 (task omat) & 0.944 & 0.180 & 0.969\,$\pm$\,0.000 & 0.153\,$\pm$\,0.001 & -- & -- & 12,075\\
\bottomrule
\end{tabular}\end{adjustbox}
\end{minipage}\par\bigskip

\par\noindent\begin{minipage}{\linewidth}\captionsetup{type=table}\centering\small
\caption{On-hull question on the WBM test half: decision heads against energy-level uncertainty propagated through the threshold, the threshold-conditioned head, and ensembles of published relaxed predictions (means $\pm$ s.d. over three seeds; same splits, training subsample and calibration slice for every method). ECE$_\mathrm{near}$: candidates with relaxed $|\hat E_\mathrm{hull}|<50$~meV/atom. $N_{0.1}$ and FDP: candidates certified by conformal selection at $q=0.1$ and their realised false-discovery proportion.}\label{tab:baselines}
\begin{adjustbox}{max width=\linewidth}\begin{tabular}{llcccccccc}
\toprule
backbone & decision rule & cost & AUROC & NLL & Brier & ECE & ECE$_\mathrm{near}$ & $N_{0.1}$ & FDP\\
\midrule
MACE-MP-0 & relaxed $\hat E_\mathrm{hull}$, isotonic map & relaxation & 0.913\,$\pm$\,0.001 & 0.248\,$\pm$\,0.001 & 0.074\,$\pm$\,0.000 & 0.006\,$\pm$\,0.003 & 0.016\,$\pm$\,0.001 & 154 & 0.070\,$\pm$\,0.042\\
 & energy UQ: heteroscedastic Gaussian, $\Phi(-\mu/\sigma)$ & 1 pass & 0.942\,$\pm$\,0.000 & 0.223\,$\pm$\,0.006 & 0.066\,$\pm$\,0.001 & 0.040\,$\pm$\,0.011 & 0.051\,$\pm$\,0.015 & 7,520 & 0.103\,$\pm$\,0.018\\
 & \quad + isotonic map & 1 pass & 0.941\,$\pm$\,0.000 & 0.211\,$\pm$\,0.001 & 0.064\,$\pm$\,0.001 & 0.007\,$\pm$\,0.002 & 0.018\,$\pm$\,0.005 & 6,841 & 0.089\,$\pm$\,0.033\\
 & energy UQ: quantile regression, $\hat F(0)$ & 1 pass & 0.924\,$\pm$\,0.001 & 0.318\,$\pm$\,0.006 & 0.062\,$\pm$\,0.001 & 0.026\,$\pm$\,0.002 & 0.044\,$\pm$\,0.004 & 8,161 & 0.103\,$\pm$\,0.018\\
 & \quad + isotonic map & 1 pass & 0.923\,$\pm$\,0.002 & 0.214\,$\pm$\,0.003 & 0.062\,$\pm$\,0.001 & 0.005\,$\pm$\,0.001 & 0.022\,$\pm$\,0.001 & 7,546 & 0.092\,$\pm$\,0.019\\
 & one head for all thresholds, shared Platt map & 1 pass & 0.948\,$\pm$\,0.001 & 0.198\,$\pm$\,0.002 & 0.060\,$\pm$\,0.001 & 0.010\,$\pm$\,0.003 & -- & 9,033 & 0.108\,$\pm$\,0.012\\
 & \quad + Platt map at the queried threshold & 1 pass & 0.948\,$\pm$\,0.001 & 0.197\,$\pm$\,0.002 & 0.059\,$\pm$\,0.001 & 0.004\,$\pm$\,0.001 & -- & -- & --\\
 & decision head for this question & 1 pass & 0.950\,$\pm$\,0.001 & 0.192\,$\pm$\,0.002 & 0.057\,$\pm$\,0.001 & 0.004\,$\pm$\,0.001 & 0.011\,$\pm$\,0.001 & 9,143 & 0.103\,$\pm$\,0.006\\
\midrule
CHGNet & relaxed $\hat E_\mathrm{hull}$, isotonic map & relaxation & 0.895\,$\pm$\,0.002 & 0.266\,$\pm$\,0.001 & 0.080\,$\pm$\,0.000 & 0.005\,$\pm$\,0.002 & 0.010\,$\pm$\,0.004 & 2,967 & 0.087\,$\pm$\,0.039\\
 & energy UQ: heteroscedastic Gaussian, $\Phi(-\mu/\sigma)$ & 1 pass & 0.935\,$\pm$\,0.000 & 0.235\,$\pm$\,0.009 & 0.070\,$\pm$\,0.002 & 0.044\,$\pm$\,0.014 & 0.055\,$\pm$\,0.018 & 6,571 & 0.094\,$\pm$\,0.011\\
 & \quad + isotonic map & 1 pass & 0.935\,$\pm$\,0.000 & 0.220\,$\pm$\,0.002 & 0.067\,$\pm$\,0.001 & 0.006\,$\pm$\,0.002 & 0.013\,$\pm$\,0.002 & 6,293 & 0.088\,$\pm$\,0.015\\
 & energy UQ: quantile regression, $\hat F(0)$ & 1 pass & 0.919\,$\pm$\,0.001 & 0.324\,$\pm$\,0.002 & 0.065\,$\pm$\,0.001 & 0.025\,$\pm$\,0.002 & 0.038\,$\pm$\,0.003 & 7,639 & 0.103\,$\pm$\,0.014\\
 & \quad + isotonic map & 1 pass & 0.918\,$\pm$\,0.000 & 0.222\,$\pm$\,0.002 & 0.064\,$\pm$\,0.001 & 0.005\,$\pm$\,0.001 & 0.019\,$\pm$\,0.004 & 7,616 & 0.103\,$\pm$\,0.014\\
 & one head for all thresholds, shared Platt map & 1 pass & 0.941\,$\pm$\,0.001 & 0.208\,$\pm$\,0.002 & 0.063\,$\pm$\,0.001 & 0.011\,$\pm$\,0.002 & -- & 7,404 & 0.091\,$\pm$\,0.008\\
 & \quad + Platt map at the queried threshold & 1 pass & 0.941\,$\pm$\,0.001 & 0.207\,$\pm$\,0.002 & 0.063\,$\pm$\,0.001 & 0.004\,$\pm$\,0.001 & -- & -- & --\\
 & decision head for this question & 1 pass & 0.946\,$\pm$\,0.002 & 0.198\,$\pm$\,0.003 & 0.059\,$\pm$\,0.001 & 0.004\,$\pm$\,0.002 & 0.008\,$\pm$\,0.002 & 7,911 & 0.085\,$\pm$\,0.007\\
\midrule
25 MP-trained & error bar: $\Phi(-\mu/\sqrt{\sigma_\mathrm{ens}^2+\mathrm{RMSE}^2})$ & 25 relax. & 0.962\,$\pm$\,0.001 & 0.260\,$\pm$\,0.005 & 0.074\,$\pm$\,0.001 & 0.124\,$\pm$\,0.005 & 0.247\,$\pm$\,0.005 & 12,200 & 0.099\,$\pm$\,0.011\\
potentials & mean $\mu$, isotonic map & 25 relax. & 0.955\,$\pm$\,0.001 & 0.161\,$\pm$\,0.002 & 0.044\,$\pm$\,0.000 & 0.004\,$\pm$\,0.002 & 0.010\,$\pm$\,0.004 & 11,714 & 0.094\,$\pm$\,0.007\\
 & mean and $\log\sigma_\mathrm{ens}$, logistic map & 25 relax. & 0.956\,$\pm$\,0.001 & 0.197\,$\pm$\,0.002 & 0.050\,$\pm$\,0.000 & 0.041\,$\pm$\,0.003 & 0.103\,$\pm$\,0.007 & 11,075 & 0.096\,$\pm$\,0.008\\
 & mean and $\log\sigma_\mathrm{ens}$, boosted map & 25 relax. & 0.977\,$\pm$\,0.000 & 0.135\,$\pm$\,0.001 & 0.040\,$\pm$\,0.000 & 0.004\,$\pm$\,0.001 & 0.010\,$\pm$\,0.002 & 13,662 & 0.109\,$\pm$\,0.016\\
 & best member (EquiformerV3+DeNS-MP), isotonic & relaxation & 0.970\,$\pm$\,0.000 & 0.128\,$\pm$\,0.002 & 0.033\,$\pm$\,0.000 & 0.003\,$\pm$\,0.001 & -- & 14,527 & 0.096\,$\pm$\,0.013\\
 & median member (DPA3-v1-MPtrj), isotonic & relaxation & 0.943\,$\pm$\,0.001 & 0.191\,$\pm$\,0.001 & 0.054\,$\pm$\,0.001 & 0.004\,$\pm$\,0.000 & -- & 7,803 & 0.088\,$\pm$\,0.036\\
\bottomrule
\end{tabular}\end{adjustbox}
\end{minipage}\par\bigskip

\par\noindent\begin{minipage}{\linewidth}\captionsetup{type=table}\centering\small
\caption{On-hull decisions on the WBM test half (means $\pm$ s.d. over seeds). ECE$_\mathrm{near}$: ECE among candidates with $|\hat E_\mathrm{hull}|<50$~meV/atom (relaxed prediction). $N_{0.1}$: candidates certified by conformal selection at FDR $\le0.1$ (not computed for the isotonic map). Unless stated otherwise, probabilities are calibrated on the same random 5\% WBM slice. Cost: one forward pass of the backbone, or a full relaxation (about 30 passes).}\label{tab:onhull}
\begin{adjustbox}{max width=\linewidth}\begin{tabular}{llccccccc}
\toprule
backbone & decision rule & cost & AUROC & NLL & Brier & ECE & ECE$_\mathrm{near}$ & $N_{0.1}$\\
\midrule
MACE-MP-0 & relax + threshold (relaxed $\hat E_\mathrm{hull}$, Platt) & relaxation & 0.912\,$\pm$\,0.001 & 0.278\,$\pm$\,0.001 & 0.079\,$\pm$\,0.000 & 0.043\,$\pm$\,0.003 & 0.086\,$\pm$\,0.003 & 182\\
 & relax + threshold (relaxed $\hat E_\mathrm{hull}$, isotonic) & relaxation & 0.913\,$\pm$\,0.001 & 0.247\,$\pm$\,0.001 & 0.074\,$\pm$\,0.000 & 0.006\,$\pm$\,0.003 & 0.015\,$\pm$\,0.002 & --\\
 & single-point $\hat E_\mathrm{hull}$, Platt & 1 pass & 0.848\,$\pm$\,0.001 & 0.329\,$\pm$\,0.002 & 0.095\,$\pm$\,0.001 & 0.054\,$\pm$\,0.004 & 0.079\,$\pm$\,0.002 & 933\\
 & head trained on MP, calibrated on MP & 1 pass & 0.838\,$\pm$\,0.003 & 0.403\,$\pm$\,0.011 & 0.101\,$\pm$\,0.001 & 0.059\,$\pm$\,0.003 & 0.139\,$\pm$\,0.002 & 0\\
 & head trained on MP, calibrated on WBM slice & 1 pass & 0.838\,$\pm$\,0.003 & 0.319\,$\pm$\,0.004 & 0.096\,$\pm$\,0.001 & 0.010\,$\pm$\,0.001 & 0.076\,$\pm$\,0.003 & 0\\
 & in-domain head, physics scalars only & 1 pass & 0.921\,$\pm$\,0.001 & 0.233\,$\pm$\,0.002 & 0.069\,$\pm$\,0.001 & 0.004\,$\pm$\,0.001 & 0.013\,$\pm$\,0.004 & 5,994\\
 & in-domain linear probe, embeddings + scalars & 1 pass & 0.895\,$\pm$\,0.000 & 0.277\,$\pm$\,0.001 & 0.082\,$\pm$\,0.001 & 0.020\,$\pm$\,0.003 & 0.019\,$\pm$\,0.001 & 2,333\\
 & in-domain head, embeddings + scalars & 1 pass & 0.951\,$\pm$\,0.001 & 0.191\,$\pm$\,0.003 & 0.057\,$\pm$\,0.001 & 0.004\,$\pm$\,0.001 & 0.011\,$\pm$\,0.001 & 9,193\\
 & in-domain head, scalars + relaxed $\hat E_\mathrm{hull}$ & relaxation & 0.937\,$\pm$\,0.001 & 0.214\,$\pm$\,0.002 & 0.064\,$\pm$\,0.001 & 0.004\,$\pm$\,0.001 & 0.006\,$\pm$\,0.003 & 6,641\\
 & in-domain head, embeddings + scalars + relaxed & relaxation & 0.962\,$\pm$\,0.001 & 0.172\,$\pm$\,0.002 & 0.052\,$\pm$\,0.001 & 0.004\,$\pm$\,0.001 & 0.007\,$\pm$\,0.002 & 10,215\\
\midrule
CHGNet & relax + threshold (relaxed $\hat E_\mathrm{hull}$, Platt) & relaxation & 0.895\,$\pm$\,0.002 & 0.284\,$\pm$\,0.001 & 0.081\,$\pm$\,0.000 & 0.030\,$\pm$\,0.004 & 0.043\,$\pm$\,0.004 & 3,021\\
 & relax + threshold (relaxed $\hat E_\mathrm{hull}$, isotonic) & relaxation & 0.895\,$\pm$\,0.002 & 0.266\,$\pm$\,0.001 & 0.080\,$\pm$\,0.000 & 0.005\,$\pm$\,0.002 & 0.011\,$\pm$\,0.004 & --\\
 & single-point $\hat E_\mathrm{hull}$, Platt & 1 pass & 0.839\,$\pm$\,0.000 & 0.336\,$\pm$\,0.002 & 0.097\,$\pm$\,0.001 & 0.053\,$\pm$\,0.003 & 0.069\,$\pm$\,0.002 & 2,267\\
 & head trained on MP, calibrated on MP & 1 pass & 0.821\,$\pm$\,0.001 & 0.418\,$\pm$\,0.006 & 0.110\,$\pm$\,0.001 & 0.066\,$\pm$\,0.002 & 0.147\,$\pm$\,0.002 & 0\\
 & head trained on MP, calibrated on WBM slice & 1 pass & 0.821\,$\pm$\,0.001 & 0.337\,$\pm$\,0.002 & 0.103\,$\pm$\,0.001 & 0.015\,$\pm$\,0.001 & 0.068\,$\pm$\,0.003 & 0\\
 & in-domain head, physics scalars only & 1 pass & 0.903\,$\pm$\,0.000 & 0.255\,$\pm$\,0.002 & 0.076\,$\pm$\,0.001 & 0.004\,$\pm$\,0.001 & 0.013\,$\pm$\,0.001 & 4,418\\
 & in-domain linear probe, embeddings + scalars & 1 pass & 0.880\,$\pm$\,0.000 & 0.292\,$\pm$\,0.001 & 0.086\,$\pm$\,0.000 & 0.023\,$\pm$\,0.002 & 0.019\,$\pm$\,0.002 & 2,332\\
 & in-domain head, embeddings + scalars & 1 pass & 0.947\,$\pm$\,0.002 & 0.196\,$\pm$\,0.003 & 0.058\,$\pm$\,0.001 & 0.004\,$\pm$\,0.002 & 0.008\,$\pm$\,0.003 & 8,432\\
 & in-domain head, scalars + relaxed $\hat E_\mathrm{hull}$ & relaxation & 0.919\,$\pm$\,0.000 & 0.238\,$\pm$\,0.002 & 0.071\,$\pm$\,0.001 & 0.004\,$\pm$\,0.001 & 0.007\,$\pm$\,0.003 & 4,748\\
 & in-domain head, embeddings + scalars + relaxed & relaxation & 0.957\,$\pm$\,0.001 & 0.180\,$\pm$\,0.003 & 0.054\,$\pm$\,0.001 & 0.003\,$\pm$\,0.001 & 0.011\,$\pm$\,0.001 & 9,537\\
\midrule
both & fusion: embeddings, scalars, relaxed $\hat E_\mathrm{hull}$ of both, disagreement & relaxation & 0.967\,$\pm$\,0.000 & 0.161\,$\pm$\,0.001 & 0.048\,$\pm$\,0.000 & 0.004\,$\pm$\,0.001 & 0.008\,$\pm$\,0.003 & 10,769\\
\bottomrule
\end{tabular}\end{adjustbox}
\end{minipage}\par\bigskip

\end{landscape}

\begin{table}[htbp]
\centering\footnotesize\setlength{\tabcolsep}{3pt}
\caption{\textbf{Probabilistic baselines on the single-pass state.} WBM test splits (three seeds; same state, labels, splits and calibration slice for every model). Top: log-loss of the stability question; middle: AUROC; bottom: log-loss minus that of the decision head ($10^{-3}$; $\pm$ half-width of the 95\% interval from a bootstrap over chemical systems pooled over seeds) for the recalibrated baselines (raw: without recalibration). Regressions of the hull distance give $P(\text{stable})=\Phi(-\mu/\sigma)$ (deep ensemble: mean over members); the conformal predictive system gives its predictive distribution at zero. The combined reader averages the logits of the head and of the recalibrated deep regression ensemble and is Platt-recalibrated on the slice (designed after the other results).}
\label{tab:mlbaselines}
\begin{adjustbox}{max width=\textwidth}
\begin{tabular}{lcccccccc}
\toprule
 & MACE-MP-0 & CHGNet & MACE-MPA-0 & MatterSim & ORB-v3 & SevenNet & eSEN & UMA\\
\midrule
\multicolumn{9}{l}{\emph{Log-loss}}\\
Decision head (boosted trees) & 0.190 & 0.196 & 0.138 & 0.130 & 0.117 & 0.128 & 0.112 & 0.152\\
Deep ens.\ of classifiers & 0.203 & 0.196 & 0.158 & 0.157 & 0.157 & 0.138 & 0.121 & 0.158\\
\quad raw & 0.203 & 0.198 & 0.158 & 0.157 & 0.157 & 0.138 & 0.121 & 0.159\\
Deep ens.\ of regressors & 0.188 & 0.190 & 0.145 & 0.132 & 0.127 & 0.142 & 0.118 & 0.159\\
\quad raw & 0.193 & 0.193 & 0.151 & 0.139 & 0.137 & 0.150 & 0.128 & 0.166\\
Conformal pred.\ system & 0.204 & 0.212 & 0.148 & 0.139 & 0.128 & 0.139 & 0.124 & 0.169\\
\quad with Platt map & 0.204 & 0.211 & 0.148 & 0.139 & 0.128 & 0.139 & 0.124 & 0.169\\
NGBoost & 0.231 & 0.244 & 0.169 & 0.161 & 0.153 & 0.158 & 0.148 & 0.210\\
\quad raw & 0.232 & 0.245 & 0.171 & 0.164 & 0.155 & 0.161 & 0.151 & 0.213\\
Gaussian process & 0.324 & 0.286 & 0.224 & 0.230 & 0.219 & 0.372 & 0.281 & 0.260\\
\quad raw & 0.338 & 0.298 & 0.245 & 0.248 & 0.237 & 0.386 & 0.299 & 0.275\\
Head + deep regressors (post hoc) & 0.178 & 0.181 & 0.131 & 0.120 & 0.109 & 0.125 & 0.104 & 0.142\\
\addlinespace
\multicolumn{9}{l}{\emph{AUROC}}\\
Decision head (boosted trees) & 0.951 & 0.947 & 0.974 & 0.977 & 0.980 & 0.976 & 0.982 & 0.969\\
Deep ens.\ of classifiers & 0.944 & 0.947 & 0.967 & 0.968 & 0.967 & 0.974 & 0.980 & 0.968\\
Deep ens.\ of regressors & 0.953 & 0.952 & 0.973 & 0.978 & 0.979 & 0.973 & 0.982 & 0.967\\
Conformal pred.\ system & 0.944 & 0.939 & 0.970 & 0.974 & 0.978 & 0.973 & 0.979 & 0.962\\
NGBoost & 0.928 & 0.917 & 0.961 & 0.965 & 0.968 & 0.966 & 0.970 & 0.942\\
Gaussian process & 0.767 & 0.884 & 0.935 & 0.933 & 0.940 & 0.640 & 0.799 & 0.915\\
Head + deep regressors (post hoc) & 0.958 & 0.956 & 0.977 & 0.981 & 0.984 & 0.978 & 0.985 & 0.973\\
\addlinespace
\multicolumn{9}{l}{\emph{Log-loss minus the decision head's ($10^{-3}$), recalibrated}}\\
Deep ens.\ of classifiers & 12.8$\pm$1.4 & 0.4$\pm$1.3 & 19.5$\pm$1.4 & 26.9$\pm$1.6 & 40.1$\pm$1.7 & 9.8$\pm$1.2 & 9.1$\pm$1.3 & 5.9$\pm$1.4\\
Deep ens.\ of regressors & $-$1.7$\pm$1.3 & $-$5.9$\pm$1.5 & 6.5$\pm$1.3 & 2.3$\pm$1.5 & 9.9$\pm$1.6 & 13.8$\pm$1.4 & 6.2$\pm$1.5 & 7.5$\pm$1.6\\
Conformal pred.\ system & 14.3$\pm$1.2 & 15.7$\pm$1.3 & 10.0$\pm$1.1 & 9.5$\pm$1.1 & 10.6$\pm$1.1 & 10.6$\pm$1.1 & 12.3$\pm$1.1 & 17.6$\pm$1.4\\
NGBoost & 41.7$\pm$1.5 & 47.9$\pm$1.5 & 30.9$\pm$1.4 & 31.3$\pm$1.4 & 35.7$\pm$1.4 & 29.6$\pm$1.3 & 36.7$\pm$1.5 & 57.9$\pm$1.8\\
Gaussian process & 134.3$\pm$2.6 & 90.2$\pm$2.5 & 85.4$\pm$2.3 & 99.7$\pm$2.5 & 101.6$\pm$2.8 & 243.3$\pm$3.5 & 169.1$\pm$2.6 & 108.1$\pm$2.8\\
Head + deep regressors (post hoc) & $-$11.8$\pm$0.7 & $-$14.9$\pm$0.8 & $-$7.1$\pm$0.7 & $-$10.3$\pm$0.8 & $-$8.1$\pm$0.8 & $-$3.4$\pm$0.8 & $-$8.1$\pm$0.8 & $-$9.4$\pm$0.9\\
\bottomrule
\end{tabular}
\end{adjustbox}
\end{table}

\snote{Value of information, escalation and DFT allocation}{sn:voi}

\paragraph{One number per potential.} If System~1's predictive distribution of the hull distance is Gaussian and System~2 observes the hull distance with independent Gaussian noise, the value of information has a closed form that depends on $p_1$, $c$ and a single \emph{information ratio} $r$, the precision of System~2 relative to the uncertainty of System~1 (Methods, Proposition~2; verified by Monte Carlo). Measuring the value of escalation from System~1 (the single-pass head) to System~2 (the head with the backbone's relaxed prediction) on the test half, with $p_2$ calibrated conditionally on $p_1$ on the calibration slice (Methods), the one-parameter form describes the measured curves of all seven potentials (Fig.~\ref{fig:voi}b; $R^2$ of the binned curves 0.89--1.00 across potentials and attempt costs), with $r$ at $c=0.5$ from 0.20 for CHGNet and 0.25 for MACE-MP-0 to 2.5 for ORB-v3, 2.5 for SevenNet-MF-ompa and 4.1 for eSEN-30M-OAM; across attempt costs from 0.1 to 0.7 the fitted $r$ of a potential varies at most 2.0-fold (Extended Data Fig.~\ref{fig:voireal}c). We fit $r$ by least squares to the gains of the individual candidates, whose expectation given $p_1$ is the value of information; fitting the closed form at the medians of wide bins of $p_1$ instead is biased where the curve is steep and gave values up to a factor of two lower. Newer potentials thus make System~2 relatively more informative, consistent with their relaxed predictions improving faster than their single passes. In information terms, System~1 captures 55\%--74\% of the 0.43~nats of label entropy and relaxation adds 0.018--0.053~nats; 46\%--78\% of that addition comes from the 20\% of candidates in the two deciles of $p_1$ below the most confident one (median $p_1$ 0.08 and 0.40), and only 2\%--12\% from the most confident decile.

\paragraph{Escalate by value, not by probability.} Under a budget of relaxations, the expected net value is maximised by relaxing the candidates with the largest value of information (Methods). Estimated on the calibration slice --- which needs System-2 outputs there, but no labels beyond those that calibrate $p_2$ --- this order recovers 78\% of the gain of relaxing every candidate when 10\% of candidates are relaxed and 93\% at 20\% (means over seven potentials and three seeds at $c=0.5$; Fig.~\ref{fig:voi}c). The order a screening funnel uses --- the most probable candidates first --- recovers 1\% and 77\% at the same budgets, because the candidates it relaxes first would be attempted anyway; random selection recovers 10\%. Ranking by distance from the decision boundary does almost as well as the value of information when $c=0.5$ (76\% at 10\%), but not when costs are asymmetric (65\% against 84\% at $c=0.7$). Where a guarantee is preferred to an optimum, the conformal screen of Fig.~\ref{fig:backbones}b discards only candidates below a threshold set on the stable structures of the slice: at a 1\% recall budget it relaxes 40\%--58\% of the candidates, keeps at least 0.988 of the stable ones and retains at least 99.7\% of the decision value at $c=0.5$ (98\%--100\% of the gain that relaxation adds), saving 41\%--58\% of the cost of relaxing every candidate.

\paragraph{Which order, and what more than $p_1$ adds.} The estimated value of information above is a function of $p_1$, so it is one of several orders that target the decision boundary. At a 10\% budget and $c=0.5$, the parameter-free perfect-information bound of Eq.~\ref{eq:oracle}, the distances $|p_1-c|$ and $|\mathrm{logit}\,p_1-\mathrm{logit}\,c|$ and the estimate recover 76\%--78\% of the gain of relaxing every candidate (means over seven potentials and three seeds), and 48\%--56\% at $c=0.1$, where the boundary sits among few stable candidates; the funnel recovers at most 4\% at $c=0.5$ (Supplementary Table~\ref{tab:voirebuild}). At $c=0.5$, targeting the boundary matters more than the estimator (at $c=0.7$ the estimate does better; above). Using more of the single pass does not help either: gradient-boosted estimates of each candidate's value of information from the backbone's forces, stresses, energies and 16 principal components of its embedding as well as $p_1$, cross-fitted over two halves of the test set grouped by chemical system, recover 78\% at $c=0.5$ against 77\% for the same regression on $p_1$ alone and 78\% for the kernel estimate (seven potentials; at $c=0.1$, 48\%, 49\% and 54\%): the calibrated single-pass probability, which is itself a function of the same state, already carries what matters for escalation. A DFT option changes the order (Fig.~\ref{fig:voi}d). If DFT at a twentieth of a hit can follow, a relaxation is worth resolving a candidate into an attempt, a calculation or a discard, $\mathbb{E}[V(p_2)\mid p_1]-V(p_1)$ with $V(p)=\max(0,p-c,p(1-c)-k)$; ordering by this estimate recovers 43\%--52\% of the relax-all gain at a 10\% budget for $c=0.3$--0.7, against 19\%--28\% for the myopic orders, which spend relaxations on candidates that DFT would settle anyway, and 26\%--33\% for the DFT-aware perfect-information bound $\min\{p_1(1-c),(1-p_1)c,k\}$.

\paragraph{An information ratio without labels.} Under the Gaussian refinement, $\Phi^{-1}(p_2)$ given $p_1$ is normal with mean $\sqrt{1+r}\,\Phi^{-1}(p_1)$ and variance $r$, so the spread of the relaxed probabilities around the single-pass ones measures $r$ without any label. Near $p_1=0.5$ this spread gives $r=0.26$ (CHGNet), 0.29 (MACE-MP-0), 0.79 (MACE-MPA-0), 0.9 (MatterSim-v1-5M), 1.5 (SevenNet-MF-ompa), 1.6 (ORB-v3) and 2.1 (eSEN-30M-OAM), in nearly the same order as the values fitted to the model-implied gains (0.20, 0.25, 0.84, 0.99, 2.5, 2.5 and 4.1 at $c=0.5$) but smaller for the newer potentials, whose relaxed probabilities have heavier tails than the model allows. With the label-free $r$, the closed form predicts the realised value of relaxing every candidate at $c=0.5$ to within a factor of 1.4 (realised over predicted 0.82--1.36): the model is a useful description, not an exact one.

\paragraph{When to run DFT.} The bound of Eq.~\ref{eq:oracle} is attained by DFT, which reveals the label. If an attempt (a synthesis, say) costs $c$ and a DFT check costs $k$, a calibrated probability should be discarded below $k/(1-c)$, attempted directly above $1-k/c$ and checked by DFT in between (Methods). On the WBM test half with $c=0.5$ and $k=0.05$ (Fig.~\ref{fig:voi}e), this band applied to the single-pass probability realises 58--66 net hits per 1,000 candidates for the seven potentials and applied to the probability after relaxation 60--72, sending 5\%--23\% of the candidates to DFT and recovering 93\%--98\% of the stable ones. The usual practice --- DFT for every candidate that the relaxed prediction calls stable --- realises 47--64 while sending 14\%--23\% of the candidates to DFT, and computing every candidate realises 27. The single pass alone beats the threshold practice for seven of the seven potentials.

\paragraph{Estimating the value of information.} Supplementary Table~\ref{tab:voi} reports the share of the gain of relaxing every candidate retained by each escalation order at four attempt costs, and Supplementary Table~\ref{tab:dftalloc} the realised value of allocating DFT with the three-way band. Two practical points matter for the estimates. First, the martingale property $\mathbb{E}[p_2\mid p_1]=p_1$ that makes the value of information non-negative holds only if $p_2$ is calibrated conditionally on $p_1$. With separately Platt-calibrated $p_1$ and $p_2$, $p_2$ was slightly under-confident relative to $p_1$ among candidates with $p_1$ between 0.6 and 0.8 (CHGNet, first seed: mean $p_2$ 0.62 and 0.71 in the two bins against realised stable fractions of 0.64 and 0.76), which made the measured value of escalation negative there; recalibrating $p_2$ by a logistic regression on $(\mathrm{logit}\,p_1,\mathrm{logit}\,p_2)$ on the calibration slice removes this and also lowers its log-loss. Second, ranking candidates by the value of information requires System-2 outputs on the calibration slice only, which is affordable because the slice is small and must be labelled anyway.

\paragraph{The value of escalation against real outcomes.} The measured curves of Fig.~\ref{fig:voi}b average the model-implied gain $u(p_2)-u(p_1)$, which is the value of escalation only if $p_2$ is calibrated. The realised gain $v=(y-c)(\mathbf{1}[p_2\ge c]-\mathbf{1}[p_1\ge c])$ needs no such assumption: it is the change in net value that the slower system's answer produced on the candidate's actual label, and $\mathbb{E}[v\mid S_1]=\mathrm{VOI}(p_1)$ when $p_2$ is calibrated. Averaged over equal-mass bins of $p_1$, realised and model-implied values agree for all seven potentials (Extended Data Fig.~\ref{fig:voireal}a); the realised value of escalating every candidate is a median 0.98 times the model-implied one (10--90\% range 0.85--1.14, full range 0.71--1.45; 53 of 84 potential--seed--cost combinations inside the bootstrap 95\% interval of the realised total), and the information ratio fitted to the realised gains is a median 0.89 times that fitted to the model-implied ones (Extended Data Fig.~\ref{fig:voireal}c): the model-implied value slightly overstates what relaxation delivers, most for SevenNet-MF-ompa and eSEN-30M-OAM (2.20 against 2.52 and 3.40 against 4.06 at $c=0.5$). On the prospective random samples the realised totals are small and noisy, and below the model-implied ones for four of five backbones (Extended Data Fig.~\ref{fig:voireal}d; Supplementary Note~\ref{sn:prospective}).

\begin{table}[t]\centering\small
\caption{Budgeted escalation from the single pass (System~1) to the head with the relaxed prediction (System~2): share of the realised gain of relaxing every candidate that is retained when a fraction $b$ of the candidates is relaxed, by escalation order and attempt cost $c$ (WBM test half; means over the backbones of Supplementary Table~\ref{tab:dftalloc} and three seeds; $p_2$ calibrated conditionally on $p_1$). Last column: the conformal screen at a 1\% recall budget (share relaxed; share retained).}\label{tab:voi}
\begin{tabular}{llccccc}
\toprule
$c$ & order & $b=5\%$ & $10\%$ & $20\%$ & $30\%$ & screen ($\alpha=1\%$)\\
\midrule
0.1 & value of information & 32\% & 54\% & 79\% & 90\% & 47\%; 92\%\\
 & closest to $c$ & 33\% & 55\% & 76\% & 86\% & \\
 & most probable first & 0\% & 0\% & 12\% & 61\% & \\
 & random & 5\% & 10\% & 20\% & 30\% & \\
\addlinespace
0.3 & value of information & 47\% & 70\% & 89\% & 95\% & 47\%; 97\%\\
 & closest to $c$ & 48\% & 70\% & 89\% & 95\% & \\
 & most probable first & 0\% & 0\% & 58\% & 88\% & \\
 & random & 4\% & 10\% & 20\% & 30\% & \\
\addlinespace
0.5 & value of information & 55\% & 78\% & 93\% & 97\% & 47\%; 99\%\\
 & closest to $c$ & 53\% & 76\% & 92\% & 97\% & \\
 & most probable first & 0\% & 1\% & 77\% & 94\% & \\
 & random & 4\% & 10\% & 20\% & 29\% & \\
\addlinespace
0.7 & value of information & 63\% & 84\% & 96\% & 99\% & 47\%; 100\%\\
 & closest to $c$ & 55\% & 65\% & 88\% & 98\% & \\
 & most probable first & 0\% & 17\% & 88\% & 98\% & \\
 & random & 4\% & 10\% & 20\% & 30\% & \\
\bottomrule
\end{tabular}
\end{table}

\begin{table}[t]\centering\small
\caption{Allocating DFT with calibrated probabilities (WBM test half, means over three seeds): realised net value per 1,000 candidates when an attempt costs $c$ and a DFT check $k$ (units of a hit), for computing every candidate, computing those whose relaxed hull distance is $\le0$, and the three-way band (discard / DFT / attempt directly) on the single-pass probability $p_1$ and on the probability after relaxation $p_2$; in brackets the share of candidates sent to DFT. $r$: information ratio of the one-parameter fit to the measured value of escalation ($c=0.5$).}\label{tab:dftalloc}
\begin{adjustbox}{max width=\textwidth}\begin{tabular}{lcccccccc}
\toprule
 & & \multicolumn{3}{c}{$c=0.5$, $k=0.05$} & \multicolumn{2}{c}{$c=0.5$, $k=0.01$} & \multicolumn{2}{c}{$c=0.7$, $k=0.05$}\\
\cmidrule(lr){3-5}\cmidrule(lr){6-7}\cmidrule(lr){8-9}
backbone & $r$ & threshold & band $p_1$ & band $p_2$ & threshold & band $p_2$ & threshold & band $p_2$\\
\midrule
MACE-MP-0 & 0.25 & 51 (21\%) & 58 (24\%) & 61 (22\%) & 59 (21\%) & 72 (40\%) & 26 (21\%) & 32 (18\%)\\
CHGNet & 0.2 & 47 (23\%) & 58 (25\%) & 60 (23\%) & 56 (23\%) & 72 (43\%) & 24 (23\%) & 31 (19\%)\\
MACE-MPA-0 & 0.84 & 58 (15\%) & 64 (15\%) & 67 (11\%) & 64 (15\%) & 74 (23\%) & 32 (15\%) & 37 (10\%)\\
MatterSim-v1-5M & 0.99 & 57 (14\%) & 64 (14\%) & 68 (10\%) & 63 (14\%) & 74 (21\%) & 31 (14\%) & 38 (9\%)\\
ORB-v3 & 2.5 & 62 (15\%) & 65 (12\%) & 70 (7\%) & 68 (15\%) & 75 (15\%) & 34 (15\%) & 40 (7\%)\\
SevenNet-MF-ompa & 2.5 & 62 (16\%) & 64 (14\%) & 70 (8\%) & 69 (16\%) & 74 (17\%) & 34 (16\%) & 39 (7\%)\\
eSEN-30M-OAM & 4.1 & 64 (15\%) & 66 (12\%) & 72 (5\%) & 70 (15\%) & 75 (12\%) & 35 (15\%) & 41 (5\%)\\
\midrule
DFT every candidate &  & 27 (100\%) &  &  & 67 (100\%) &  & -4 (100\%) & \\
\bottomrule
\end{tabular}\end{adjustbox}
\end{table}

\begin{table}[htbp]
\centering\footnotesize\setlength{\tabcolsep}{3pt}
\caption{\textbf{Escalation orders, a DFT option and an information ratio without labels} (WBM test sets, three seeds, $c=0.5$). Share (\%) of the gain of relaxing every candidate retained when 10\% are relaxed: in order of decreasing $p_1$ (funnel), of the perfect-information bound, of $|p_1-c|$, of the estimated value of information and of a learned estimate that also uses the forces, stresses and embedding of the pass; with a DFT option at $k=0.05$, of the myopic estimate and of the estimate that looks ahead to the calculation. Information ratio $r$ from the spread of $p_2$ near $p_1=0.5$ (no labels) and fitted to the model-implied gains (mean of the fits per seed; the pooled fits of Fig.~\ref{fig:voi}b differ by at most 0.2); realised value of relaxing every candidate and the value predicted by the closed form with the label-free $r$ (net hits per 1,000 candidates).}
\label{tab:voirebuild}
\begin{tabular}{lrrrrrrrrrrr}
\toprule
Backbone & Funnel & Bound & $|p_1-c|$ & VOI & Learned & Myopic & \shortstack{Look-\\ahead} & \shortstack{$r$\\(no labels)} & \shortstack{$r$\\(fitted)} & Realised & Predicted\\
\midrule
MACE-MP-0 & 3 & 75 & 75 & 77 & 78 & 3 & 42 & 0.29 & 0.27 & 3.9 & 3.9\\
CHGNet & 4 & 78 & 78 & 79 & 81 & 1 & 44 & 0.26 & 0.2 & 3.0 & 3.6\\
MACE-MPA-0 & $-$1 & 74 & 74 & 78 & 78 & 16 & 44 & 0.79 & 0.87 & 6.7 & 5.4\\
MatterSim-v1-5M & 0 & 77 & 77 & 79 & 79 & 23 & 46 & 0.9 & 1 & 6.6 & 5.4\\
ORB-v3 & 0 & 77 & 77 & 79 & 80 & 35 & 50 & 1.6 & 2.5 & 9.0 & 6.7\\
SevenNet-MF-ompa & 0 & 73 & 73 & 75 & 75 & 29 & 45 & 1.5 & 2.7 & 10.0 & 7.4\\
eSEN-30M-OAM & 0 & 77 & 77 & 78 & 78 & 40 & 48 & 2.1 & 3.9 & 10.3 & 7.7\\
\bottomrule
\end{tabular}
\end{table}

\snote{Typed questions beyond stability and beyond crystals}{sn:typed}

\paragraph{Ordinal, prototype and choice questions.} A System-1 layer is useful in proportion to the questions it can answer from the same state. On WBM, the single pass answers the five-level hull-band Score question better than an ordinal model on the relaxed prediction for six of the six backbones, predicts whether DFT relaxation will preserve the input prototype --- whether the input geometry can be trusted at all --- with AUROC 0.86 (MACE-MP-0), and chooses the lowest-energy polymorph among candidates of one composition, the one question on which relaxation clearly wins (0.67 against 0.73 top-1 accuracy for MACE-MP-0; Supplementary Tables~\ref{tab:typed} and~\ref{tab:backbones_typed}).

\paragraph{Electronic structure from a potential's single pass.} The backbones were trained on energies, forces and stresses (and, for CHGNet, magnetic moments), yet their single pass on the unrelaxed input carries information about the electronic structure of the relaxed crystal (Extended Data Fig.~\ref{fig:typed}a; Supplementary Table~\ref{tab:breadth}). Whether DFT will find a metal is answered with AUROC 0.977--0.989 across the six backbones (composition alone: 0.951), whether the gap falls in the 1--2~eV window of single-junction photovoltaic absorbers (3.4\% of candidates) with AUROC 0.971--0.984 (composition: 0.951), and the five-level band-gap Score with a ranked probability score of 0.013--0.019 (composition alone 0.032).

\paragraph{Compound questions should be calibrated as compounds.} Discovery campaigns ask conjunctions: stable \emph{and} in the photovoltaic window, stable \emph{and} non-metallic. Multiplying two calibrated marginal answers assumes the two events are conditionally independent given the state, and they are not (Extended Data Fig.~\ref{fig:typed}b): the product of the marginals under-forecasts the number of stable non-metals by 3\%--13\% and of stable photovoltaic absorbers by up to 14\% (most for the MP-trained backbones) --- the signature of events that are positively associated given what the layer knows (Methods) --- whereas a one-dimensional recalibration of the product on the slice (ECE 0.001 for every backbone, against 0.001--0.005 for ``stable and non-metallic'') or a head trained on the compound question itself (0.001--0.002) is calibrated. The recalibrated product has the lowest log-loss for both compound questions (0.018--0.024 for ``stable and photovoltaic'' against 0.023--0.029 for the dedicated head), because the marginal heads are each trained on more positives. A typed interface makes the compound explicit, so its calibration can be checked rather than assumed.

\paragraph{Beyond the hull.} We posed 12 screening questions on seven Matbench property datasets\cite{dunn2020matbench} --- metallicity, a photovoltaic window and a wide gap from PBE band gaps; stiff, soft, hard and compliant crystals from elastic moduli; high refractive index; soft and stiff lattices from the highest phonon peak; easy exfoliation of layered crystals; and low formation energy of perovskites --- and answered them from one MatterSim or ORB-v3 pass with five-fold cross-validation (Extended Data Fig.~\ref{fig:typed}c,d; Methods). Deciding directly gives a lower log-loss than both regress-then-threshold baselines on the same features (a regression with a calibrated threshold, and a regression with its residual distribution propagated through the threshold) for 20 of the 24 question--backbone pairs, by more than 0.005~nats for 16 of them, and significantly for 13 (paired $t$-test over the five folds, $P<0.05$; 9 after Holm correction for the 24 pairs; only fold-level results were stored, so folds are the finest pairing); the better baseline's log-loss is 1.15 times the decision head's (median). Regression does better for a low perovskite formation energy (significantly for both backbones, one after correction), a smooth target on which it uses the full label, and in log-loss for stiff phonon spectra, where a few confident errors in a small dataset dominate the decision head's log-loss while its Brier score is lower (0.022--0.024 against 0.030--0.031). The largest differences arise where the question sits at an edge of the property's distribution or where the target is heavy-tailed --- where the noise of a regression changes shape from one input to another, which no monotone map of its output can absorb (Methods): whether a material is a metal (a gap of exactly zero) is answered with log-loss 0.162--0.190 against 0.262--0.295 for the calibrated regression, whose propagated-residual variant is badly miscalibrated (ECE 0.22 against 0.007), and a high refractive index with 0.163 against 0.30--0.34 for the regressions and 0.25 for composition alone. Conformal selection certifies 766--1,089 candidates per test fold for the photovoltaic window at a nominal 10\% false-discovery rate from the decision head (realised 0.11--0.12), against at most 31 from either regression. Composition alone reaches AUROC 0.70--0.98 against 0.91--0.997 for the single pass; exfoliability, a question about how layers bind, is where structure matters most (0.91--0.93 against 0.70).

\paragraph{Beyond crystals.} Nothing in the layer is specific to crystals: it needs the state of a frozen atomistic model, labels and a question. On 130,831 QM9 molecules\cite{ramakrishnan2014qm9} --- whose geometries, unlike the crystal inputs, are already relaxed at the level of theory of the labels, so that these questions test reading the state rather than escalation --- one pass of a molecular foundation potential (MACE-OFF23\cite{kovacs2025maceoff}) or of the crystal-trained ORB-v3 on the molecule in a box answers whether the B3LYP HOMO--LUMO gap is wide ($\geq8$~eV) or small ($\leq5$~eV) and whether the dipole moment exceeds 4~D with AUROC 0.970--0.998 (composition alone: 0.766--0.875), and deciding directly beats both regress-then-threshold baselines in log-loss for all six question--backbone pairs (the better baseline's log-loss is 1.08--1.39 times the decision head's; Extended Data Fig.~\ref{fig:typed}e). A Choice question --- which of five constitutional isomers has the lowest energy $U_0$ --- shows how a typed layer builds on a backbone's own physics rather than replacing it. The backbone's energy alone picks the lowest-energy isomer in 90.2\% (MACE-OFF23) and 85.0\% (ORB-v3) of the test sets; neither model was trained at the level of theory of the labels. A Choice head --- a softmax over the backbone energy plus a learned correction, with its temperature fitted on calibration sets --- raises this to 98.0\% and 96.8\% and states how sure it is (log-loss 0.05 and 0.08 against 1.61 for a uniform guess; Extended Data Fig.~\ref{fig:typed}f). That the crystal-trained potential, whose training data contain the same elements but no molecules, serves almost as well as the molecular one suggests that the layer does not need a backbone trained on the system class it is asked about.

\paragraph{Datasets and thresholds.} Matbench property questions\cite{dunn2020matbench} use the datasets' structures as inputs (one MatterSim-v1-5M or ORB-v3 pass) and thresholds on their targets: PBE gap $=0$, $1\le$ gap $\le2$~eV, gap $\ge3$~eV (mp\_gap, 106,113 structures); $K_\mathrm{VRH}\ge150$ or $\le30$~GPa and $G_\mathrm{VRH}\ge100$ or $\le15$~GPa (10,987); refractive index $\ge2.5$ (4,764); highest phonon density-of-states peak $\le300$ or $\ge800$~cm$^{-1}$ (1,265); exfoliation energy $\le100$~meV/atom (636); perovskite formation energy $\le0.5$~eV per cell (18,928). Five-fold cross-validation holds out 10\% of each training fold as the calibration slice.

\begin{landscape}
\par\noindent\begin{minipage}{\linewidth}\captionsetup{type=table}\centering\small
\caption{Ordinal, choice and prototype questions for MACE-MP-0 and CHGNet (WBM test set, means $\pm$ s.d. over three seeds). Ordinal: five-level hull-distance band (ranked probability score, RPS; NLL; accuracy). Choice: lowest-energy polymorph among WBM candidates of one composition (top-1 accuracy; NLL of the true choice; top-choice ECE). Prototype kept: DFT relaxation keeps the input prototype (AUROC; NLL).}\label{tab:typed}
\begin{adjustbox}{max width=\linewidth}\begin{tabular}{lllccc}
\toprule
question & backbone & decision rule & \multicolumn{3}{c}{metrics}\\
\midrule
Ordinal & MACE-MP-0 & ordinal model on relaxed $\hat E_\mathrm{hull}$ & RPS 0.108\,$\pm$\,0.000 & NLL 1.00\,$\pm$\,0.00 & acc. 0.60\,$\pm$\,0.00\\
Ordinal & MACE-MP-0 & ordinal head (1 pass) & RPS 0.077\,$\pm$\,0.000 & NLL 0.82\,$\pm$\,0.00 & acc. 0.69\,$\pm$\,0.00\\
Ordinal & MACE-MP-0 & ordinal head + relaxed & RPS 0.069\,$\pm$\,0.000 & NLL 0.75\,$\pm$\,0.00 & acc. 0.71\,$\pm$\,0.00\\
Ordinal & CHGNet & ordinal model on relaxed $\hat E_\mathrm{hull}$ & RPS 0.115\,$\pm$\,0.000 & NLL 1.04\,$\pm$\,0.00 & acc. 0.58\,$\pm$\,0.00\\
Ordinal & CHGNet & ordinal head (1 pass) & RPS 0.079\,$\pm$\,0.000 & NLL 0.85\,$\pm$\,0.00 & acc. 0.68\,$\pm$\,0.00\\
Ordinal & CHGNet & ordinal head + relaxed & RPS 0.072\,$\pm$\,0.000 & NLL 0.78\,$\pm$\,0.00 & acc. 0.70\,$\pm$\,0.00\\
\midrule
Choice & MACE-MP-0 & relaxed $\hat E_\mathrm{hull}$, softmax & acc. 0.727\,$\pm$\,0.001 & NLL 0.685\,$\pm$\,0.003 & ECE 0.096\,$\pm$\,0.011\\
Choice & MACE-MP-0 & single-point $\hat E_\mathrm{hull}$, softmax & acc. 0.609\,$\pm$\,0.001 & NLL 0.826\,$\pm$\,0.003 & ECE 0.086\,$\pm$\,0.001\\
Choice & MACE-MP-0 & single point + learned residual (1 pass) & acc. 0.671\,$\pm$\,0.003 & NLL 0.716\,$\pm$\,0.003 & ECE 0.045\,$\pm$\,0.007\\
Choice & MACE-MP-0 & relaxed + learned residual & acc. 0.700\,$\pm$\,0.001 & NLL 0.671\,$\pm$\,0.004 & ECE 0.054\,$\pm$\,0.008\\
Choice & CHGNet & relaxed $\hat E_\mathrm{hull}$, softmax & acc. 0.709\,$\pm$\,0.003 & NLL 0.700\,$\pm$\,0.006 & ECE 0.081\,$\pm$\,0.011\\
Choice & CHGNet & single-point $\hat E_\mathrm{hull}$, softmax & acc. 0.602\,$\pm$\,0.003 & NLL 0.833\,$\pm$\,0.002 & ECE 0.086\,$\pm$\,0.002\\
Choice & CHGNet & single point + learned residual (1 pass) & acc. 0.661\,$\pm$\,0.002 & NLL 0.720\,$\pm$\,0.002 & ECE 0.033\,$\pm$\,0.007\\
Choice & CHGNet & relaxed + learned residual & acc. 0.694\,$\pm$\,0.001 & NLL 0.676\,$\pm$\,0.004 & ECE 0.039\,$\pm$\,0.009\\
\midrule
Prototype kept & MACE-MP-0 & logistic on force/stress magnitudes & AUROC 0.615\,$\pm$\,0.002 & NLL 0.305\,$\pm$\,0.001 & \\
Prototype kept & MACE-MP-0 & head on embeddings + scalars & AUROC 0.862\,$\pm$\,0.004 & NLL 0.224\,$\pm$\,0.001 & \\
Prototype kept & CHGNet & logistic on force/stress magnitudes & AUROC 0.622\,$\pm$\,0.001 & NLL 0.304\,$\pm$\,0.001 & \\
Prototype kept & CHGNet & head on embeddings + scalars & AUROC 0.877\,$\pm$\,0.004 & NLL 0.214\,$\pm$\,0.002 & \\
\bottomrule
\end{tabular}\end{adjustbox}
\end{minipage}\par\bigskip

\par\noindent\begin{minipage}{\linewidth}\captionsetup{type=table}\centering\small
\caption{Typed questions and the shift ladder across backbones (WBM test set, means over three seeds). Ordinal: ranked probability score of the five-level hull band (ordinal model on the relaxed prediction; single-pass ordinal head; head with the relaxed prediction). Choice: top-1 accuracy of the lowest-energy polymorph (relaxed energies; single pass with a learned residual; relaxed energies with a learned residual). Prototype kept: AUROC of ``relaxation keeps the prototype'' (logistic model on force and stress magnitudes; head). Shift: boosted head trained on MP and evaluated on WBM rounds 2--5, calibrated on MP or on the WBM slice (coverage of 90\% sets, near-hull ECE), and realised FDR of conformal selection at $q=0.1$ with the linear MP head.}\label{tab:backbones_typed}
\begin{adjustbox}{max width=\linewidth}\begin{tabular}{lccccccccccccc}
\toprule
 & \multicolumn{3}{c}{Ordinal (RPS)} & \multicolumn{3}{c}{Choice (top-1)} & \multicolumn{2}{c}{Prototype kept (AUROC)} & \multicolumn{2}{c}{coverage} & \multicolumn{2}{c}{ECE$_\mathrm{near}$} & FDR\\
\cmidrule(lr){2-4}\cmidrule(lr){5-7}\cmidrule(lr){8-9}\cmidrule(lr){10-11}\cmidrule(lr){12-13}\cmidrule(lr){14-14}
backbone & relaxed & head & head+rel. & relaxed & 1 pass & rel.+res. & $|F|,\sigma$ & head & MP cal. & slice & MP cal. & slice & MP / slice\\
\midrule
MACE-MP-0 & 0.108 & 0.077 & 0.069 & 0.727 & 0.671 & 0.700 & 0.615 & 0.862 & 0.87 & 0.90 & 0.139 & 0.076 & 0.24 / 0.10\\
CHGNet & 0.115 & 0.079 & 0.072 & 0.709 & 0.661 & 0.694 & 0.622 & 0.877 & 0.86 & 0.90 & 0.149 & 0.071 & 0.21 / 0.11\\
MACE-MPA-0 & 0.059 & 0.054 & 0.039 & 0.823 & 0.727 & 0.789 & 0.616 & 0.881 & 0.88 & 0.90 & 0.172 & 0.096 & 0.07 / 0.12\\
MatterSim-v1-5M & 0.056 & 0.050 & 0.036 & 0.834 & 0.739 & 0.796 & 0.625 & 0.898 & 0.88 & 0.90 & 0.141 & 0.092 & 0.12 / 0.13\\
ORB-v3 & 0.051 & 0.045 & 0.027 & 0.858 & 0.750 & 0.820 & 0.617 & 0.895 & 0.88 & 0.90 & 0.178 & 0.089 & 0.06 / 0.11\\
SevenNet-MF-ompa & 0.050 & 0.050 & 0.030 & 0.862 & 0.740 & 0.821 & 0.617 & 0.876 & 0.88 & 0.90 & 0.174 & 0.088 & 0.06 / 0.10\\
\bottomrule
\end{tabular}\end{adjustbox}
\end{minipage}\par\bigskip

\par\noindent\begin{minipage}{\linewidth}\captionsetup{type=table}\centering\small
\caption{Electronic-structure and compound questions on WBM from one forward pass on the unrelaxed input (test half, means over three seeds; same splits and calibration slice as the stability question). Metal: DFT band gap of the relaxed structure is zero; PV: $1\le E_g\le2$~eV; gap band: five levels (RPS). Single pass: embeddings and physics scalars; comp.: element fractions only. Compound questions (stable and non-metallic; stable and PV): ECE of the product of the two calibrated marginal answers, of that product recalibrated on the slice, and of a head trained on the compound question; NLL of the recalibrated product and of the dedicated head.}\label{tab:breadth}
\begin{adjustbox}{max width=\linewidth}\begin{tabular}{lccccccccccc}
\toprule
 & \multicolumn{2}{c}{Metal (AUROC)} & \multicolumn{2}{c}{PV (AUROC)} & \multicolumn{2}{c}{Gap band (RPS)} & \multicolumn{3}{c}{Stable \& non-metal (ECE)} & \multicolumn{2}{c}{Stable \& PV (NLL)}\\
\cmidrule(lr){2-3}\cmidrule(lr){4-5}\cmidrule(lr){6-7}\cmidrule(lr){8-10}\cmidrule(lr){11-12}
backbone & single pass & comp. & single pass & comp. & single pass & comp. & product & recal. & dedicated & recal. & dedicated\\
\midrule
MACE-MP-0 & 0.977 & 0.951 & 0.972 & 0.951 & 0.0192 & 0.0323 & 0.0050 & 0.0007 & 0.0013 & 0.0245 & 0.0287\\
CHGNet & 0.977 & 0.951 & 0.974 & 0.951 & 0.0190 & 0.0323 & 0.0049 & 0.0008 & 0.0019 & 0.0244 & 0.0277\\
MACE-MPA-0 & 0.983 & 0.951 & 0.978 & 0.951 & 0.0160 & 0.0323 & 0.0023 & 0.0007 & 0.0010 & 0.0213 & 0.0259\\
MatterSim-v1-5M & 0.984 & 0.951 & 0.979 & 0.951 & 0.0154 & 0.0323 & 0.0021 & 0.0006 & 0.0011 & 0.0205 & 0.0257\\
ORB-v3 & 0.989 & 0.951 & 0.984 & 0.951 & 0.0129 & 0.0323 & 0.0011 & 0.0010 & 0.0014 & 0.0182 & 0.0234\\
SevenNet-MF-ompa & 0.979 & 0.951 & 0.971 & 0.951 & 0.0180 & 0.0323 & 0.0019 & 0.0007 & 0.0012 & 0.0224 & 0.0261\\
\bottomrule
\end{tabular}\end{adjustbox}
\end{minipage}\par\bigskip

\end{landscape}

\snote{Where guarantees hold: shift, chemistries and closed loops}{sn:guarantees}

\paragraph{Shift between where labels are and where decisions are.} Training and calibrating on MP, where labels are plentiful, and deciding on WBM fails twice (Extended Data Fig.~\ref{fig:decision}). Discrimination degrades (AUROC 0.838 for a MACE-MP-0 head trained on MP, against 0.848 for its raw single-point hull distance), and MP-fitted calibration does not transfer: on the WBM rounds 2--5 structures of the test split the 90\% conformal sets of the MP-calibrated head cover 0.87 (MACE-MP-0) and 0.86 (CHGNet) of the true labels, and conformal selection with an MP-calibrated linear head realises a false-discovery rate of 0.24 and 0.21 at a nominal 0.1. Moving the calibration source towards the candidates repairs coverage and false-discovery control without retraining (random slice: coverage 0.90 and 0.90; FDR 0.10 and 0.11) but cannot repair ranking. The same holds for the leaderboard models (Fig.~\ref{fig:audit}e): calibrating conformal selection on WBM rounds 1--2 and deciding on rounds 4--5 breaks false-discovery control for 20 of 25 MP-trained potentials, and for the conformal screen of the main text, a threshold set on MP structures with an MP-trained head discards 19\% of the stable candidates at a nominal 1\%.

\paragraph{Remedies for shift.} We tested the standard corrections on the shift between substitution rounds, which is the shift a campaign meets when it calibrates on its first generation of candidates and decides on later ones (Extended Data Fig.~\ref{fig:shift}). For each of the 25 MP-trained potentials and each of 20 draws, a calibration set of 5\% of WBM was drawn from rounds 1--2 and a test batch of 2,000 structures from rounds 4--5. A logistic domain classifier on element fractions, the numbers of elements and sites and the model's predicted hull distance, cross-fitted on 20,000 structures of each side, separates the rounds with AUROC 0.74 (median); its odds give importance weights, clipped at their 99th percentile. Weighted Platt calibration raises the median expected calibration error of the probabilities from 2.5 to 3.7 percentage points, and label-shift re-estimation by expectation--maximisation\cite{saerens2002adjusting} gives 2.8: the probabilities were barely shifted to begin with, and neither correction improves them. Conformal selection is more sensitive, because it relies on the tail of the calibration scores: the median realised false-discovery rate at $q=0.1$ is 0.10 unweighted and 0.12 weighted, and the mean over draws exceeds $q$ significantly ($z>2$) for 8 and 15 of the 25 potentials. Weighted conformal selection with heterogeneous pruning\cite{jin2023weighted} is valid only for the true weights, and estimated weights that separate the rounds this weakly add variance rather than remove bias. Calibrating instead on $n$ random labelled structures of the later rounds restores control for every potential at $n=$ 50 to 500 (median realised rates 0.02, 0.05 and 0.07 at $n=100$, 200 and 500), at the price of power: the procedures certify 0.09, 0.20 and 0.34 of the stable candidates, against 0.48 for the invalid early-round calibration. Adding the later-round labels to the early-round calibration set does not help (8 potentials above $q$ with 100 added labels): the guarantee needs exchangeability with the decisions, not more data. The weighted and unweighted remedies were registered for the third prospective stage (Supplementary Note~\ref{sn:prospective}).

\paragraph{Guarantees hold on average over chemistries, not within each.} A map and a selection rule fitted to the whole pool are calibrated and controlled marginally. Within anion families (halides, oxides, chalcogenides, pnictides, intermetallics) the global map of the median MP-trained potential misstates the stable fraction of its worst family by 1.7 percentage points, and conformal selection realises false-discovery proportions of 0.114 among oxides. One map and one selection procedure per family (Mondrian calibration\cite{vovk2005alrw}) reduce these to 0.5 points and 0.11, redistributing selection power between families rather than losing it (Supplementary Table~\ref{tab:mondrian}). A typed answer has to state the population over which its guarantee is claimed.

\paragraph{Closed loops select their own calibration data.} In simulated budgeted DFT campaigns on WBM (Extended Data Fig.~\ref{fig:loop}; 30,000 candidates, 5,000 DFT calls in batches of 250, heads refitted every round), greedy acquisition finds 3.7--3.8 times as many stable structures as random sampling, but refitting the calibration map on the latest acquisitions --- which sit at the top of the ranking and on which the head has just been trained --- leaves the remaining pool miscalibrated (ECE 0.058 for CHGNet against 0.016 under random acquisition) and underestimates the stable structures it still contains by 76\%--80\%. Out-of-sample calibration --- cross-fitting over all acquisitions, the random seed, or a 10\% random ``calibration tax'' in every batch --- reduces the error to 13\%--25\% but does not remove it, and no strategy forecast the next batch without bias. Inside a loop, honest probabilities require labels the loop did not choose.

\begin{table}[t]\centering\small
\caption{Chemistry-conditional calibration of the leaderboard models (test part of five 5\% calibration slices; medians over models). CITL: calibration in the large, mean calibrated probability minus stable fraction (percentage points). FDP and power of conformal selection at $q=0.1$ within each family, with one isotonic map and one Benjamini--Hochberg procedure for all candidates (global) or one per family (Mondrian).}\label{tab:mondrian}
\begin{adjustbox}{max width=\textwidth}\begin{tabular}{llcccccccc}
\toprule
 & & & \multicolumn{2}{c}{CITL (points)} & \multicolumn{2}{c}{FDP at $q=0.1$} & \multicolumn{2}{c}{power} \\
\cmidrule(lr){4-5}\cmidrule(lr){6-7}\cmidrule(lr){8-9}
models & family & stable & global & Mondrian & global & Mondrian & global & Mondrian\\
\midrule
MP-trained potentials (25) & halide & 28\% & -1.7 & +0.3 & 0.060 & 0.102 & 0.56 & 0.72\\
 & oxide & 12\% & -0.6 & -0.2 & 0.114 & 0.053 & 0.43 & 0.21\\
 & chalcogenide & 19\% & -0.4 & -0.2 & 0.069 & 0.083 & 0.52 & 0.51\\
 & pnictide & 17\% & -0.5 & +0.0 & 0.072 & 0.088 & 0.56 & 0.54\\
 & intermetallic & 13\% & +0.6 & +0.1 & 0.112 & 0.100 & 0.43 & 0.37\\
 & all & 15\% & +0.1 & +0.0 & 0.097 & 0.100 & 0.48 & 0.45\\
\midrule
potentials, larger sets (30) & halide & 28\% & -0.9 & +0.2 & 0.074 & 0.098 & 0.91 & 0.94\\
 & oxide & 12\% & +0.1 & -0.3 & 0.106 & 0.080 & 0.91 & 0.88\\
 & chalcogenide & 19\% & -0.3 & +0.0 & 0.079 & 0.088 & 0.91 & 0.92\\
 & pnictide & 17\% & -0.0 & +0.1 & 0.081 & 0.096 & 0.93 & 0.94\\
 & intermetallic & 13\% & +0.1 & -0.0 & 0.110 & 0.099 & 0.90 & 0.88\\
 & all & 15\% & -0.0 & -0.0 & 0.097 & 0.096 & 0.91 & 0.89\\
\midrule
non-potential models (10) & halide & 28\% & -6.9 & +0.5 & 0.026 & 0.057 & 0.02 & 0.02\\
 & oxide & 12\% & +1.0 & -0.1 & 0.162 & 0.000 & 0.00 & 0.00\\
 & chalcogenide & 19\% & -3.2 & -0.9 & 0.002 & 0.065 & 0.00 & 0.01\\
 & pnictide & 17\% & -0.8 & -0.3 & 0.041 & 0.000 & 0.00 & 0.00\\
 & intermetallic & 13\% & +1.6 & +0.1 & 0.024 & 0.025 & 0.00 & 0.00\\
 & all & 15\% & -0.0 & -0.0 & 0.039 & 0.101 & 0.00 & 0.01\\
\bottomrule
\end{tabular}\end{adjustbox}
\end{table}

\snote{DFT labels on the Materials Project energy scale}{sn:dft}

\paragraph{Calculations.} Calculations use VASP\cite{kresse1996efficient} with MP's GGA(+U) protocol, start, as in WBM, from the unrelaxed candidate structure, and use a double relaxation. They differ from MP's legacy calculations in the VASP version and in the release of the PBE projector-augmented-wave potentials\cite{blochl1994paw,perdew1996pbe}: the potentials available to us carry the symbols and titles of MP's, but their data identify them as a later release, and prospective chemistry is restricted to the 50 elements covered. Total energies therefore differ from MP's by amounts that depend on the elements: recomputing 23 MP ground states from their MP structures gave energies that differ from MP's by a median of 26~meV/atom (up to 235).

\paragraph{Per-element shifts.} Such differences are dominated by a constant per atom of each element. We measured it by recomputing MP's elemental phases (the smallest cells within 0.1~eV/atom of the elemental ground state, two for 45 of the 50 elements, started from MP's relaxed structures and magnetic moments) and, for Ni, V and W, which MP treats with $+U$ in oxides and fluorides, a binary oxide. One shift $\delta_\mathrm{el}$ per element, and one per $+U$ element in oxides and fluorides, was fitted by robust least squares to these calculations and to a first control set of 23 MP ground states and 8 WBM structures (129 calculations, median residual 2.9~meV/atom; NiO and Si, both references and controls, enter twice), and every energy is shifted by $-\sum_\mathrm{el}x_\mathrm{el}\delta_\mathrm{el}$, with $x_\mathrm{el}$ the atomic fraction (shifts from $-$26 to 349~meV/atom).

\paragraph{Validation.} A second control set, chosen before its results were known, contains 36 MP ground states covering all 50 elements and 9 $+U$ compounds of Ni, V and W, started from MP's structures, and 20 WBM structures started, like WBM, from their initial geometry. Aligned energies reproduce MP's to a median of 4.1~meV/atom (40 before the alignment; 95th percentile 13, at most 21; $+U$ compounds 3.5, at most 13), and aligned hull distances reproduce WBM's to a median of 3.7~meV/atom (56 before) with WBM's stability label for all 20 (Extended Data Fig.~\ref{fig:align}). MP ground states lie on the hull by construction, so the sign of their residual decides their label, and 20 of 36 keep it. The residuals are not those of a pure constant per element: the tellurides of the two control sets lie below their reference after alignment (nine of nine, median 7~meV/atom), and the largest residuals are those of intermetallics. Because the first two sets contain many MP ground states and intermetallics and the candidates are ionic substitutions, a third control set matched to the candidate pool was chosen and computed after a review of the alignment, before any prospective label was collected: 100 WBM structures from the pool's anion families in proportion to their share of it (hydroxides over-sampled), two thirds ternary and one third quaternary, with at most 12 sites and WBM hull distances between $-0.05$ and $+0.10$~eV/atom, started from their initial geometry. On it, aligned hull distances reproduce WBM's to a median of 3.4~meV/atom with WBM's label for 92 of 98 (11 of the 16 within 10~meV/atom of WBM's hull), but the residuals have a heavy tail (95th percentile 44~meV/atom, mean +6.4): ten structures differ by more than 25~meV/atom, up to 301, eight of them with our energy above WBM's, and seven of them are structures whose prototype WBM's relaxation changed, against 7 of the other 88. The tail therefore reflects relaxations from the same unrelaxed start that end in different minima, not the alignment. These residuals define the label-noise model of the sensitivity analyses below.

\paragraph{Labels.} The MP2020 corrections\cite{wang2021mp2020} depend on the composition, on the oxide type of the relaxed structure and on an oxidation-state assignment, but not on the energy, so the aligned hull distance is the raw one minus the shift. We compute them with pymatgen\cite{ong2013pymatgen}, as for WBM's labels: all 30 WBM structures of the first two control sets receive WBM's correction, and of the 68 MP compounds only TiSiOs$_2$ differs, carrying in MP's database a Si anion correction that the current implementation does not assign. Hull distances are taken against the patched phase diagram of the MP 2023-02-07 entries, which reproduces the WBM labels of Matbench Discovery exactly (400 random WBM entries, maximum deviation below 0.1~meV/atom); a candidate is stable if its hull distance, rounded to 1~meV/atom as in WBM, is at most zero. The shifts were fixed, and the SHA-256 digest of the file that holds them recorded (\texttt{b88706bf90887082}, 2026-10-02 23:08:28 UTC), before any prospective calculation was collected. A secondary set of shifts adds to the fit MP's elemental ground states, which the selection of the smallest cells missed for 20 elements (by up to 54~meV/atom, for Sn), and the second control set; it was fixed before the third control set and any prospective label were known (\texttt{1932829c2a3ce264}). On the third set it reproduces WBM's hull distances to a median of 3.7~meV/atom (primary shifts: 3.4) with WBM's label for 94 of 98, no material change; the prospective analyses are repeated with its labels. Calculations that did not finish are reported and excluded (five of the stage-1 sample); stage-1 calculations that failed or were interrupted while a fault of the MPI launcher serialised the runs sharing a node were restarted once after the fault was fixed. A stopping time for the last stage-1 calculation still running was recorded in the decision log before any stage-1 label was collected (it would then have counted as not finished); the calculation ended, unconverged, before that time, so the rule was not applied. All 100 shortlisted calculations finished.

\paragraph{How constant are the shifts?} The differences from MP are not explained by LASPH, which changes the energies of the first control set much less than they differ from MP, except for NiO. To first order they are constants of the elements: the two polymorphs computed for 45 elements differ by a median of 7.4~meV/atom, and shifts fitted on the elements and the first control set carry over to the second set (median residual 3.9~meV/atom). The polymorphs differ by more (24--44~meV/atom) for six elements (H, Nd, Sr, Sb, Ir, Lu), whose shifts are less certain; the 18 compounds of the second set that contain them are aligned to a median of 3.1~meV/atom (at most 9). The constant is not exact: tellurides lie systematically below their reference after alignment, intermetallics carry the largest residuals, and the shift implied by a compound can depend on the bonding of the element (199~meV per atom of Sn in SnO$_2$, 140 in SmSn$_2$, against a fitted 162). This is why the third control set, matched to the candidates, is the one that measures the label noise of the prospective test. On it (Supplementary Table~\ref{tab:control3}), residuals are smallest for oxides, fluorides and pnictides and larger for chalcogenides, halides and hydroxides, and its tail is set by structures that WBM's relaxation carried to another prototype. The elemental ground states that the first selection missed confirm that the shift of an element is only approximately a constant: their energies differ from the fitted shifts by a median of 14~meV per atom of the element (at most 36). Adding them and the second control set to the fit (the secondary shifts) changes the shifts by a median of 1.6~meV/atom (at most 14) and does not materially change the agreement on the third set.

\FloatBarrier

\begin{table}[t]\centering\small
\caption{Third control set, matched to the prospective pool: WBM structures from their initial geometry, near the hull, by anion family (hydroxides over-sampled). $\Delta E_\mathrm{hull}$: our aligned hull distance minus WBM's (meV/atom); label: WBM's stability label reproduced.}\label{tab:control3}
\begin{tabular}{lrrrrrr}
\toprule
family & computed & finished & median $|\Delta E_\mathrm{hull}|$ & mean $\Delta E_\mathrm{hull}$ & max $|\Delta E_\mathrm{hull}|$ & label\\
\midrule
oxide & 34 & 33 & 2.7 & +3.3 & 73 & 31/33\\
chalcogenide & 21 & 21 & 7.0 & +15.0 & 301 & 19/21\\
fluoride & 16 & 15 & 2.8 & +0.9 & 38 & 15/15\\
halide & 16 & 16 & 6.0 & +9.4 & 58 & 14/16\\
pnictide & 7 & 7 & 2.0 & +1.2 & 7 & 7/7\\
hydroxide & 6 & 6 & 8.3 & +5.3 & 23 & 6/6\\
\midrule
all & 100 & 98 & 3.4 & +6.4 & 301 & 92/98\\
\bottomrule
\end{tabular}
\end{table}

\snote{Training data and reference phases}{sn:overlap}

\paragraph{WBM is inside Alexandria.} The Alexandria PBE database\cite{schmidt2024alexandria} grew from 4.61 to 5.95 million entries between its 2023.12.29 and 2025.07.02 releases (5.78~million distinct entries over the three releases we indexed). We kept the structures of every entry whose reduced formula occurs in WBM or in our pool, and matched WBM's DFT-relaxed unique prototypes against Alexandria's relaxed structures of the same formula with pymatgen's StructureMatcher (default tolerances): 99.4\% of them have a counterpart, in every release separately as in their union, and 99.8\% share a formula with an Alexandria entry, because Alexandria absorbed WBM's structures. For a model trained on Alexandria without removing them, as ORB-v3 was\cite{rhodes2025orbv3}, the WBM test set is therefore no test of generalisation; sAlex and OMat24 removed every structure that shares a prototype label with a WBM structure\cite{barroso2024omat24}, and MP-trained models never saw them. The audit, the single-pass heads and the value-of-information analyses on WBM compare backbones on candidates some of them may have seen, which is one reason we test prospectively.

\paragraph{Our candidates are mostly new.} 17\% of our pool has a counterpart in some Alexandria release, through its initial structure or, for computed candidates, its DFT-relaxed one --- substitutions that Alexandria's own searches also made --- and 83\% has none. The random samples match the pool (18\% in stage~1, 18\% in stage~3), whereas the shortlist is enriched (54\%): the candidates a backbone ranks highest are often ones that Alexandria also found stable. Counterparts concentrate the stable candidates: 14.9\% of the random candidates with a counterpart are stable, against 3.7\% of those without one.

\paragraph{Calibration without the overlap.} Restricted to the 1,635 labelled random candidates without a counterpart, the registered single-pass forecasts over-state the stable fraction by 0.5--2.5 percentage points, with 95\% intervals clustered by parent prototype that include zero for one of the eight, and rank the candidates with AUROC 0.921--0.966 (Extended Data Fig.~\ref{fig:overlap}b): the over-statement of Supplementary Note~\ref{sn:prospective} is not an effect of the overlap. Calibration is worse on the 349 candidates with a counterpart (expected calibration error 2.9--4.8 points, against 0.5--2.6 without one), which are more often stable and concentrate at the top of the rankings. The registered analysis by protostructure label (AFLOW prototype and chemical system; analysis 6(b)) agrees: 16\% of the pool has an Alexandria entry with its protostructure, and on the 1,667 labelled random candidates without one (4.5\% stable) the forecasts over-state the stable fraction by 0.7--2.3 points (intervals excluding zero for all eight).

\paragraph{Our labels against Alexandria's.} Alexandria's energies are on MP's scale: per-element offsets fitted to WBM--Alexandria pairs of the same structure are 0.4~meV/atom in the median. Ours, computed with a different PAW release, are not: for the 391 candidates whose DFT-relaxed structure matches an Alexandria entry, our energy lies 40~meV/atom above Alexandria's in the median before our per-element alignment to MP's scale, and Alexandria's minus ours is $-$0.3~meV/atom in the median after it (median absolute difference 3.7~meV/atom; Extended Data Fig.~\ref{fig:overlap}c). The shifts we froze before the prospective calculations thus put our labels on the common scale of MP and Alexandria, as intended.

\paragraph{Reference phases.} Hull distances are measured against MP's phases, as in WBM and Matbench Discovery, and gaps in MP's reference make some candidates look more stable than they are; the stable shortlisted erbium bromides and osmium iodides lie in systems in which MP holds no binary compound of the cation with the anion (Supplementary Note~\ref{sn:prospective}). Completing the reference with Alexandria's phases within 0.1~eV/atom of its hull in the candidates' subsystems, put on MP's scale by per-element shifts fitted to 255,499 WBM--Alexandria pairs of the same structure (median absolute residual 0.4~meV/atom) and given the MP2020 corrections, and leaving out each candidate's own counterparts, 84 of the 211 labelled candidates that are stable against MP remain stable (Extended Data Fig.~\ref{fig:overlap}d). The forecasts were made, and are evaluated, against MP's hull; against the completed hull the stable fraction of the second random sample falls from 5.8\% to 1.7\% and the registered forecasts over-state it by 5.2--6.2 percentage points: a forecast inherits the reference it was calibrated against.

\snote{The registered prospective test in detail}{sn:prospective}

\paragraph{Candidates and registration.} Following WBM\cite{wang2021wbm}, candidates are data-mined ionic substitutions\cite{hautier2011substitution} into near-hull MP prototypes with two to four elements and small cells, within the elements for which DFT labels can be put on MP's energy scale (below); only compositions absent from MP and from WBM are kept, and the pool is the 8,000 most probable candidates (5,780 compositions in 4,221 chemical systems, 98\% with three or four elements, median 10 sites; 8,000 with forecasts from all backbones).
\emph{Forecasts and registration.} For each of the six backbones, heads are trained on all WBM unique prototypes except a random 5\% calibration slice and applied unchanged to the pool; System-2 forecasts use the backbone's own relaxation of each candidate (MACE-MP-0, CHGNet, MatterSim-v1-5M and ORB-v3). All forecasts, the sampling seed, the stage-1 DFT sample --- a uniform random sample of the eligible pool, split at random into a calibration slice of 200 and a test sample of 400 candidates --- and the stage-2 shortlist --- the 100 remaining candidates with the highest single-pass probability of the primary backbone (ORB-v3, fixed in advance) --- were written to a registration file whose SHA-256 digest (\texttt{8f359667a6acda7c}\ldots) was recorded outside the computing cluster at 2026-10-02 21:12:07 UTC, before any DFT calculation of a pool member, together with every backbone's forecast of the shortlist's yield. After the stage-1 labels, and before the shortlist's DFT, the primary backbone's single-pass probability was recalibrated (Platt) on all stage-1 labels and the updated forecast of the yield was registered in the same way (\texttt{07d33c3f6708c63a}\ldots, 2026-10-03 16:12:04 UTC).

\paragraph{Analysis plan.} As registered --- stage 1, on the test sample: calibration in the large (forecast minus realised stable fraction), ECE, log-loss and Brier score of every backbone's registered System-1 forecast and, for the four backbones that relax, System-2 forecast, against the same forecasts recalibrated on the calibration slice (Platt and isotonic); realised net value at $c=0.1$, 0.3, 0.5 and 0.7 of the System-1 rule $p_1\ge c$, the System-2 rule and relax-and-threshold; recall of the conformal screen calibrated on the slice ($\alpha=0.05$, 0.1); and the false-discovery proportion of conformal selection ($q=0.1$, 0.2) among the test sample with the slice as calibration set. Stage 2: the forecast against the realised number of stable structures in the shortlist, for the forecasts registered with the pool and for the updated forecast. Secondary: the stage-1 analyses without the $+U$ chemistries (oxides and fluorides of Ni, V and W), whose shifts rest on fewer reference calculations. In addition to the registered plan we report bootstrap 95\% intervals of calibration in the large, AUROC and the ECE with five equal-mass bins, as implemented in the evaluation script fixed before the registration. Not registered, and added after the registration but before any stage-1 label was collected, are sensitivity analyses of stage~1 to the DFT labels: the expected number of labels that WBM's protocol would give differently, from the residuals of the third control set (a candidate with aligned hull distance $h$ has WBM-protocol hull distance $h-r$, with $r$ drawn from those residuals); calibration in the large with such a residual resampled for every candidate in each bootstrap replicate; the analyses without the candidates within 10~meV/atom of the hull; and labels from the secondary shifts. Also not registered are analyses of the shortlist: the probability of the realised yield under each forecast if the outcomes were independent (a Poisson-binomial distribution); every backbone's forecast and single-pass hull distance for the unstable candidates; their DFT hull distance at the unrelaxed geometry (the final one plus the DFT relaxation energy, from the first ionic step); and DFT relaxations of them without symmetry ($\mathrm{ISYM}=0$, otherwise the same settings) from MACE-MP-0's relaxed structure and from the unrelaxed structure with Gaussian random displacements ($\sigma=0.05$~\AA{} per Cartesian component).

\paragraph{Calibration on new candidates.} Of the 600 stage-1 calculations, five did not finish (Methods); 23 of the 397 test candidates (5.8\%) and nine of the 198 candidates of the calibration slice (4.5\%) are stable. Calibrated on WBM alone, the registered single-pass forecasts over-stated the stable fraction of the test sample by 0.8--2.1 percentage points, with bootstrap 95\% intervals that include zero for four of the six backbones (ORB-v3: 0.8 points, interval $-$0.5 to 2.0), ranked the candidates with AUROC 0.965--0.982 and had an ECE of 0.009--0.021 (Fig.~\ref{fig:prospective}b,c; Supplementary Table~\ref{tab:prospective}); the forecasts after relaxation erred similarly ($-$0.1 to 2.4 points, intervals including zero for two of four; AUROC 0.976--0.995). Part of the over-statement may lie in the labels rather than in the forecasts. On the control set matched to the pool, the differences between our hull distances and WBM's are centred on zero but have a tail in which ours are higher --- structures that WBM's relaxation carried to a lower minimum from the same start; resampling these differences, WBM's protocol would be expected to give 9.3 test candidates (2.3\%) the other label, mostly towards stable (net +1.4 points), and with this label noise propagated the intervals of all six backbones include zero (ORB-v3: $-$2.5 to +1.1 points; a sensitivity analysis that was not registered, Methods). Excluding the 16 test candidates within 10~meV/atom of the hull, or labelling with the secondary shifts (which change two labels), gives +1.0 and +0.3 points for ORB-v3; without the $+U$ chemistries (the registered secondary analysis, 338 test candidates) the single-pass forecasts over-stated the stable fraction by 0.2--2.2 points. Recalibrating on the random slice moved calibration in the large closer to zero for five of the six backbones (median absolute error 1.3 points before, 0.6 after; range $-$1.6 to +0.8) but lowered the log-loss for only three: with nine stable candidates, the slice can correct the level of the forecasts but not their shape. Calibrated on WBM alone, the forecasts thus carried over to new candidates generated the same way to within about two percentage points of the stable fraction; whether the remaining over-statement belongs to the forecasts or to the labels, this test cannot settle.

\paragraph{The probabilities were worth acting on.} At an attempt cost of $c=0.5$ (a failed attempt costs half a stable structure), attempting the test candidates with $p_1\ge c$ realised 8--19 net hits per 1,000 candidates across the six backbones (Fig.~\ref{fig:prospective}d; Supplementary Table~\ref{tab:prospvalue}). For CHGNet and MACE-MP-0, whose relaxations call too many candidates stable (precision 57\% and 49\%), the single pass realised 14 and 8 against 6 and $-$1 for relax-and-threshold, matched or beat it in seven of the eight comparisons at the registered costs and kept a positive value at $c=0.7$, where relax-and-threshold realised $-$11 and $-$23. Paired bootstrap intervals of the differences (post hoc; 10,000 resamples of the test candidates) exclude zero at $c=0.7$ (CHGNet 17, 95\% interval 5 to 31; MACE-MP-0 26, 11 to 42) but not at $c=0.5$ (8, $-$4 to 19; 9, $-$4 to 21). For MatterSim and ORB-v3, whose relaxed calls are precise (83\% and 90\%), the two rules differed by at most 4 net hits per 1,000 (1.6 among the 397 test candidates), mostly in favour of relax-and-threshold, as the retrospective comparison anticipated (main text). Calibrating the relaxed prediction (the System-2 rule) matched or beat relax-and-threshold at $c=0.5$ for all four backbones that relax.

\paragraph{Guarantees need larger calibration samples.} The conformal procedures were consistent with their guarantees, which hold on average, but with a slice this small they guaranteed little (Supplementary Table~\ref{tab:prospconf}). A screen at $\alpha=0.05$ cannot discard any candidate until the slice holds 19 stable ones, and it kept all; at $\alpha=0.1$ it discarded 53\%--83\% of the test candidates and kept 87\%--100\% of the stable ones. Because a conformal p-value cannot be smaller than one over the number of calibration candidates plus one, selection with 198 calibration candidates can certify, at $q=0.1$, only sets of at least 20 of the 397 test candidates --- nearly all 23 stable ones; it did so for SevenNet alone (21 candidates, false-discovery proportion 0.14), and at $q=0.2$ certified 0--23 candidates with false-discovery proportions of 0.07--0.17. Useful guarantees therefore need several times more random labels than calibration does: at a stable fraction of 5\%, about 400 for a screen at $\alpha=0.05$, and, whatever the stable fraction, about 800 for selection at $q=0.1$ to certify sets as small as five among 400 candidates.

\paragraph{The shortlist.} The 100 candidates that a screening campaign would have computed next yielded 98 stable structures, most of them far below the hull (median hull distance of the stable ones $-$123~meV/atom; Fig.~\ref{fig:prospective}e; Supplementary Table~\ref{tab:prospstage2}). Stability is defined against MP's hull, as in WBM, and two features of the shortlist qualify its yield. It contains 79 distinct compositions, because several were generated from more than one parent prototype, and ten of its stable candidates (erbium bromides and osmium iodides) lie in chemical systems in which MP holds no binary compound of a cation with one of its anions, so that their hull distances are measured against an incomplete reference rather than against the phases into which they would decompose. The forecasts of this yield registered with the pool ranged from 91 to 100. Those of CHGNet and MACE-MP-0 under-stated it (for independent candidates, a yield of 98 or more had a probability below 0.008); those of MACE-MPA-0, MatterSim and SevenNet were consistent with it; and the primary backbone's forecast --- 99.9 $\pm$ 0.3 at registration, 99.8 $\pm$ 0.5 after recalibration on the stage-1 labels --- was over-confident: two failures had a probability of 0.005 under it. The update could not have caught this, because only three stage-1 candidates fell in the shortlist's probability range. The two failures, SrWSe$_{2}$F and CaWSe$_{2}$F (0.53~eV/atom above the hull), are substitutions into the same prototype, SrSbSe$_{2}$F (W$^{3+}$ for Sb$^{3+}$, and in CaWSe$_2$F also Ca$^{2+}$ for Sr$^{2+}$), which ORB-v3 rated stable with probability 0.999. At the unrelaxed geometry, DFT places them 0.66~eV/atom above the hull, whereas the single passes of MACE-MP-0, MACE-MPA-0 and ORB-v3 placed them below it ($-$0.17 to $-$0.06~eV/atom) and those of CHGNet, MatterSim and SevenNet above it (0.32--0.60~eV/atom), and the latter gave both a probability of at most 0.01; the lowest probability that any backbone gave any other shortlisted candidate, before or after relaxation, was 0.09. At the structures to which MACE-MP-0 relaxed them, which it places 0.53--0.57~eV/atom below the hull, DFT places them 0.53--0.56~eV/atom above it, and DFT relaxations without symmetry from there, or from the unrelaxed structure with random displacements, end within 1.2~meV/atom of the registered calculations: the low-energy structure that MACE-MP-0 found does not exist in DFT, and the labels are not an artefact of a relaxation held in place by symmetry. A binomial forecast treats candidates as independent, but a shortlist built from one backbone's ranking concentrates that backbone's errors, and they come in groups --- here two substitutions into one prototype --- so the yield of a shortlist is less certain than its binomial spread suggests, and the single pass of a second backbone is a cheap way to find such groups (analyses not registered; Methods).

\paragraph{Both failures and the $+U$ convention.} MP's input set computes both failures with $+U$ on tungsten only because each contains one fluoride ion per formula unit, and the MP2020 correction for tungsten in oxides and fluorides supplies about three quarters of their total correction ($-0.89$ of $-1.17$~eV/atom). Their labels, and the potentials' disagreement with them, may therefore reflect the GGA/GGA$+U$ mixing convention as much as the chemistry; the labels follow the benchmark's protocol, which WBM's labels share.

\paragraph{Stage 3: a larger random sample.} The stage-1 sample was too small for the conformal procedures to say much (above), and calibration estimates from 397 test candidates are imprecise. After the stage-2 labels we therefore registered a third stage (\texttt{117beb36d621d968}\ldots, 2026-10-04 07:10:06 UTC): a uniform random sample of 1,400 of the 7,300 pool members outside stage~1 and the shortlist (seed 20261004), every backbone's decisions for it frozen on the stage-1 labels --- the Platt and isotonic recalibration maps, the conformal screens at $\alpha=0.05$, 0.1 and 0.2 and conformal selection at $q=0.1$ and 0.2 with the 600 stage-1 candidates as calibration set --- and an analysis plan that names the same calibration, value and conformal analyses on stage~3 and on all random labels, with intervals clustered by parent prototype, and as secondary analyses the candidates without an Alexandria counterpart, those without $+U$ chemistries and labels against a hull completed with Alexandria's phases. Inputs were written once, on one cluster, and the calculations ran on three clusters with byte-identical PAW files: each candidate's cluster was drawn at random, and unstarted calculations were later moved to the fastest cluster to balance the load, as the plan allowed. Before pooling, 24 stage-1 structures were recomputed from fresh inputs on every cluster: the largest median absolute difference from the original final energies was 0.15~meV/atom and the largest single difference 12.3~meV/atom, and the number of changed stability labels was zero, so labels from all clusters were pooled as registered. Three addenda were registered before any stage-3 label was collected: the shift remedies of Supplementary Note~\ref{sn:guarantees} with WBM's slice as source (\texttt{748d1f689b3c740a}\ldots), the single-pass forecasts of eSEN-30M-OAM and UMA-s-1p1 with the same frozen decisions (\texttt{4913d95a7c30cc86}\ldots), and their relaxations of the labelled candidates (\texttt{0788e409ad786f12}\ldots). 11 of the 1,400 calculations are excluded: eight failed after the maximum number of error corrections and three when VASP stopped mid-run, leaving an incomplete output.

\paragraph{Calibration on 2,000 new candidates.} Of the 1,389 stage-3 candidates, 81 (5.8\%) are stable; of all 1,984 random candidates with labels, 113 (5.7\%). Calibrated on WBM alone, the registered single-pass forecasts of the eight backbones over-state the stable fraction of all random candidates by 0.9--2.2 percentage points, with intervals clustered by parent prototype that exclude zero for all eight, rank them with AUROC 0.931--0.971 and have an expected calibration error of 1.0--2.3 points (Fig.~\ref{fig:prospective}c; Supplementary Table~\ref{tab:prospective3}). On stage~3 alone the registered forecasts over-state the stable fraction by 1.0--2.0 points; the stage-1 recalibration maps, frozen before stage~3 was computed, bring this to $-$0.9 to 0.8 points (Platt) and $-$0.9 to 1.1 (isotonic). The remedies fitted on WBM alone do so less reliably: the label-shift re-estimate errs by $-$2.5 to $-$0.7 points (intervals excluding zero for seven of the eight) and the importance-weighted map by 0.3 to 1.4 (excluding zero for three of the eight; Extended Data Fig.~\ref{fig:shift}d). Paired, the stage-1 recalibration and the weighted map do not differ significantly in the absolute calibration in the large for any backbone; recalibration reduced it by 0.3--1.6 points, significantly for no backbone, and the weighted map significantly for six, and the 90\% interval of the recalibrated forecast lies within $\pm1$ point (two one-sided tests) for three of the eight (post hoc). The importance weights rest on a classifier that separates the pool from WBM's slice with AUROC 0.975--0.976 and leave an effective sample of 541--550 of the slice's 10,774 structures. Without the $+U$ chemistries (1,238 candidates; registered), the registered forecasts over-state the stable fraction of stage~3 by 1.2--2.5 points; on stage~3 their calibration slopes are 0.82--1.01, log-losses 0.088--0.138 and Brier scores 0.024--0.039. The registered System-2 forecasts (relaxed predictions with their WBM-calibrated maps) over-state it by $-$0.8 to 2.9 points for the six backbones that relax. The forecasts of eSEN-30M-OAM and UMA-s-1p1 for the stage-1 candidates were made after the stage-1 labels, by the registered protocol and without free choices; on stage~3 alone, for which they were registered before any label, they over-state the stable fraction by 1.7 (interval 0.8 to 2.6) and 1.1 points (0.2 to 2.1). Resampling parent prototypes rather than candidates barely changes the intervals of the random samples (design effects 0.8--1.1 on stage~1, whose 600 candidates come from 448 parents, and 1.1--1.3 on all random labels), but doubles the variance of the shortlist's yield (below).

\paragraph{Labels or population?} The stage-1 sensitivity analysis found that WBM's protocol would give part of the candidates the other label (above). Repeated on stage~3 and on all random labels (post hoc), the residuals of the third control set (median 0.0, mean 6.4~meV/atom, 98 structures), resampled as $h_\mathrm{WBM}=h_\mathrm{ours}-r$ as in the stage-1 analysis, predict 1.8 points more stable candidates under WBM's protocol on stage~3 and 1.8 on all random labels; with this label noise propagated, the calibration in the large of the registered forecasts is $-$0.9 to 0.4 points, with intervals that include zero for all eight backbones. Treated instead as noise added to WBM's hull distances ($h_\mathrm{ours}=h_\mathrm{WBM}+r$), the same residuals predict 0.0 points. The two forms differ because the tail of the residuals comes from relaxations that reached different minima from the same start, and how that depends on the true hull distance a sample of 98 structures cannot settle. For the 11 registered stage-3 calculations without a label (all failed), the forecasts expect 0.3--1.0 stable ones; had all been stable the over-statement on stage~3 would be 0.2--1.3 points, and had none, 1.0--2.1.

\paragraph{The bridging set.} The registered bridging set (Methods) measures the difference between the protocols directly. Of its 1,105 WBM structures, 1,099 were labelled with our protocol; six calculations failed (VASP stopped mid-run, leaving an incomplete output), and had every structure without a label been stable or unstable under our protocol, the difference $D$ below would be $-$0.2 or 0.4 points. Our labels called stable 92\% of the structures that WBM's labels call stable (sensitivity; 95\% interval 88\%--96\%) and 1.5\% of those it calls unstable (false-positive rate; 0.7\%--2.3\%): 17 structures moved from stable to unstable and 13 the other way, so that WBM's stable fraction exceeded ours by $D=0.4$ points ($-$0.6 to 1.4; 0.3 points post-stratified to the random candidates by anion family and forecast group). Within 50~meV/atom of WBM's hull (338 structures) the sensitivity was 88\% and the false-positive rate 5.5\%. The residuals of the hull distance (ours minus WBM's) have median 0.2 and mean 6.7~meV/atom, with 10.3\% beyond 25~meV/atom. Where WBM's relaxation changed the space group of its initial structure (136 structures; spglib, symprec 0.1), our hull distance was higher by a median of 15.2~meV/atom and by more than 25~meV/atom for 39\% of them, against 3\% of the others: from the same start, our relaxations more often stayed in a higher minimum. Where our relaxed structure matches WBM's (pymatgen's StructureMatcher, default tolerances; 1,005 structures), the residuals have median $-$0.3 and mean 0.3~meV/atom, with 6\% beyond 25~meV/atom; where it does not (94 structures), our hull distance is higher by a median of 36.8~meV/atom. Re-estimated on these residuals, the Berkson form predicts 2.0 points more stable random candidates under WBM's protocol and the classical form $-$0.2; the direct measurement, 0.4 points, is close to the latter. The registered adjustment converts the stable fraction of the random candidates under our protocol, 5.7\%, to 5.2\% under WBM's (3.8\%--6.6\%), lower rather than higher because, where stable candidates are rare, the unstable structures that our protocol calls stable outnumber the stable ones it misses, and leaves the over-statement of the registered forecasts at 1.4--2.8 points (lowest lower bound 0.3), with intervals that exclude zero for all eight backbones (1.6--2.6 points on stage~3 alone). On the bridging structures themselves, the registered forecasts are calibrated against both labels: their calibration in the large is $-$0.6 to 0.4 points against WBM's labels and $-$0.3 to 0.8 against ours. By the registered rule the over-statement on the new candidates is attributed to a shift of the population. The registered B2 calculations measure, in addition, how much our own labels near the hull move when the relaxation starts elsewhere, without symmetry or with antiferromagnetic order.

\paragraph{Protocol-robust labels (B2).} Of the 1,129 registered B2 calculations, 1,120 finished and nine failed, all before the registered stopping time; failed calculations are reported and excluded as registered. For the 461 candidates of Part~A, the lowest of the three energies came from the eSEN-started relaxation for 62\%, from the rattled one for 19\% and from the original for 19\%; the rattled relaxation was lower by more than 1~meV/atom for 27\% of the candidates and the eSEN-started one for 31\%. The hull distance fell by more than 1, 5, 10 and 25~meV/atom for 39.3\%, 15.0\%, 8.7\% and 3.5\% of them (median $-$0.5, fifth percentile $-$20.5, largest $-$314~meV/atom), by more than 10~meV/atom for 5\%, 10\% and 10\% of the candidates of stages 1, 2 and 3; no MP2020 correction changed, 13 candidates labelled unstable became stable and, by construction, none moved the other way. With these labels the stable fraction of the random samples rose from 5.7\% to 6.4\% (+0.7~points; 95\% interval 0.3 to 1.1, parent prototypes resampled) and from 12.1\% to 12.9\% at 25~meV/atom; at 50~meV/atom no call can change, because Part~A contains every random candidate within 50~meV/atom. On stage~3 alone the rise was +0.6~points (0.2 to 1.0). The forecasts are unchanged, so the calibration in the large of every backbone falls by the same amount, from 0.9--2.2 to 0.2--1.6 points, with intervals that exclude zero for CHGNet, MACE-MP-0 and eSEN-30M-OAM (0.4--1.5 points on stage~3 alone). This shift is an upper bound on what the effect contributes to the over-statement: WBM's labels, on which the forecasts were calibrated, also come from single relaxations, and to the extent that they understate stability near the hull in the same way, the shift cancels. The shortlist yields 98 hits with these labels, as with the original ones. Antiferromagnetic starts gave an energy lower by more than 1~meV/atom for 28 of the 75 spin-polarised candidates (37\%; median change $-$0.6, tenth percentile $-$8.2, largest $-$61~meV/atom); these candidates were already stable or shortlisted, so no call changed. SCAN local hulls (Part~C; Supplementary Table~\ref{tab:b2scan}): of the 18 candidates, 16 are stable by their MP2020-corrected PBE energies and four against their SCAN local hulls; the label changed for 12 of them at 0~meV/atom and for ten at 50~meV/atom, all from stable to unstable, and the SCAN hull distance exceeded the MP2020-corrected one by a median of 159~meV/atom. For four candidates a competing phase failed, so their SCAN hull distances are lower bounds: the two changes among them are certain; the stable SCAN labels of the other two are not.

\begin{table}[htbp]\centering\small
\caption{\textbf{SCAN local hulls of the B2 Part-C candidates (registered secondary analysis C1).} Hull distances in meV/atom: the MP2020-corrected PBE label as collected, and the SCAN hull distance against the convex hull of the SCAN energies of the candidate's competing phases (MP entries of its chemical system within 50~meV/atom of the MP2020 hull, lowest per composition, at most 40 sites), without empirical corrections. Where a competing phase failed or did not finish, the hull is that of the computed phases, as registered, and the SCAN value is a lower bound: an unstable call is then certain, a stable one is not. The last column names the phases that define the SCAN hull at the candidate's composition.}\label{tab:b2scan}
\begin{tabular}{lllrrl}
\toprule
candidate & identifier & group & MP2020 & SCAN & note\\
\midrule
CaWSe$_{2}$F & cj-06119 & mixed anion, $+U$ & +531 & +547 & \parbox[t]{5.9cm}{\raggedright vs CaF$_{2}$ + CaSe + W + WSe$_{2}$\strut}\\
RbWIF$_{3}$ & cj-05937 & mixed anion, $+U$ & $-$28 & +172 & \parbox[t]{5.9cm}{\raggedright lower bound; RbI failed\strut}\\
SrWSe$_{2}$F & cj-02481 & mixed anion, $+U$ & +533 & +572 & \parbox[t]{5.9cm}{\raggedright vs W + WSe$_{2}$ + SrF$_{2}$ + SrSe\strut}\\
VI$_{3}$F & cj-04418 & mixed anion, $+U$ & $-$2 & +58 & \parbox[t]{5.9cm}{\raggedright vs I + VF$_{3}$ + VI$_{2}$\strut}\\
I$_{5}$Br & cj-02319 & halogens only & $-$377 & $-$83 & \parbox[t]{5.9cm}{\raggedright vs I + IBr$_{2}$\strut}\\
IBr$_{5}$ & cj-02854 & halogens only & $-$266 & $-$6 & \parbox[t]{5.9cm}{\raggedright vs Br + IBr$_{2}$\strut}\\
IrBr$_{4}$ & cj-02152 & 5d iodide or bromide & $-$21 & +72 & \parbox[t]{5.9cm}{\raggedright vs Br + IrBr$_{3}$\strut}\\
IrBr$_{5}$ & cj-02092 & 5d iodide or bromide & $-$121 & +40 & \parbox[t]{5.9cm}{\raggedright vs Br + IrBr$_{3}$\strut}\\
IrBr$_{5}$ & cj-04182 & 5d iodide or bromide & $-$121 & +40 & \parbox[t]{5.9cm}{\raggedright vs Br + IrBr$_{3}$\strut}\\
IrI$_{5}$ & cj-03434 & 5d iodide or bromide & $-$57 & +61 & \parbox[t]{5.9cm}{\raggedright vs I + IrI$_{3}$\strut}\\
IrI$_{6}$ & cj-00924 & 5d iodide or bromide & $-$30 & +128 & \parbox[t]{5.9cm}{\raggedright vs I + IrI$_{3}$\strut}\\
OsBr$_{5}$ & cj-00025 & 5d iodide or bromide & $-$138 & $-$247 & \parbox[t]{5.9cm}{\raggedright lower bound; OsBr$_{4}$ failed\strut}\\
OsBr$_{5}$ & cj-00155 & 5d iodide or bromide & $-$136 & $-$247 & \parbox[t]{5.9cm}{\raggedright lower bound; OsBr$_{4}$ failed\strut}\\
OsI$_{4}$ & cj-00256 & 5d iodide or bromide & $-$210 & +74 & \parbox[t]{5.9cm}{\raggedright vs I + Os\strut}\\
OsI$_{5}$ & cj-00036 & 5d iodide or bromide & $-$211 & +82 & \parbox[t]{5.9cm}{\raggedright vs I + Os\strut}\\
OsI$_{5}$ & cj-00091 & 5d iodide or bromide & $-$208 & +82 & \parbox[t]{5.9cm}{\raggedright vs I + Os\strut}\\
OsI$_{6}$ & cj-07398 & 5d iodide or bromide & $-$154 & +94 & \parbox[t]{5.9cm}{\raggedright vs I + Os\strut}\\
ReI$_{6}$ & cj-00318 & 5d iodide or bromide & $-$53 & +87 & \parbox[t]{5.9cm}{\raggedright lower bound; Re failed\strut}\\
\bottomrule
\end{tabular}
\end{table}

\paragraph{Decisions on stage~3.} As registered, the single-pass rule was compared with the System-2 rule (the relaxed prediction with its WBM-calibrated map) and with relax-and-threshold, the hard rule on the relaxed prediction. Against the System-2 rule it realised $-$10 to $-$2 net hits per 1,000 stage-3 candidates at $c=0.1$ (significantly negative for three of the six backbones with registered relaxations), $-$11 to 2 at 0.3 (two), $-$7 to 7 at 0.5 (one significantly negative, one positive) and $-$3 to 3 at 0.7 (none significant). Against relax-and-threshold, at an attempt cost of 0.7 the registered single-pass rule $p_1\ge c$ of MACE-MP-0 and CHGNet realised 28 and 21 more net hits per 1,000 stage-3 candidates than relax-and-threshold (paired intervals clustered by parent prototype 18 to 37 and 13 to 29) and, at 0.5, 10 (3 to 17) and 5 ($-$1 to 13). For MatterSim-v1-5M and UMA-s-1p1 the two rules did not differ significantly at any cost; relax-and-threshold was better for ORB-v3 at $c=0.1$ and 0.3 ($-$8, $-$12 to $-$3; $-$7, $-$12 to $-$2) and for eSEN-30M-OAM at 0.1, 0.3 and 0.5 ($-$10, $-$14 to $-$6; $-$12, $-$17 to $-$7; $-$6, $-$11 to $-$2) (Fig.~\ref{fig:prospective}d). Calibrating the relaxation recovers the gains over relax-and-threshold at $c=0.7$ (25 and 20 net hits per 1,000 for MACE-MP-0 and CHGNet; intervals 16 to 34 and 13 to 29). The registered forecasts of the heads given the relaxed prediction over-state the stable fraction of stage~3 as the single-pass ones do (1.1--2.0 points for the five backbones that have them). On the random candidates of stages 1 and 3 the realised value of escalating from System~1 to System~2 is close to its model-implied value for eSEN-30M-OAM (6.7 against 6.9 net hits per 1,000 candidates, mean over the four attempt costs) and below it for ORB-v3 (4.5 against 7.6), MatterSim-v1-5M (3.0 against 6.3), MACE-MP-0 (0.7 against 4.9) and CHGNet (0.7 against 6.9); the model-implied value lies inside the bootstrap interval of the realised one for 12 of the 20 backbone--cost combinations (Extended Data Fig.~\ref{fig:voireal}d).

\paragraph{Conformal guarantees from one calibration set.} With the 595 stage-1 labels (32 stable) as calibration set, the conformal screen at $\alpha=0.05$, 0.1 and 0.2 realised recalls of 90\%--98\%, 88\%--95\% and 70\%--91\% on the stage-3 candidates (Supplementary Table~\ref{tab:prospective3}). The screens calibrated on WBM's slice, registered in the shift addendum, realised recalls of 74\%--85\% at $\alpha=0.1$ unweighted and 74\%--86\% with importance weights, below the 90\% target for every backbone, and WBM-calibrated selection at $q=0.2$ realised false-discovery proportions of up to 0.32 unweighted and 0.29 weighted, above $q$ for two and five backbones. Conformal selection certified no candidate at $q=0.1$ for CHGNet, MACE-MP-0, MACE-MPA-0 and MatterSim-v1-5M and returned 28--61 candidates for the others, with false-discovery proportions of 0.07--0.24; at $q=0.2$ it certified 31--95 candidates for seven backbones with false-discovery proportions of 0.24--0.35, above $q$ for seven of them. Both guarantees hold in expectation over calibration sets. To see how far one calibration set of this size can stray, we pooled the random labels of stages 1 and 3, split them 2,000 times at random into a calibration set of 595 and a test set of the rest, and applied the same procedures (not registered): the mean recall of the screen was 0.97, 0.91 and 0.81 at $\alpha=0.05$, 0.1 and 0.2, and the mean false-discovery proportion of selection 0.04--0.08 at $q=0.1$ and 0.16--0.19 at $q=0.2$, as guaranteed. The realised stage-3 values are within the spread of single splits (between 0.05 and 0.29 of the splits gave false-discovery proportions at least as high at $q=0.2$), and all backbones err in the same direction because they share the calibration set: with some thirty stable calibration examples, the guarantee of any one set is weak. Guarantees that hold for the calibration set in hand, rather than on average over sets, are correspondingly conservative (post hoc). A training-conditional screen\cite{vovk2012conditional} that misses at most 10\% of stable candidates with probability 0.9 over calibration sets must discard only below the 1st smallest of the 32 stable stage-1 probabilities: it kept 90\%--98\% of the stable stage-3 candidates but passed on 21\%--64\% of all candidates, against 13\%--27\% for the marginal screen. Conformal selection with $p$-values valid conditionally on the calibration set\cite{bates2023testing} --- exact beta quantiles of the calibration order statistics with a Bonferroni correction over the first 100 ranks, or the Dvoretzky--Kiefer--Wolfowitz bound, each at probability 0.9 --- certified no candidate for any backbone at $q=0.1$ or 0.2: with 595 calibration labels, the smallest attainable $p$-value exceeds what the Benjamini--Hochberg procedure needs among 1,389 candidates.

\paragraph{How many labels the fix needs.} Post hoc, we recalibrated every registered single-pass forecast on random subsets of the stage-1 labels and evaluated it on stage~3 (200 subsets of each size; Extended Data Fig.~\ref{fig:shift}e). Refitting the intercept of the logit on 100, 200 and 400 labels gave median absolute errors in the stable fraction of 1.1--1.6, 0.6--1.2 and 0.3--0.9 points, against 1.0--2.0 without recalibration, and improved the log-loss for eight backbones from 200 labels on; with 25 labels the median error was 2.2--3.1 points, worse than leaving the WBM calibration alone for eight backbones, and with 50 labels 1.6--1.9 points, worse for six. Refitting the full Platt map behaved alike. At the stable fractions of substitution campaigns, recalibration pays from about two hundred random labels. Small samples need not do harm, however. Estimating the intercept with a normal prior of standard deviation 0.5 on the logit shift (the posterior mode), which falls back to the registered forecast when labels are few, left 20\%--42\% of the 25-label draws worse than no recalibration, against 58\%--83\% for the unregularised fit, with median absolute errors of 0.7--1.5 points (0.3--0.9 with 200 labels). A prediction-powered estimate of the pool's stable fraction\cite{angelopoulos2023prediction}, the mean forecast over the unlabelled stage-3 pool plus the mean residual of the labelled subset, is unbiased but no more precise with 25 labels (58\%--82\% of draws worse than no recalibration), because the variance of the residual $y-p$, not the bias, dominates at these sample sizes (Supplementary Table~\ref{tab:singleset}; post hoc).

\begin{table}[htbp]
\centering\footnotesize\setlength{\tabcolsep}{3pt}
\caption{\textbf{Recalibration with few labels and guarantees for one calibration set} (stage~3, post hoc). Registered calibration in the large (CITL, percentage points); median absolute CITL over 200 random subsets of $n$ stage-1 labels (in brackets: \% of subsets worse than no recalibration) for the refitted logit intercept (shift), the same with a normal prior of standard deviation 0.5 (shrunk) and the prediction-powered estimate (PPI); recall/share passed on (\%) of the conformal screen at $\alpha=0.1$ with the marginal guarantee and with a guarantee for this calibration set with probability 0.9 (PAC); candidates certified by conformal selection at $q=0.2$ with marginal and with calibration-conditional $p$-values.}
\label{tab:singleset}
\begin{tabular}{lrrrrrrrrrrr}
\toprule
 & & \multicolumn{3}{c}{$n=25$} & \multicolumn{3}{c}{$n=200$} & \multicolumn{2}{c}{Screen} & \multicolumn{2}{c}{Selection}\\
\cmidrule(lr){3-5}\cmidrule(lr){6-8}\cmidrule(lr){9-10}\cmidrule(lr){11-12}
Backbone & CITL & Shift & Shrunk & PPI & Shift & Shrunk & PPI & Marginal & PAC & Marginal & Conditional\\
\midrule
CHGNet & 1.5 & 2.4 (69) & 1.1 (24) & 2.6 (71) & 0.9 (30) & 0.5 (4) & 0.7 (17) & 93/24 & 96/33 & 31 & 0\\
MACE-MP-0 & 2.0 & 2.3 (58) & 1.5 (20) & 2.6 (58) & 1.2 (16) & 0.5 (0) & 0.8 (7) & 88/19 & 94/40 & 0 & 0\\
MatterSim-v1-5M & 1.2 & 3.1 (83) & 0.9 (25) & 2.8 (71) & 0.8 (31) & 0.6 (8) & 0.6 (17) & 95/20 & 96/38 & 63 & 0\\
MACE-MPA-0 & 1.0 & 2.4 (80) & 0.7 (26) & 2.5 (78) & 0.8 (46) & 0.3 (6) & 0.7 (38) & 91/17 & 93/21 & 65 & 0\\
ORB-v3 & 1.0 & 2.2 (74) & 0.8 (26) & 2.4 (71) & 0.7 (28) & 0.5 (12) & 0.6 (22) & 95/21 & 98/30 & 91 & 0\\
SevenNet-MF-ompa & 1.4 & 2.5 (72) & 1.1 (26) & 2.5 (65) & 0.6 (16) & 0.6 (6) & 0.5 (8) & 93/24 & 98/64 & 82 & 0\\
eSEN-30M-OAM & 1.7 & 2.4 (64) & 1.4 (23) & 2.3 (58) & 0.6 (10) & 0.8 (4) & 0.5 (5) & 90/13 & 95/23 & 90 & 0\\
UMA-s-1p1 & 1.1 & 2.6 (76) & 1.0 (42) & 3.0 (82) & 1.0 (46) & 0.9 (38) & 1.0 (42) & 89/27 & 90/31 & 54 & 0\\
\bottomrule
\end{tabular}
\end{table}

\paragraph{Deviations from the registered plan.} (i) The label pipeline was completed after the registration and before any prospective label was collected: the primary per-element shifts were fixed, and their digest recorded, two hours after the registration; the secondary shifts and the third control set followed. (ii) A stopping time for the last unfinished stage-1 calculation was recorded before any stage-1 label was collected; the calculation ended unconverged before that time, and the rule was not applied. (iii) Stage-1 calculations that failed or were interrupted while a fault of the MPI launcher serialised the runs sharing a node were restarted once, before any label was collected. (iv) Bootstrap intervals and AUROC were computed by the evaluation script fixed before the registration but are not part of the registered plan, nor is the five-bin implementation of the registered expected calibration error; for forecasts with tied probabilities its value depends on the order of ties by up to 0.01, and we report the registered evaluation. (v) The sensitivity analyses of the labels were added after the registration and before any stage-1 label was collected. (vi) All analyses of the shortlist's failures, including the diagnostic DFT calculations, are post hoc. (vii) In stage~3, calculations that crashed on one cluster before its stack limit was raised, or that ran with misplaced processor pinning, were reset and rerun before any stage-3 label was collected. (viii) The hull completed with Alexandria's phases leaves out each candidate's own counterparts; this refinement of the registered secondary analysis was decided after seeing the version without it on the stage-1 and stage-2 labels and before any stage-3 label. (ix) Clustered intervals and the deterministic expected calibration error were added after stage~2 and registered for stage~3; for stages~1 and~2 they are post hoc. (x) A first collection of the stage-3 labels, made when all but a few long calculations had finished, showed that 48 others had never run: the load-balancing script had moved them back to a cluster they had left earlier, where the claim of their first move still blocked them. They were computed with the same inputs and PAW files on free nodes and are included; the forecasts and decisions, frozen at registration, were not affected. (xi) For stage~3, the recalibration on subsets of the stage-1 labels and the resampling of calibration sets for the conformal procedures are post hoc. (xii) The stage-3 labels reported here were collected after all 1,400 calculations had finished, before the registered stopping time (1,389 labelled, 11 failed); two interim evaluations preceded this one (a check of the pipeline on 1,382 labels and an evaluation on 1,388 labels while one calculation was still running; no label changed since), and none changed a forecast or a decision. (xiii) The forecasts of eSEN-30M-OAM and UMA-s-1p1 for the stage-1 candidates were made after the stage-1 labels; results on stage~3 alone are reported for them. (xiv) The first stage-3 evaluation omitted the registered System-2 rule of the four original backbones that relax (a merging error in the evaluation script, since corrected). (xv) The registered secondary analysis 6(b), counterparts by protostructure label, was run after the first report of the stage-3 results. (xvi) The bridging set was registered after the stage-3 labels were known, to answer the question of the labels; its analysis and decision rule were fixed before any of its calculations, and the digests of all registration files were posted publicly (\url{https://gist.github.com/kpleo/69d0855be72bc7830025aec1dfbbc116}, 2026-10-05 05:35 UTC) before any of its labels was collected. The digests of stages 1--3 were recorded only by the authors before their labels. The registration files, their full digests and the decision log will be deposited with the data. (xvii) On one cluster the four bridging-set pack jobs ran every lane of a node on the same eight cores (a fix to the processor pinning made there on 4 October had been overwritten by a later synchronisation of the job scripts), so their calculations barely progressed; the jobs were cancelled, the 25 started calculations reset to their inputs and rerun with the corrected scripts, and on another cluster one node that could not start MPI was excluded and its calculations rerun (four that its cancelled job had left half-started were found, and rerun the same way, only once every other bridging calculation had finished), all before any bridging label was collected. (xviii) B2 was registered after the labels of stages 1--3 and before any of its calculations; its SCAN inputs were written in two further steps after a fault in the input writer (the PAW files and settings are those registered), and the 39 competing phases chosen by the registered rule were logged with their digest before any calculation ran; the digests of all B2 files were posted publicly (\url{https://gist.github.com/kpleo/fa7e8aec40b9006de808b3ffca38aae0}, 2026-10-05 08:57 UTC) before any B2 calculation started. On one cluster the disk quota was exhausted while B2 ran; the 135 calculations that failed for this reason or ended without a status from then on (including all 46 SCAN calculations there) were reset and rerun from inputs rewritten by the registered input writer and checked against the inputs as generated (a first rerun, started from custodian's input backups that the full disk had truncated, was stopped within minutes and none of its outputs is used); failures before the fault are reported as registered, and no B2 label had been collected. (xix) A first collection of the bridging-set labels, made while two calculations were still running, served to test the collection and analysis code; it found that the code compared WBM's full protostructure labels in secondary analysis (6) where the registration names the space group, which was corrected before the final labels. The registered analysis reported here uses the final labels. (xx) The B2 pack jobs reached the clusters' four-day wall-time limit before the registered stopping time, and the limit of a running job cannot be raised there; as planned and logged before any of their labels, the five calculations they left unfinished, all of Part~C, were continued from their last geometry with the registered inputs in new jobs, on 24 or 32 cores each instead of 8 (only the number of MPI processes differs; one competing phase that had finished its first relaxation, restarted by an error of the continuation script from its original structure, was restarted from its last geometry six minutes later), and all of them finished before the stopping time. (xxi) A preliminary collection made after every PBE calculation of B2 had finished was used only to test the analysis code and to draft the text; it showed that the code reported SCAN calculations that had not finished as failed and did not flag local hulls missing a competing phase, both corrected before the final collection.

\FloatBarrier
{\footnotesize\setlength{\tabcolsep}{4pt}
\begin{longtable}{llllrrrrr}
\caption{Prospective test, stage 1 (registered analysis): forecasts of the 397 randomly sampled test candidates against new DFT labels. CITL: forecast minus realised stable fraction in percentage points (bootstrap 95\% interval); value: realised net hits per 1,000 candidates at $c=0.5$. Registered forecasts are calibrated on WBM only; slice forecasts are recalibrated on the 198 candidates of the random calibration slice.}\label{tab:prospective}\\
\toprule
backbone & forecast & calibration & CITL (95\% CI) & ECE & log-loss & AUROC & value\\
\midrule
\endfirsthead
\caption[]{(continued)}\\
\toprule
backbone & forecast & calibration & CITL (95\% CI) & ECE & log-loss & AUROC & value\\
\midrule
\endhead
MACE-MP-0 & System 1 & registered & +2.1 (+0.5, +3.7) & 0.021 & 0.097 & 0.970 & +8\\
MACE-MP-0 & System 1 & slice, Platt & $-$1.3 ($-$3.0, +0.3) & 0.024 & 0.102 & 0.970 & +10\\
MACE-MP-0 & System 1 & slice, isotonic & $-$1.2 ($-$2.9, +0.5) & 0.013 & 0.111 & 0.941 & +6\\
MACE-MP-0 & System 2 & registered & +2.4 (+0.8, +4.2) & 0.024 & 0.095 & 0.979 & +5\\
MACE-MP-0 & System 2 & slice, Platt & $-$1.0 ($-$2.8, +0.6) & 0.024 & 0.096 & 0.979 & +5\\
MACE-MP-0 & System 2 & slice, isotonic & $-$0.9 ($-$2.7, +0.8) & 0.012 & 0.098 & 0.975 & +9\\
MACE-MP-0 & relaxed, isotonic & registered & +2.6 (+0.8, +4.5) & 0.026 & 0.143 & 0.970 & $-$3\\
MACE-MP-0 & relaxed, isotonic & slice, Platt & $-$0.9 ($-$2.8, +0.9) & 0.019 & 0.135 & 0.970 & +9\\
MACE-MP-0 & relaxed, isotonic & slice, isotonic & $-$1.1 ($-$2.9, +0.6) & 0.015 & 0.096 & 0.969 & +6\\
MACE-MP-0 & relax, $\hat E_\mathrm{hull}\le0$ & -- & -- & -- & -- & -- & $-$1\\
\addlinespace
CHGNet & System 1 & registered & +1.4 ($-$0.1, +2.8) & 0.014 & 0.085 & 0.980 & +14\\
CHGNet & System 1 & slice, Platt & $-$1.6 ($-$3.3, 0.0) & 0.023 & 0.094 & 0.980 & +10\\
CHGNet & System 1 & slice, isotonic & $-$1.3 ($-$3.1, +0.3) & 0.025 & 0.117 & 0.973 & +10\\
CHGNet & System 2 & registered & +2.1 (+0.6, +3.6) & 0.021 & 0.085 & 0.984 & +14\\
CHGNet & System 2 & slice, Platt & $-$1.3 ($-$2.9, +0.2) & 0.019 & 0.084 & 0.984 & +6\\
CHGNet & System 2 & slice, isotonic & $-$0.8 ($-$2.6, +1.0) & 0.008 & 0.133 & 0.969 & +5\\
CHGNet & relaxed, isotonic & registered & +1.7 (+0.2, +3.2) & 0.017 & 0.092 & 0.984 & +10\\
CHGNet & relaxed, isotonic & slice, Platt & $-$1.6 ($-$3.3, 0.0) & 0.027 & 0.094 & 0.984 & +11\\
CHGNet & relaxed, isotonic & slice, isotonic & $-$1.4 ($-$3.1, +0.1) & 0.014 & 0.081 & 0.982 & +11\\
CHGNet & relax, $\hat E_\mathrm{hull}\le0$ & -- & -- & -- & -- & -- & +6\\
\addlinespace
MACE-MPA-0 & System 1 & registered & +1.2 ($-$0.1, +2.5) & 0.012 & 0.072 & 0.982 & +19\\
MACE-MPA-0 & System 1 & slice, Platt & $-$0.5 ($-$1.9, +0.7) & 0.018 & 0.075 & 0.982 & +15\\
MACE-MPA-0 & System 1 & slice, isotonic & $-$0.3 ($-$1.9, +1.1) & 0.004 & 0.089 & 0.981 & +15\\
\addlinespace
MatterSim-v1-5M & System 1 & registered & +0.9 ($-$0.5, +2.2) & 0.009 & 0.076 & 0.969 & +18\\
MatterSim-v1-5M & System 1 & slice, Platt & $-$0.3 ($-$1.6, +1.0) & 0.008 & 0.074 & 0.969 & +19\\
MatterSim-v1-5M & System 1 & slice, isotonic & $-$0.1 ($-$1.6, +1.2) & 0.009 & 0.113 & 0.938 & +19\\
MatterSim-v1-5M & System 2 & registered & +0.5 ($-$0.8, +1.8) & 0.008 & 0.078 & 0.976 & +19\\
MatterSim-v1-5M & System 2 & slice, Platt & $-$1.0 ($-$2.4, +0.4) & 0.013 & 0.076 & 0.976 & +15\\
MatterSim-v1-5M & System 2 & slice, isotonic & $-$1.5 ($-$3.0, $-$0.2) & 0.015 & 0.108 & 0.928 & +11\\
MatterSim-v1-5M & relaxed, isotonic & registered & +0.8 ($-$0.5, +2.2) & 0.011 & 0.082 & 0.966 & +19\\
MatterSim-v1-5M & relaxed, isotonic & slice, Platt & $-$1.3 ($-$2.8, +0.1) & 0.016 & 0.081 & 0.966 & +13\\
MatterSim-v1-5M & relaxed, isotonic & slice, isotonic & $-$1.5 ($-$3.0, $-$0.2) & 0.016 & 0.114 & 0.927 & +13\\
MatterSim-v1-5M & relax, $\hat E_\mathrm{hull}\le0$ & -- & -- & -- & -- & -- & +19\\
\addlinespace
ORB-v3 & System 1 & registered & +0.8 ($-$0.5, +2.0) & 0.012 & 0.071 & 0.975 & +19\\
ORB-v3 & System 1 & slice, Platt & +0.1 ($-$1.2, +1.3) & 0.006 & 0.069 & 0.975 & +19\\
ORB-v3 & System 1 & slice, isotonic & $-$0.2 ($-$1.5, +1.0) & 0.008 & 0.104 & 0.941 & +19\\
ORB-v3 & System 2 & registered & $-$0.1 ($-$1.4, +1.0) & 0.004 & 0.047 & 0.995 & +21\\
ORB-v3 & System 2 & slice, Platt & $-$0.4 ($-$1.7, +0.7) & 0.005 & 0.048 & 0.995 & +21\\
ORB-v3 & System 2 & slice, isotonic & $-$0.2 ($-$1.5, +0.8) & 0.003 & 0.056 & 0.994 & +21\\
ORB-v3 & relaxed, isotonic & registered & +0.1 ($-$1.2, +1.3) & 0.002 & 0.053 & 0.994 & +21\\
ORB-v3 & relaxed, isotonic & slice, Platt & $-$1.0 ($-$2.3, +0.3) & 0.011 & 0.057 & 0.994 & +20\\
ORB-v3 & relaxed, isotonic & slice, isotonic & $-$1.1 ($-$2.6, +0.2) & 0.011 & 0.121 & 0.909 & +18\\
ORB-v3 & relax, $\hat E_\mathrm{hull}\le0$ & -- & -- & -- & -- & -- & +21\\
\addlinespace
SevenNet-MF-ompa & System 1 & registered & +1.5 (+0.3, +2.7) & 0.018 & 0.069 & 0.965 & +18\\
SevenNet-MF-ompa & System 1 & slice, Platt & +0.8 ($-$0.4, +2.0) & 0.011 & 0.067 & 0.965 & +20\\
SevenNet-MF-ompa & System 1 & slice, isotonic & +1.1 ($-$0.1, +2.3) & 0.016 & 0.091 & 0.961 & +20\\
\addlinespace
\bottomrule
\end{longtable}
\begin{longtable}{llrrrr}
\caption{Prospective test, stage 1 (registered analysis): realised net value per 1,000 test candidates, $1000\,\overline{(y-c)\,\mathbf{1}[\text{attempt}]}$, at the four registered attempt costs $c$, for the System-1 rule $p_1\ge c$ and the System-2 rule $p_2\ge c$ with the registered (WBM-calibrated) forecasts, and for relax-and-threshold (attempt if the backbone's relaxed, MP2020-corrected hull distance is at most zero; precision and recall in brackets).}\label{tab:prospvalue}\\
\toprule
backbone & rule & $c=0.1$ & $c=0.3$ & $c=0.5$ & $c=0.7$\\
\midrule
\endfirsthead
\caption[]{(continued)}\\
\toprule
backbone & rule & $c=0.1$ & $c=0.3$ & $c=0.5$ & $c=0.7$\\
\midrule
\endhead
MACE-MP-0 & System 1, $p_1\ge c$ & +40.3 & +20.9 & +7.6 & +2.8\\
MACE-MP-0 & System 2, $p_2\ge c$ & +44.3 & +21.7 & +5.0 & $-$1.5\\
MACE-MP-0 & relax-and-threshold (49\%, 91\%) & +42.1 & +20.4 & $-$1.3 & $-$22.9\\
\addlinespace
CHGNet & System 1, $p_1\ge c$ & +43.1 & +25.2 & +13.9 & +5.5\\
CHGNet & System 2, $p_2\ge c$ & +43.6 & +24.9 & +13.9 & +4.3\\
CHGNet & relax-and-threshold (57\%, 87\%) & +41.6 & +23.9 & +6.3 & $-$11.3\\
\addlinespace
MACE-MPA-0 & System 1, $p_1\ge c$ & +40.1 & +28.2 & +18.9 & +8.8\\
\addlinespace
MatterSim-v1-5M & System 1, $p_1\ge c$ & +38.0 & +28.0 & +17.6 & +10.3\\
MatterSim-v1-5M & System 2, $p_2\ge c$ & +42.1 & +30.0 & +18.9 & +5.8\\
MatterSim-v1-5M & relax-and-threshold (83\%, 83\%) & +42.1 & +30.5 & +18.9 & +7.3\\
\addlinespace
ORB-v3 & System 1, $p_1\ge c$ & +40.6 & +32.0 & +18.9 & +9.3\\
ORB-v3 & System 2, $p_2\ge c$ & +45.8 & +30.5 & +21.4 & +8.6\\
ORB-v3 & relax-and-threshold (90\%, 83\%) & +42.6 & +32.0 & +21.4 & +10.8\\
\addlinespace
SevenNet-MF-ompa & System 1, $p_1\ge c$ & +41.3 & +29.7 & +17.6 & +11.8\\
\addlinespace
\bottomrule
\end{longtable}
\begin{longtable}{lrrrrrr}
\caption{Prospective test, stage 1 (registered analysis): conformal screen and conformal selection with the 198 candidates of the random calibration slice (nine of them stable) as calibration set and the 397 test candidates (23 stable) as candidates, from the registered System-1 scores. Screen at level $\alpha$: fraction of test candidates kept and recall of the stable ones (the guarantee: recall at least $1-\alpha$ in expectation; with $n$ stable calibration candidates the screen can discard candidates only if $\alpha(n+1)\ge1$). Selection at false-discovery rate $q$: number of candidates certified and the realised false-discovery proportion.}\label{tab:prospconf}\\
\toprule
& \multicolumn{2}{c}{screen $\alpha=0.05$} & \multicolumn{2}{c}{screen $\alpha=0.1$} & selection $q=0.1$ & selection $q=0.2$\\
backbone & kept & recall & kept & recall & $n$ (FDP) & $n$ (FDP)\\
\midrule
\endfirsthead
\caption[]{(continued)}\\
\toprule
backbone & kept & recall & kept & recall & $n$ (FDP) & $n$ (FDP)\\
\midrule
\endhead
MACE-MP-0 & 100\% & 100\% & 20\% & 96\% & 0 & 0\\
CHGNet & 100\% & 100\% & 32\% & 100\% & 0 & 12 (0.08)\\
MACE-MPA-0 & 100\% & 100\% & 19\% & 100\% & 0 & 14 (0.07)\\
MatterSim-v1-5M & 100\% & 100\% & 17\% & 87\% & 0 & 15 (0.07)\\
ORB-v3 & 100\% & 100\% & 19\% & 91\% & 0 & 18 (0.11)\\
SevenNet-MF-ompa & 100\% & 100\% & 47\% & 96\% & 21 (0.14) & 23 (0.17)\\
\bottomrule
\end{longtable}
\begin{longtable}{llrrr}
\caption{Prospective test, stage 2: the shortlist a screening campaign would compute (the 100 candidates with the highest registered single-pass probability of ORB-v3 outside stage 1) yielded 98 stable structures. Expected yield $\pm$ binomial standard deviation for every forecast registered before the shortlist's DFT; the last two columns (not registered) give the probability of the realised yield or less, and of the realised yield or more, if the candidates' outcomes were independent with the forecast probabilities.}\label{tab:prospstage2}\\
\toprule
forecast & registered & expected yield & $P(\le\text{realised})$ & $P(\ge\text{realised})$\\
\midrule
\endfirsthead
\caption[]{(continued)}\\
\toprule
forecast & registered & expected yield & $P(\le\text{realised})$ & $P(\ge\text{realised})$\\
\midrule
\endhead
MACE-MP-0, System 1 & with the pool & 92.6 $\pm$ 2.2 & 1.00 & 0.008\\
MACE-MP-0, System 2 & with the pool & 92.2 $\pm$ 2.3 & 1.00 & 0.005\\
CHGNet, System 1 & with the pool & 91.7 $\pm$ 1.9 & 1.00 & $<$0.001\\
CHGNet, System 2 & with the pool & 91.2 $\pm$ 1.9 & 1.00 & $<$0.001\\
MACE-MPA-0, System 1 & with the pool & 99.1 $\pm$ 0.9 & 0.21 & 0.95\\
MatterSim-v1-5M, System 1 & with the pool & 97.1 $\pm$ 0.7 & 1.00 & 0.30\\
MatterSim-v1-5M, System 2 & with the pool & 98.6 $\pm$ 0.9 & 0.42 & 0.88\\
ORB-v3, System 1 & with the pool & 99.9 $\pm$ 0.3 & 0.005 & 1.00\\
ORB-v3, System 2 & with the pool & 99.5 $\pm$ 0.7 & 0.10 & 0.98\\
SevenNet-MF-ompa, System 1 & with the pool & 97.8 $\pm$ 0.5 & 0.99 & 0.81\\
ORB-v3, System 1, recalibrated (Platt) on stage 1 & after stage 1 & 99.8 $\pm$ 0.5 & 0.026 & 1.00\\
ORB-v3, System 1, recalibrated (isotonic) on stage 1 & after stage 1 & 100.0 $\pm$ 0.0 & $<$0.001 & 1.00\\
\bottomrule
\end{longtable}
}

\begin{landscape}\par\noindent\begin{minipage}{\linewidth}\captionsetup{type=table}\centering\small
\caption{The registered prospective test with the random extension (stage~3). All random labels: stages~1 and~3 pooled (1,984 candidates, 113 stable); stage~3: 1,389 candidates, 81 stable. Calibration in the large (CITL): forecast minus realised stable fraction in percentage points, with 95\% bootstrap intervals that resample parent prototypes. Registered: calibrated on WBM only; Platt: recalibrated on the stage-1 labels (frozen before stage~3); EM: label-shift re-estimate of the stable fraction; weighted: importance-weighted Platt map on the WBM slice. Screen: recall of the stable stage-3 candidates by the conformal screen at $\alpha=0.1$ (share of candidates kept); selection: candidates certified by conformal selection at $q=0.1$ (false-discovery proportion). $^\ast$Registered before any stage-3 label (addendum); their stage-1 forecasts were made after the stage-1 labels were known but by the same fixed protocol.}\label{tab:prospective3}
\begin{adjustbox}{max width=\linewidth}\begin{tabular}{lrrlrrllllrr}
\toprule
 & \multicolumn{5}{c}{all random labels, registered} & \multicolumn{4}{c}{stage 3: CITL} & \multicolumn{2}{c}{stage 3: conformal}\\
\cmidrule(lr){2-6}\cmidrule(lr){7-10}\cmidrule(lr){11-12}
backbone & stable & forecast & CITL & ECE & AUROC & registered & Platt & EM & weighted & screen & selection\\
\midrule
MACE-MP-0 & 113 & 157 & 2.2 [1.3, 3.3] & 2.3 & 0.931 & 2.0 [0.9, 3.2] & $-$0.9 [$-$2.0, 0.1] & $-$1.8 [$-$2.9, $-$0.8] & 0.9 [$-$0.2, 2.0] & 88\% (19\%) & 0 (--)\\
CHGNet & 113 & 146 & 1.7 [0.8, 2.6] & 1.8 & 0.945 & 1.5 [0.5, 2.6] & $-$0.7 [$-$1.8, 0.3] & $-$1.8 [$-$2.9, $-$0.8] & 1.0 [$-$0.1, 2.0] & 93\% (24\%) & 0 (--)\\
MACE-MPA-0 & 113 & 136 & 1.2 [0.4, 1.9] & 1.2 & 0.965 & 1.0 [0.1, 1.9] & $-$0.7 [$-$1.7, 0.1] & $-$1.7 [$-$2.6, $-$0.8] & 0.3 [$-$0.7, 1.2] & 91\% (17\%) & 0 (--)\\
MatterSim-v1-5M & 113 & 137 & 1.2 [0.5, 2.0] & 1.2 & 0.963 & 1.2 [0.4, 2.0] & $-$0.5 [$-$1.4, 0.3] & $-$1.4 [$-$2.3, $-$0.6] & 0.7 [$-$0.2, 1.5] & 95\% (20\%) & 0 (--)\\
ORB-v3 & 113 & 133 & 1.0 [0.3, 1.7] & 1.0 & 0.971 & 1.0 [0.1, 1.9] & 0.0 [$-$0.9, 0.8] & $-$1.1 [$-$2.0, $-$0.3] & 1.0 [0.1, 1.9] & 95\% (21\%) & 28 (0.07)\\
SevenNet-MF-ompa & 113 & 139 & 1.3 [0.7, 2.0] & 1.6 & 0.956 & 1.4 [0.5, 2.2] & $-$0.1 [$-$1.0, 0.7] & $-$1.1 [$-$2.0, $-$0.3] & 1.0 [0.1, 1.8] & 93\% (24\%) & 61 (0.18)\\
eSEN-30M-OAM$^{\ast}$ & 113 & 145 & 1.6 [0.9, 2.3] & 1.8 & 0.965 & 1.7 [0.8, 2.6] & 0.1 [$-$0.8, 1.0] & $-$0.7 [$-$1.6, 0.1] & 1.4 [0.5, 2.3] & 90\% (13\%) & 51 (0.14)\\
UMA-s-1p1$^{\ast}$ & 113 & 131 & 0.9 [0.1, 1.7] & 1.3 & 0.933 & 1.1 [0.2, 2.1] & 0.8 [$-$0.2, 1.8] & $-$2.5 [$-$3.5, $-$1.5] & 0.7 [$-$0.3, 1.7] & 89\% (27\%) & 50 (0.24)\\
\bottomrule
\end{tabular}\end{adjustbox}
\end{minipage}\par\bigskip
\end{landscape}

\end{document}